\documentclass[12pt]{article}

\usepackage{amsmath,amssymb,bm,epsf,epsfig,graphicx,dsfont}
\usepackage{dsfont,caption,subcaption}
\usepackage[section]{placeins}
\usepackage{multirow}

\input epsf.sty
\numberwithin{equation}{section}

\newcommand{\Tr}{\mathop{\rm Tr}\nolimits}
\newcommand{\re}{\mathop{\rm Re}\nolimits}
\newcommand{\im}{\mathop{\rm Im}\nolimits}

\def\bra#1{\langle #1 |}

\def\ket#1{|#1 \rangle}

\def\aver#1{\left\langle\, #1 \,\right\rangle}

\def\half{\frac{1}{2}}

\def \be {\begin{equation}}
\def \ee {\end{equation}}
\def \bea {\begin{eqnarray}}
\def \eea {\end{eqnarray}}
\def \bdm {\begin{displaymath}}
\def \edm {\end{displaymath}}

\def \zz {{\mathbb Z}}

\def \rr {{\mathbb R}}

\def \hh {{\cal H}}

\def \del {\partial}

\def \la {\langle}
\def \ra {\rangle}

\def \ps {\phantom{+}}

\def\hsp{\hspace{0pt}}
\def\W3J#1#2#3#4#5#6{\mbox{${\protect\begin{pmatrix} #1 &\hsp #2 &\hsp #3\\ #4 &\hsp #5 &\hsp #6 \end{pmatrix}}$}}

\begin{document}
\vskip 2.1cm

\centerline{\Large \bf  Universal Solutions in Open String Field Theory}
\vspace*{8.0ex}

\centerline{\large \rm Mat\v{e}j Kudrna$^{(a)}$\footnote{Email: {\tt matej.kudrna at email.cz}},
Martin Schnabl$^{(a)}$\footnote{Email: {\tt schnabl.martin at gmail.com}}
}
\vspace*{6.0ex}

\centerline{\large \it CEICO, Institute of Physics of the Czech Academy of Sciences}
\vspace*{1.0ex}

\centerline{\large \it Na Slovance 2, Prague 8, Czech Republic}

\vspace*{8.0ex}

\centerline{\bf Abstract}
\bigskip

An outstanding problem in open bosonic string field theory is the existence of nontrivial classical solutions in its universal sector. Such solutions independent of the underlying CFT would be relevant for any background, but aside of the celebrated tachyon vacuum, no well behaved solutions have been found so far. In this work we revisit the problem using the old fashioned level truncation technique, but armed with much more sophisticated techniques and greater computer power. In particular, using the homotopy continuation method we construct all solutions at level 6 (or 5) with twist symmetry imposed (or not), and improve the viable ones by Newton's method to levels 24--30. Surprisingly, a handful of solutions survive. One of them is tantalizingly close to the elusive double brane solution, although the fits to infinite level seem to invalidate this conclusion. A better behaved solution matches unexpectedly the properties of a ghost brane. This does not contradict anything, since the solution violates the reality condition of string field theory. Relaxing $SU(1,1)$ singlet condition two more exotic solutions with the rough characteristics of half brane and ghost half brane are found. For the tachyon vacuum, by explicit numerical computations up to level 30, we confirm the Gaiotto and Rastelli's prediction about the turning point of the tachyon potential minimum as a function of the level.

%In this work we attempt to systematically explore the space of all solutions of bosonic open string field in the universal sector. Such solutions are independent of the underlying CFT and are thus shared by any background. We search for possible solutions using the level truncation method. For the resulting systems of quadratic equations we apply a combination of Newton and linear homotopy continuation method. Gaiotto and Rastelli's prediction by  about the turning point of the critical value of the tachyon potential in its minimum as a function of the level is confirmed by explicit numerical computations up to level 30. Systematically scanning all solutions at level 6 (or 5) with twist symmetry imposed (or not), and improving the viable ones
%to levels 24--28 we find handful of  We also search for solutions without fixing the gauge and/or imposing twist symmetry. Potential candidates for the elusive double-brane solution are discussed.

 \vfill \eject

\baselineskip=16pt

\tableofcontents

\setcounter{footnote}{0}

\section{Introduction an summary}

While a lot of recent research in open string field theory is concerned with the exciting prospects of classifying possible conformal boundary conditions in a given CFT's, see e.g. \cite{KRS,EM}, string field theory received its great popularity around the turn of the millennium for its remarkable universal properties. In the pioneering work \cite{SZ} a level truncation method was used to find an approximate classical solution describing the endpoint of the open string tachyon condensation. This solution, independent of the details of the underlying BCFT, offered a strong evidence in favor of Sen's conjectures \cite{SenCon} stating that the open string tachyon triggers D-brane decay with all of its logical consequences. In particular the string field energy difference between the initial and final vacuum configuration should exactly match the D-brane energy, and that at the minimum no perturbative excitations should survive. Shortly after, a question arose \cite{ET-private} whether solutions describing the opposite process exist, but no solutions in level truncation were found. Early arguments \cite{Sch-B} suggested that it should indeed be possible, but it was not until the first exact analytic solution was found \cite{Sch-Analytic} when it became feasible. In a series of papers \cite{MS1,MS2,HK1,HK2,HK3,H-BV,MNT}\footnote{See also \cite{EM} in which however the solution does not stay in the universal sector.} this issue was studied for a class of ansatzes and indeed the correct quantization of energy was established. However finding a proper analytic multibrane solution free of any ambiguities or anomalies still remains a challenge.

With the goal of settling many of the outstanding issues we return in this work to the early attempts armed with more sophisticated techniques and more computational power.
To put it in perspective, we have managed to repeat the impressive early computation by Gaiotto and Rastelli \cite{GaiottoRastelli} of the tachyon vacuum to level 18 in Siegel gauge now within just a few seconds and further improved it to level 30. This has allowed us to prove their prediction that the turning point of the critical value of the tachyon potential in its minimum as a function of the level is reached at level 28. This particular computation to level 30 involved only 21 times more fields compared to Gaiotto and Rastelli's results at level 18, since we managed to impose the SU(1,1) singlet condition \cite{ZwiebachSU11} on the string field throughout our computations. This alone has reduced the required number of fields at level 30 by more than a half. The number of contributing cubic vertices grows however with the third power of the number of fields, so we still had to tackle about $10\,000$ times more vertices. To find the solution in a reasonable time we had to resort to parallel computing on a multicore machine capable of storing all the vertices (in their matter-ghost factorized form) in the computer memory.

More interestingly, the newly gained power allows us to start analyzing the space of {\em all} possible solutions of open string field theory truncated to a given level. This is by no means easy. For instance, the number of twist-even SU(1,1)-singlet solutions at level 30 in the universal sector is estimated to be $2^{85\,604} \approx 10^{26\,000}$.  One can however find as many solutions as one wishes (and at some lower levels possibly all of them) using the homotopy continuation method which we have successfully implemented. Essentially the idea of this method is to start with a simple system of equations of the same order in the same variables, such as $t_i (\rho_i - t_i) = 0$ for some $\rho_i \ne 0$ where one knows all the solutions, and then deform continuously---in practise in small steps---the initial system into the one of interest while tracking all the individual solutions as they evolve under this deformation.
%\footnote{In reality this number will be just slightly smaller due to some accidental coincidences in the ghost sector discovered in \cite{KRS}.For instance the number of missing solutions at level 6 is at most 1294 which is less than one per mille of the generically expected number of $2^{21}$ solutions at this level.}

\begin{figure}
\centering
\includegraphics[width=16cm]{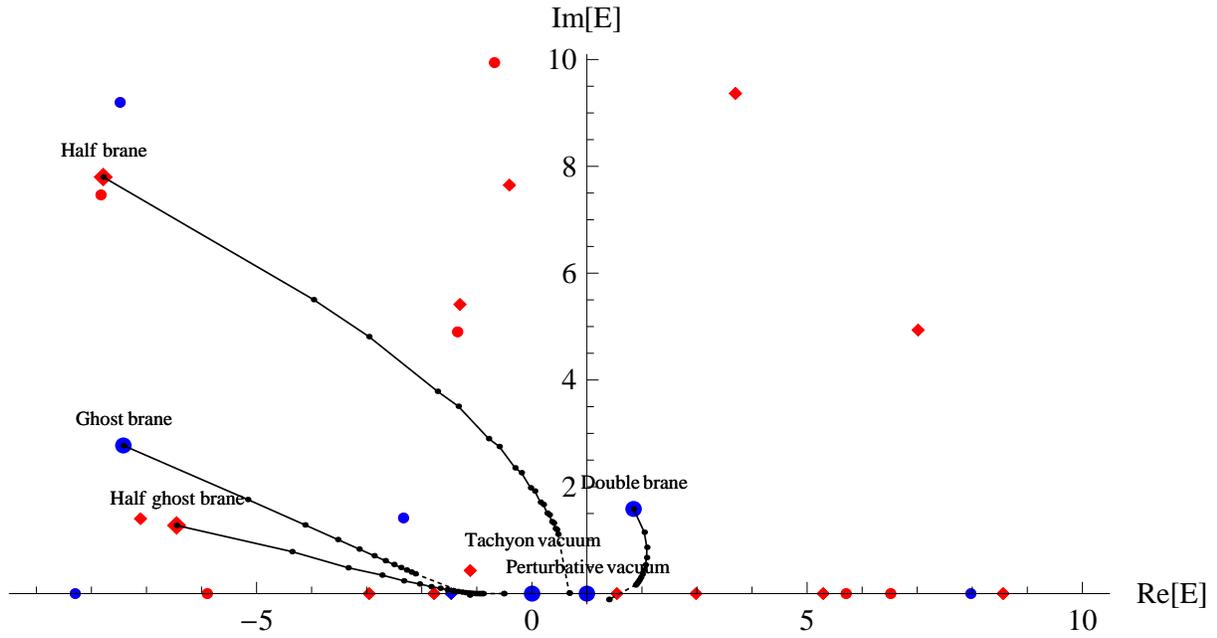}
\caption{Twist even solutions at level 6 and twist non-even solutions at level 5 represented as dots or diamonds respectively in the energy complex plane. SU(1,1) singlet solutions are represented by blue color, non-singlet solutions by red. By black line and dots we show the improvement of the interesting solutions to higher levels, the dashed part of the line is infinite level extrapolation. }
\label{Fig:intro}
\end{figure}

In this work we have used the linear homotopy method to construct all complex or real solutions at level 6 with twist symmetry imposed, and all solutions at level 5 without such constraint.
We then focused on the viable ones with the absolute value of energy $|E|<50$. Taking these solution as initial seeds and applying Newton's method successively we studied whether they persist at higher levels. While in most cases convergence was lost, for few we were able to reach levels 24--30. The figure \ref{Fig:intro} shows some of our starting points.

Our most surprising results are two new universal solutions: one which can be interpreted as a ghost-brane with minus the tension of the usual D-brane and another solution which at finite levels is quite close to the conjectured double-brane with twice the tension of the usual D-brane. Both solutions were found from two complex starting points at level 4 and further improved to level 28 and extrapolated to the infinite level. Normally one would not expect stacks of negative number of branes to appear in string theory, in fact the ghost-brane solution turns out to have nontrivial imaginary part which does not seem to vanish in the limit of infinite level. Therefore this solution does not represent a valid classical OSFT vacuum, but may play a role in the quantum theory as a complex saddle point.

The status of the would-be double-brane is different. Given our results from level truncation, the solution will probably become real at very high level, but unfortunately it does not seem to obey even asymptotically the out-of-Siegel-gauge equations. However, around level 20, the real part of the energy of the solution is so close to the double-brane value of $+2$ that it leads us to believe that perhaps somewhere nearby in the coefficient space, there is a proper real solution which would obey the full set of equations of motion.
%necessary extra out-of-Siegel-gauge equations.
Such a conjectured nearby double brane solution may not necessarily exist in Siegel gauge, or its domain of attraction might be too small, so that we were not able to find it. While one could speculate about other interpretations of the solution (one-and-half brane or a Gribov copy of the perturbative vacuum would both be roughly consistent with the infinite level fits of the energy), all such attempts are equally plagued by asymptotically non-vanishing violation out-of-Siegel-gauge equations.

Relaxing the SU(1,1) singlet condition we were able to find two more convergent twist non-even solutions\footnote{Surprisingly these solutions turn out to be even under a modified twist symmetry $\Omega (-1)^J$, see section~\ref{sec:OSFT}.} for which the energies computed from the action and the Ellwood invariant are comparable, and the violation of the out-of-Siegel-gauge equations is moderate. We have nicknamed these solutions as "ghost half brane" and "half-brane" respectively. These rather peculiar objects are from level truncation point of view
as good as the "double brane" discussed above. A priori we did not expect existence of such solutions, although Erler had constructed puzzling half brane analytic solutions  \cite{Erler-exotic} in the so called modified superstring field theory, which however were widely believed to be artifacts of this theory. We are now tempted to interpret these solutions as meronic D-branes in analogy with gauge theory.

%Their physical significance, if any, is not clear. While one could simply dismiss these solutions, it must be stressed, that from level truncation point of view these are quite as good as the "double brane" discussed above.

\begin{table}\nonumber
\centering\footnotesize
{\renewcommand{\arraystretch}{1.2} %<- modify value to suit your needs
\begin{tabular}{|l|llllll|}\hline
\multirow{2}{6em}{Solution}          & \multirow{2}{6em}{\ps Energy$^{L=\infty}$} & \multirow{2}{6em}{\ps $E_0^{L=\infty} $} & \multirow{2}{4em}{\ps $\Delta_S^{L=\infty}$} & \multirow{2}{6em}{Reality}   & \multirow{2}{2em}{Twist even} & \multirow{2}{3em}{SU(1,1) singlet} \\
&&&&&& \\\hline
tachyon vacuum     & $   -8\times 10^{-6} $ & $\ps 0.0004     $ & $-7\times 10^{-6} $ & yes               & yes & yes \\
single brane       & $\ps 1               $ & $\ps 1          $ & $\ps 0            $ & yes               & yes & yes \\
"ghost brane"      & $   -1.13+0.024i     $ & $   -1.01+0.11i $ & $\ps 0.08         $ & no                & yes & yes \\
"double brane"     & $\ps 1.40+0.11i      $ & $\ps 1.23+0.04i $ & $\ps 0.20         $ & possibly$^*$      & yes & yes \\
"half ghost brane" & $   -0.51            $ & $   -0.66       $ & $\ps 0.17         $ & pseudoreal$^{**}$ & no  & no  \\
"half brane"       & $\ps 0.68-0.01i      $ & $\ps 0.54+0.1i  $ & $\ps 0.23         $ & no                & no  & no  \\\hline
\multicolumn{4}{c}{} & \multicolumn{1}{c}{$^*$ as $L\to\infty$}  &\multicolumn{2}{c}{}\\
\multicolumn{4}{c}{} & \multicolumn{1}{c}{$^{**}$ for $L\ge 22$} &\multicolumn{2}{c}{}\\
\end{tabular}
}
\caption{Summary of all viable solutions found from starting points at level 6 for twist even or level 5 for non-even. The second and third column give the fit to infinite level of energy computed from the action and the Ellwood invariant respectively. $\Delta_S^{L=\infty} $ gives a simple measure of violation of out-of-Siegel-gauge equations of motion. }
\label{tab:solutions}
\end{table}

This paper is organized as follows. We start by setting up our notation and presenting the basic strategies for solving string field theory in level truncation. In particular we explain how to efficiently work with the SU(1,1) singlet sector of the ghost BCFT which allows for substantial savings in computer resources.
%compute the string vertex in the SU(1,1)-singlet basis and review ....  and numerical techniques in detail. {\bf [I have put the numerics in appendix]}
Then in Section \ref{sec:Siegel} we present the computation of the tachyon vacuum in Siegel gauge up to level 30 and discuss the multitude of other solutions of the equations of motion that are numerically stable in level truncation. Among them are candidates for the double-brane and ghost-brane (i.e. "minus-one brane") as well as half-brane and ghost-half-brane in the nonsinglet sector. Their relevance for physics needs to be further established.

%possibility of finding other solutions such as the double-brane or ghost D-brane (i.e. "minus-one brane") solutions. We present two complex solution, which are stable under level truncation, but we are not sure whether they are physical and what is their interpretation. {\bf Add non-singlet solutions.}

In Section \ref{sec:noGF}, we also look for solutions without fixing their gauge. The exact degeneracy of the equations of motion is broken by level truncation, so still only discrete solutions exist, nevertheless the approximate gauge symmetry leads to enormous proliferation of solutions corresponding to the same discretized gauge orbit. A rather unwelcome consequence is the extreme instability of the Newton's method. Especially passing from one level to the next one, the solution often changes dramatically and the result depends on how exactly the step is performed. We can partially circumvent the problem by using the homotopy continuation method to find a family of solutions at higher level related to the one at lower level. We display our results as points in the complex energy plane. Without gauge fixing there is essentially no notion of stability under level truncation, so the only clue about the meaningfulness of our solutions is how the two ways of computing their energy---one from the action and another from the so called Ellwood invariant---match each other. The full complex plane ends up covered by some solutions, but the good solutions tend to cluster around the perturbative and tachyon vacua. Whether there exist other fundamentally distinct solutions in the universal sector remains unfortunately unclear.

%Unfortunately we find that the usual level truncation scheme fails, so we can search for solutions only using the homotopy continuation method. We can clearly see clusters of solutions around perturbative and tachyon vacuum, which correspond to approximate gauge transformations. The existence of other solutions remains unclear.

\section{String field theory in the universal sector}
\label{sec:OSFT}

Small fluctuations of D-brane systems are described by the attached open string strings governed by a boundary conformal field theory (BCFT). This BCFT in general is given by the tensor product BCFT$^m\otimes$BCFT$^{gh}$ of matter BCFT$^m$ with $c^m=26$ and reparametrization ghost BCFT$^{gh}$ with $c^{gh}=-26$. To study possibly large deformations such as tachyon condensation (i.e. D-brane decay triggered by some open string tachyon)  one  can conveniently use open string field theory (OSFT), which in the bosonic case is given by the Witten's action
\be\label{action}
S = - \frac{1}{g^2} \Tr \int \left( \frac{1}{2}\Psi * Q\Psi + \frac{1}{3} \Psi*\Psi*\Psi \right).
\ee
The string field $\Psi$ is a generic element of a given BCFT in the case of a single D-brane, or more generally an element of a collection of bimodules $\bigoplus_{a,b} \hh^{(a,b)}$, where $a$ and $b$ label the two boundary conditions on the two ends of the open string. Since in this work we are interested in maximally universal solutions, which are relevant for any open string background, our string field will take values in the Verma module of identity of a single BCFT$^m$ (tensored with BCFT$^{gh}$).

The string field in the universal subsector takes the general form
\be
\label{SFbc}
\Psi = \sum_{I,N,M} t_{I,N,M} L_{-I}^{m} b_{-N} c_{-M} \ket{0},
\ee
where the sum is over multi-indices $I, N, M$ defined as $I=\{\ldots, i_3, i_2, i_1\}$ with $i_{k+1} \ge i_k \ge 2, \forall k$, and $L_{-I} = \ldots L_{-i_2} L_{-i_1}$, and analogously for $b_{-N}$ and $c_{-M}$ with the only modification $n_{k+1} > n_k \ge 2$ and $m_{k+1} > m_k \ge -1, \forall k$ for the $b$ and $c$ ghosts respectively. The vacuum $\ket{0}$ is taken to be the $SL(2,\rr)$-invariant vacuum of the total BCFT. For classical solutions the total ghost number of the string field should equal to one, and hence the number of $c$ ghosts should be higher by one than the number of $b$ ghosts. For practical computations, however, other basis choices in the ghost sector may be more suitable. We shall discuss few other options in the remainder of this section.

The sum $|I|=\sum i_k$ is the level of a state in the matter sector, while in the ghost sector we define the level as $|N|+|M|+1$, so that the level of lowest lying state $c_1\ket{0}$ is zero. The total level of a state is given by $|I|+|N|+|M|+1$ and is equal to the eigenvalue of $L_0 + 1$. Restricting the string field only to states up to total level $L$ and solving the corresponding truncated equations of motion is at the heart of the {\em level truncation} method of string field theory.

A fundamental property of the OSFT action (\ref{action}) is its enormous gauge symmetry given by $\delta\Psi = Q \Lambda + \Psi*\Lambda - \Lambda*\Psi$, for any ghost number zero string field $\Lambda$. This symmetry is however broken by truncating to a finite level. Therefore solutions to the truncated classical equations of motion do not have (at least in the universal sector) continuous degeneracy corresponding to gauge orbits, but they appear as isolated discrete points. In principle, there are thus two possibilities to study possible classical solutions in level truncation: we may choose to fix a gauge in order to reduce a number of unknown field values, or we may choose to proceed without gauge fixing.  The former approach is much more efficient from the numerical standpoint, but there is a risk that some solutions might be missed \cite{ET}. In the latter approach, the correspondence between solutions at various levels is practically lost, and the method produces far more spurious solutions as we will show in section \ref{sec:noGF}.

In the universal sector---and often more generally when the underlying worldsheet theory is parity invariant---there is another symmetry, which does survive level truncation.
This so called {\em twist symmetry} can be defined in the universal sector simply as
\be
\Omega \Psi = (-1)^{L_0+1} \Psi
\ee
when acting on states of any ghost number. It commutes with $Q_B$ and obeys
\bea
\aver{\Omega \Psi_A , \Omega \Psi_B } &=& \aver{\Psi_A , \Psi_B },
\\
\Omega \left( \Psi_A * \Psi_B \right) &=& (-1)^{AB+1} (\Omega \Psi_B) * (\Omega \Psi_A),
\eea
where the sign is given by the Grassmann parity of the two fields. These conditions imply
\be\label{OmegaABC}
\aver{\Omega \Psi_A, \Omega \Psi_B * \Omega \Psi_C} = (-1)^{BC+1} \aver{\Psi_A, \Psi_C * \Psi_B}
\ee
and guarantee that the action (\ref{action}) is invariant under the twist symmetry. The equations of motion are invariant too, even when truncated to finite level. For any classical solution its twist conjugate is also a solution of the equations of motion. Particularly nice, especially from a numerical standpoint are twist self-conjugate, i.e. twist even solutions.

From a computational point of view an important consequence of the twist property (\ref{OmegaABC}) and the cyclic symmetry
\be
\aver{\Psi_A, \Psi_B * \Psi_C} = \aver{\Psi_B, \Psi_C * \Psi_A}
\ee
 is that the three-vertex $\aver{\Psi_A, \Psi_B * \Psi_C}$ need not be computed for all six permutations of $\Psi_A, \Psi_B, \Psi_C$, but only for one canonical ordering. This fact reduces the computation time and memory requirements by up to a factor of $1/6$.

\subsection{String field theory in Siegel gauge --- SU(1,1) symmetry}

%\subsubsection{SU(1,1) symmetry}

The most popular gauge fixing for numerical computations in OSFT is the Siegel gauge $b_0 \Psi = 0$. In this gauge there is an extra continuous symmetry $SU(1,1)$ generated by
\be
J_3 = \frac{1}{2}\sum_{n=1}^\infty \left(c_{-n} b_{n}-b_{-n} c_{n}\right),
\qquad J_{+} = \sum_{n=1}^\infty n c_{-n} c_{n},
\qquad J_{-} = \sum_{n=1}^\infty \frac{1}{n} b_{-n} b_{n},
\ee
which commute both with $b_0$ and $L_0$ operators.
Therefore all states can be decomposed into irreducible representations of the SU(1,1) group. The representations are clearly finite-dimensional because there is only a limited number of states at given level. The representations are labeled by a half-integer spin $j$ and its projection $m$, which is given by the $J_3$ eigenvalue. The $J_3$ generator is related to the zero mode of the ghost current by $j_0^{gh}=2J_3+c_0 b_0+1$ and therefore in Siegel gauge we get $g=2m+1$.

There is a very convenient basis in the ghost sector generated by the 'twisted' Virasoro generators introduced in \cite{GRSZ} and the corresponding primaries. The 'twisted' Virasoro generators are given by
\begin{equation}\label{twisted Virasoro}
L'^{gh}_n=L^{gh}_n + n j^{gh}_n + \delta_{n,0} = \sum_{m=-\infty}^{\infty}(n-m):\!b_m c_{n-m}\!:,
\end{equation}
where $j^{gh}_n$ are modes of the ghost current $j^{gh}=-:\!b c\!:$. The operators $L'^{gh}_{n}$ form a Virasoro algebra with central charge $c'^{gh}=-2$ and commute both with $b_0$ and $J_{3,\pm}$. The primary fields $b$ and $c$ acquire new conformal weights of $1$ and $0$ respectively. Interesting results about this theory have been found already in \cite{Kausch}.

To construct the twisted Virasoro primaries note that the SU(1,1) highest weight states
\begin{equation}
\ket{j,j} \equiv c_{-2j}\dots c_{-1}c_1 \ket{0}
\end{equation}
are also primaries of weight $h_j' = (2j+1)j$ with respect to the 'twisted' Virasoro generators by virtue of the explicit form (\ref{twisted Virasoro}). Notice that odd ghost numbers correspond to integer spins and even ghost numbers to half-integer spins.
By the repeated action of $J_-$ and convenient choice of normalization one obtains the whole orthonormal basis of the twisted Virasoro primaries (of the same weight $h_j'$)\footnote{Up to an overall normalization, they can be written explicitly in terms of antisymmetrized products of $j+m+1$ modes of $\partial c$ and $j-m$ modes of $b$ with mode numbers $-1, -2, \ldots, -2j$ acting on $c_1 \ket{0}$. }
\begin{equation}\label{spin primary}
\ket{j,m} \equiv N_{j,m} (J_-)^{j-m} \ket{j,j},
\end{equation}
where $N_{j,m}=\prod_{k=m+1}^j \left(j(j+1)-k(k-1)\right)^{-\frac{1}{2}}$.
These states obey the standard SU(2) recursion relations
\be
J_{\pm} \ket{j,m} = \sqrt{(j \mp m) (j \pm m +1)} \ket{j,m \pm 1},
\ee
but in our context they are normalized with respect to the BPZ inner product so that
\be
\bra{j,m} c_0 \ket{j,-m} = (-1)^{j-m}.
\ee

A generic universal-sector string field in the Siegel-gauge is therefore given by
\begin{equation}\label{PsiLLtw}
\Psi = \sum_{K,L,j,m} t_{K,L,j,m} L_{-K}^{m} L_{-L}'^{gh} \ket{j,m},
\end{equation}
where the action of the 'twisted' Virasoros does not change the SU(1,1) representation of the state. Because the classical string field should have ghost number one, the sum over $j$ runs over integers only.  As we will discuss more in the following subsection, the Verma modules constructed using $L_{-n}'^{gh} $ over the $\ket{j,m}$ primaries contain null vectors. Eliminating these, we provide a rigorous counting argument showing the equivalence of this basis to the one constructed from the $b$ and $c$ ghosts.

In \cite{ZwiebachSU11} Zwiebach observed that the Witten's three vertex restricted to Siegel gauge obeys
\be\label{Vsym}
\bra{V_3} \left( J_{\pm,3}^{(1)}+ J_{\pm,3}^{(2)}+ J_{\pm,3}^{(3)} \right) = 0,
\ee
which is equivalent to the property that $J_{\pm,3}$ behave as derivatives of star product of Siegel gauge states projected back to the Siegel gauge
\begin{equation}
J_{\pm,3} c_0 b_0 (\Psi_1\ast \Psi_2)= c_0 b_0 (J_{\pm,3}\Psi_1\ast \Psi_2)+c_0 b_0 (\Psi_1\ast J_{\pm,3} \Psi_2).
\end{equation}
By following the usual steps, we can derive from (\ref{Vsym}) the Wigner-Eckart theorem for the ghost 3-vertex
%\begin{equation}
%\la V_3|L'^{gh}_{-I_1}|j_1,m_1\ra L'^{gh}_{-I_2}|j_2,m_2\ra L'^{gh}_{-I_3}|j_3,m_3\ra=
%\la j_1 m_1, j_2 m_2 |j_3 -m_3\ra C(j_1,j_2,j_3,I_1,I_2,I_3),
%\end{equation}
\be
\la V_3|L'^{gh}_{-I_1}|j_1,m_1\ra L'^{gh}_{-I_2}|j_2,m_2\ra L'^{gh}_{-I_3}|j_3,m_3\ra= \W3J{j_1}{j_2}{j_3}{m_1}{m_2}{m_3} C(j_1,j_2,j_3,I_1,I_2,I_3),
\ee
where the first factor is the familiar SU(2) 3-j symbol, and $C(j_1,j_2,j_3,I_1,I_2,I_3)$ is the $m$-independent {\em reduced} vertex.
The 3-j symbol nicely captures the cyclic property of the Witten vertex
\be
\W3J{j_1}{j_2}{j_3}{m_1}{m_2}{m_3}  = \W3J{j_2}{j_3}{j_1}{m_2}{m_3}{m_1}  = \W3J{j_3}{j_1}{j_2}{m_3}{m_1}{m_2},
\ee
so that the reduced vertex
\be
C(j_1,j_2,j_3,I_1,I_2,I_3) = C(j_2,j_3,j_1,I_2,I_3,I_1)= C(j_3,j_1,j_2,I_3,I_1,I_2)
\ee
is also cyclic.

Further constraints arise from considering the twist symmetry of the Witten vertex.
Within the ghost sector\footnote{Note however, that twist symmetry defined to act purely on ghost (or matter) degrees of freedom is not a symmetry of the OSFT action, not even when restricted to the Siegel gauge.} the twist operator $\Omega$ acts as
\be
\Omega L'^{gh}_{-I} \ket{j,m} = (-1)^{|I|+h_j'} L'^{gh}_{-I} \ket{j,m},
\ee
which together with another well known property of the 3-j symbol
\be
\W3J{j_1}{j_2}{j_3}{m_1}{m_2}{m_3}  = (-1)^{j_1+j_2+j_3} \W3J{j_2}{j_1}{j_3}{m_2}{m_1}{m_3}
\ee
implies
\be
C(j_1,j_2,j_3,I_1,I_2,I_3) = (-1)^{|I_1| + |I_2| +|I_3|} C(j_2,j_1,j_3,I_2,I_1,I_3).
\ee
Showing this requires an identity
\be
(-1)^{j_1+j_2+j_3} = (-1)^{(2j_1+1)j_1+(2j_2+1)j_2+(2j_3+1)j_3+ (2j_1+1)(2j_2+1)+1}
\ee
which holds trivially for integer spins, and slightly less obviously also when two of the spins are half integer. The last two terms in the exponent on the right hand side account for the Grassmann sign from commuting two string fields with ghost numbers $2m_1+1$ and $2m_2+1$ (see (\ref{OmegaABC}) and note that $2j+1 \equiv 2m+1 \mod 2$).

The well known "time-reversal" symmetry of the 3-j symbol
\be
\W3J{j_1}{j_2}{j_3}{-m_1}{-m_2}{-m_3} = (-1)^{j_1+j_2+j_3}  \W3J{j_1}{j_2}{j_3}{m_1}{m_2}{m_3}
\ee
implies an unexpected $\zz_4$ symmetry of the Witten's string field theory in Siegel gauge \cite{HataShinohara,ZwiebachSU11}
\be
(-1)^J: \quad L'^{gh}_{-I}\ket{j,m} \to (-1)^j L'^{gh}_{-I}\ket{j,-m}
\ee
and in particular of the ghost vertex
\bea
&& \la V_3|L'^{gh}_{-I_1}|j_1,-m_1\ra L'^{gh}_{-I_2}|j_2,-m_2\ra L'^{gh}_{-I_3}|j_3,-m_3\ra  =
\nonumber\\
&& \qquad\qquad =(-1)^{j_1+j_2+j_3} \la V_3|L'^{gh}_{-I_1}|j_1,m_1\ra L'^{gh}_{-I_2}|j_2,m_2\ra L'^{gh}_{-I_3}|j_3,m_3\ra.
\eea
For classical solutions at ghost number one, which corresponds to $m_i=0$, the interaction vertex is thus zero unless the sum of all three spins is even. Classical solutions found in Siegel gauge come either in pairs related by the action $(-1)^J$, or they are self-conjugate, i.e. they contain only even spins $j=0,2,...$. A fully generic solution of the equations of motion will have thus four-fold degeneracy in energy given by $(-1)^J$ and the twist symmetry $\Omega$. Together with complex conjugation symmetry, the absolute value of the energy has an eight-fold degeneracy.

Up to a handful of interesting exceptions, most of the well behaved string field theory solutions in Siegel gauge that we found in this paper obey also the $SU(1,1)$ singlet condition $J_{+} \Psi =J_{-} \Psi = J_{3} \Psi =0$.\footnote{Interestingly, at ghost number one the condition $J_{+} \Psi =0$ alone guarantees both the Siegel gauge condition $b_0 \Psi=0$ as well as the other relations $ J_{-} \Psi= J_3 \Psi=0$.
The last statement $J_3 \Psi=0$ follows trivially from the ghost-number restriction. To prove the rest observe that $\left(J_{-}\right)^N \Psi =0$ for sufficiently large $N$ for any fixed level. Acting repeatedly with $J_+$ the claim follows.}
While in the non-singlet case it is technically easiest to use the basis (\ref{SFbc}), for the singlet sector it pays off significantly to use the basis (\ref{PsiLLtw}) restricted to $j=0$.
The null states can be eliminated simply by omitting the $-1$ from the multiindices $-I$ and $-J$ in
\begin{equation}
\label{SFtw}
\Psi = \sum_{I,J} t_{I,J} L_{-I}^{m} L_{-J}'^{gh} c_{1} \ket{0},
\end{equation}
since $L'^{gh}_{-1}c_1 \ket{0} = 0$. The universal matter and singlet ghost sectors have thus formally the same structure \cite{GaiottoRastelli}. This basis will make the computation of the ghost three vertex more technically complicated --- fortunately not more time consuming --- the pay off being the reduced number of string field components entering the equations.

%In section \ref{sec:noGF} we shall analyze classical solutions in level truncation obtained without imposing the Siegel gauge condition. In this case we can use the basis generated by the modes of $b$ and $c$ ghosts as in (\ref{SFbc}). Alternatively, at ghost number one, the basis can be equivalently generated by ghost Virasoro generators $L^{gh}_{-n}$ with $n\geq1$ acting on $c_1 \ket{0}$. The Virasoro basis is more convenient, since a simpler algorithm can be used to compute cubic vertices.\footnote{Alternative approach would be to use the basis made out of ghost current modes $j^{gh}_{-n}$. At ghost number one these two approaches are fully equivalent, but at ghost numbers zero and two {\bf !!! incorrect !!!} the ghost current approach would be better, since for example $L_{-1}^{gh} \ket{0}$ is a null state.}

%Finally we mention that in all the used Hilbert spaces we assume that the states are canonically ordered. The precise form of the ordering is not important, but it must include ordering by level. We will label the states just by their number $|i\ra$ with respect to this ordering instead of the multiindex.

\subsection{Characters and state counting}
\label{s-Characters}
%\subsubsection{Characters and state counting}

In order to estimate the complexity and computer requirements for our calculations let us  count the states level by level in the various Hilbert spaces used.
For the case of the Siegel gauge we are also going to demonstrate the equivalence of the conventional description of the ghost CFT Hilbert space using the modes of $b$ and $c$ ghosts, with an alternative picture given by a sum over representations labeled by $j$ and $m$ of the $c=-2$ twisted Virasoro theory.

The multiplicity of states at different levels in any given CFT  is encoded in the modular characters
\begin{equation}
\chi^\hh(q)=q^{-\frac{c}{24}} {\Tr}_{\hh} q^{L_0}.
\end{equation}
In this work we are interested only in OSFT in the universal sector, so that only the identity Verma module of the $c=26$  matter bosonic worldsheet CFT appears. Here the number of states is given by the generating function
\be
{\rm ch}^{\mathrm{matter}}(q) = {\Tr}_{\hh^{\rm matter}} q^{L_0} =\prod_{n=2}^\infty\frac{1}{1-q^n}.
\ee
For the ghost sector the situation is a bit more interesting. Let us introduce the ghost number counting variable $y$, then the generating function for the ghost CFT is given by
\be
{\rm ch}^{\rm ghost}(q, y) = {\rm{Tr}}_{\hh^{\rm ghost}} q^{L_0 +1} y^{j_0^{gh}} = \sum_{g=-\infty}^\infty {\rm ch}_g^{\rm ghost}(q) y^g.
\ee
Notice the convenient shift of $L_0$ by unity which guarantees that the power of $q$ will count the level above $c_1 \ket{0}$, as customary in OSFT.
As has been already shown in \cite{RastelliZwiebach} the generating function given by counting $b_{-n}$ and $c_{-n}$ excitations equals
\begin{eqnarray}
\label{ch_ghost}
{\rm ch}^{\mathrm{gh}}(q,y)&=&y(1+y)\prod_{n=1}^\infty (1+q^n y)\left(1+\frac{q^n}{y}\right)\\
     &=&\prod_{m=1}^\infty \frac{1}{1-q^m}\sum_{g=-\infty}^\infty y^g q^{\frac{g^2-3g+2}{2}}
               ={\rm ch}^{\mathrm{gen}}(q) \sum_{g=-\infty}^\infty y^g q^{\frac{g^2-3g+2}{2}},
\nonumber
\end{eqnarray}
where ${\rm ch}^{\rm gen}(q)  = \prod_{n=1}^\infty \frac{1}{1-q^n}$ denotes the generating function for a generic Virasoro representation.

Note that the generating function for a fixed ghost number is equal to the generic Virasoro character up to a power of $q$ . This means that the basis states given in terms of $bc$ descendants of the vacuum can be replaced with Virasoro descendants of a ghost-primary of the given ghost number as long as there are no null states. In the ghost CFT, as follows from Kac determinant analysis, null states appear only at ghost numbers $g$ given by multiples of three starting at level $(-2/3 g+1)^2$ . For ghost numbers $g= 0, -3, -6,\ldots$ the null states are explicitly zero when expressed in terms of $b$,$c$ ghosts, such as $L_{-1}^{gh} \ket{0} =0$. On the other hand, for positive ghost numbers the null states are given by a non-vanishing combination of the $b$,$c$ oscillators, however they have the property that they vanish when contracted with an arbitrary Virasoro descendant of an arbitrary primary. These states can still be nonzero, since in the nonunitary ghost CFT, there are states which are neither primary nor descendant. For instance $L_{-1}^{gh} c_{-1} c_0 c_1  \ket{0} \ne 0$ has a nonvanishing contraction with $b_{-2} c_1 \ket{0}.$ In practise, in string field theory, for a given ghost number one needs the dual states as well, so the Virasoro basis is a practical option only for ghost numbers nondivisible by three. For classical solutions this is not a issue, and indeed the basis formed by $L^{gh}_{-n}$ is the one we used for computations in section \ref{sec:noGF}. Alternatively, we could have constructed the basis by the modes of the ghost current $j^{gh}_{-n}$. The number of such states at a given level is also given by the generic Virasoro character.

To obtain the generating function in Siegel gauge it is sufficient to divide by $1+y$, which accounts for elimination of the $c_0$ mode
\bea
{\rm ch}^{\mathrm{Siegel}} (q, y) &=& \frac{1}{1+y} {\rm ch}^{\mathrm{gh}}(q,y)  \nonumber\\
&=&  {\rm ch}^{\mathrm{gen}}(q) \sum_{g=-\infty\phantom{|}}^\infty \; \sum_{s=|g-1|}^\infty (-1)^{s+g-1} y^g q^{\frac{s^2+s}{2}}
\nonumber\\
&=&  {\rm ch}^{\mathrm{gen}}(q) \sum_{g=-\infty\phantom{|}}^\infty \; \sum_{s=|g-1|\; \mathrm{mod}\, 2}^\infty  y^g \left( q^{\frac{s^2+s}{2}} - q^{\frac{(s+1)^2+s+1}{2}} \right).
\label{ch_Siegel}
\eea
From the second or third line, which follows by substituting (\ref{ch_ghost}) and simple manipulations, we can easily read off the number of states at a given ghost number $g$.

Now we shall show how to reinterpret this formula as a sum over SU(1,1) representations labeled by $j$ and $m$. From the Kac determinant for $c=-2$ we find that null states appear
for the representations with highest weights of the form
\be
h'_j = (2j+1)j
\ee
with the first one at level $2j+1$. The irreducible character takes up the form
\be\label{chj}
{\rm ch}_j (q) = {\rm ch}^{\mathrm{gen}}(q) q^{h'_j} \left( 1 - q^{2j+1} \right) = {\rm ch}^{\mathrm{gen}}(q) \left( q^{h'_j}- q^{h'_{j+1/2}}\right).
\ee
The weights $h'_j$ coincide with the $L_0'$ weights of the twisted Virasoro primaries $\ket{j,m}$ defined in (\ref{spin primary}).
Now, by setting $s=2j$ and $g=2m+1$ we can rewrite the generating function (\ref{ch_Siegel}) as
\bea
{\rm ch}^{\mathrm{Siegel}} (q, y)  &=& \sum_{g=-\infty\phantom{|}}^\infty \;  \sum_{j=|g-1|/2\; \mathrm{mod}\, 1}^\infty y^{g} {\rm ch}_j (q) \\
&=&  \sum_{j=0\; \mathrm{mod}\, 1/2\phantom{|}}^\infty \;  \sum_{m=-j\phantom{|}}^{j} y^{2m+1} {\rm ch}_j (q),
\eea
which is what we wanted to show.

As we have already alluded to, most solutions that we have found live in the SU(1,1) singlet sector with $j=0$. The corresponding Verma module at ghost-number one contains a single primary null state at level one, ${L'}_{-1} c_1 \ket{0} =0$. The generating function for this subspace
is given by
\be
{\rm ch}^{\mathrm{singlet}} (q) \equiv {\rm ch}_{j=0} (q) = {\rm ch}^{\mathrm{gen}}(q) \left( 1 - q \right) =\prod_{n=2}^\infty\frac{1}{1-q^n},
\ee
and takes exactly the same form as in the matter sector. The subspace has particularly simple description, one can simply forget ${L'}_{-1}$ when building up the Hilbert space.

To construct nondegenerate basis in the non-singlet sectors labeled by $j$, it is tempting to omit $L_{-(2j+1)}'^{gh}$. From (\ref{chj}) we see that we get the correct number of states, but one still has to verify, that the inner product is nondegenerate in this basis. We do not have a proof, but it is straightforward to test this numerically to any required level.
Alternatively, by writing
\be
\frac{1-q^{2j+1}}{1-q^{\phantom{2j+1}}} = 1+ q+ \cdots + q^{2j}
\ee
we observe that we can omit states which contain $L_{-1}'^{gh}$, for power higher or equal to $2j+1$. In this case it is now possible to prove that the inner product becomes nondegenerate. From the character of the twisted ghost CFT with $c=-2$ we know that all null states in the Verma module for the highest weight $h_j=j(2j+1)$ state with Kac labels $(r,s)=(2j+1,1)$ are given as Virasoro descendants of the level $r=2j+1$ null state. For this null state we can use the Benoit-Saint-Aubin formula \cite{BSA, francesco}
\bea
\ket{\chi_r} &=& \sum_{\substack{p_i \ge 1\\ p_1+\cdots +p_k =r}}  \frac{[(r-1)!]^2 (-t)^{r-k}}{\prod_{i=1}^{k-1} (p_1+\cdots +p_i)(r-p_1 -\cdots -p_i)} L_{-p_1} \ldots L_{-p_k} \ket{h_{r,1}(t)} \qquad \\
&=& \biggl( (L_{-1})^r -t \biggl( (r-1) L_{-2} (L_{-1})^{r-2} + 2(r-2) L_{-1} L_{-2} (L_{-1})^{r-3} + \cdots +
\nonumber
\\
&& \qquad + (r-1) (L_{-1})^{r-2} L_{-2} \biggr) + \cdots +[(r-1)!]^2 (-t)^{r-1} L_{-r} \biggr) \ket{h_{r,1}(t)},
\eea
where $t=2$ is related to the central charge $c$ by $c=13-6(t+1/t)$. Now it is clear that the coefficient in front of $(L_{-1})^r$ is equal to $1$ regardless of any reordering. On the contrary, upon canonical reordering, the coefficient in front of $L_{-r}$ receives contributions of both signs from a large number of terms, so that it is difficult to prove that it is nonzero in general.

Before closing this subsection, let us make few more comments. As we will explain later in section \ref{sec:cubic action}, direct computation of ghost vertices in the singlet sector is somewhat nontrivial. Most economical way that we found requires introduction of an auxiliary ghost sector, where a single mode of $j^{gh}$ is allowed, so that we wish to count the number of states of the form  $j_{-k}^{gh} {L'}_{-M}^{gh} c_1 \ket{0}$. The counting function is clearly given by
\begin{equation}
{\rm ch}^{\mathrm{aux}}(q)=(q+q^2+q^3+\dots)\ {\rm ch}^{\mathrm{singlet}}(q) = \frac{q}{1-q}\ {\rm ch}^{\mathrm{singlet}}(q)=q\ {\rm ch}^{\mathrm{gen}}(q).
\end{equation}
Finally, and trivially, to obtain the total number of states in the combined matter and ghost CFT up to some level, one just has to multiply the respective generating functions. To impose then the twist even condition, one can insert a projector $(1+(-1)^{L_0+1})/2$ which produces
\begin{equation}
\chi_{\mathrm{even}}(q)=\frac{1}{2}\left(\chi(q)+\chi(-q)\right).
\end{equation}

%In table \ref{tab:states 2} we show the numbers of states in the physical string fields.

%\FloatBarrier

\subsection{Singlet sector ghost three vertex}
\label{sec:cubic action}

One of the challenges we took up in this work has been to reach level 30 in the computation of the tachyon vacuum. Previous record of level 26 by Kishimoto \cite{Kishimoto} came at a cost of several months of computer time, spent mostly on tedious computations of the ghost sector vertices. We shall now describe our progress and improvements on this front.

The most convenient and efficient approach developed for the matter sector vertices uses the conservation laws \cite{RastelliZwiebach,GaiottoRastelli}
\begin{equation}\label{Lconslaw}
\bra{V_3} L_{-m}^{(2)} =\bra{V_3} \left( \alpha^m c +\sum_{r=1}^3 \sum_{n \ge 0}  \alpha_n^{m(r)} L_n^{(r)} \right),
\end{equation}
where $c=26$ is the matter sector central charge, and $\alpha$'s are known coefficients. In the ghost sector, however, it is a priori not clear what is the most efficient strategy.
When one does not impose the Siegel gauge (see our results in Section \ref{sec:noGF}), it is most convenient to use the basis of states $L_{-I}^{gh} c_1\ket{0}$, and apply the same conservation laws as in the matter sector keeping in mind the difference $L_{-1}^{gh} c_1\ket{0}  \ne 0$.

In practise, however, level truncation in string field theory becomes useful only when the Siegel gauge is imposed, mostly thanks to the highly improved numerical stability,  but also due to significant reduction of the number of states (see Table \ref{tab:states 2}) and fewer number of vertices required. Previous studies have used conservation laws analogous to (\ref{Lconslaw}) for the $b$ and $c$ oscillators separately. This however requires to compute vertices at intermediate stages at other ghost numbers as well --- though one can limit themselves to ghost numbers $\aver{0,1,2}$ --- which increases memory (or storage) and time requirements.
To make further progress one has to exploit the SU(1,1) symmetry discussed above. At higher levels more than half of the vertices become zero and need not be computed. While we were not motivated enough to design an efficient algorithm which would compute all the Siegel gauge vertices in the basis (\ref{PsiLLtw}) we at least
found and implemented efficient computation for the singlet sector which is of interest not only for the tachyon vacuum.

The main idea is in fact quite simple. To derive conservation law for the twisted Virasoro generator
\be\label{L'def}
L'^{gh}_n=L^{gh}_n + n j^{gh}_n + \delta_{n,0}
\ee
one can use the conservation law (\ref{Lconslaw}) and analogous one for the ghost current
\begin{equation}\label{Jconslaw}
\bra{V_3} j_{-m}^{(2)} =\bra{V_3} \left( \beta^m q +\sum_{r=1}^3 \sum_{n \ge 0}  \beta_n^{m(r)} j_n^{(r)} \right),
\end{equation}
where $q = -3/2$ is related to the ghost number anomaly.

The problem is that the coefficients $\alpha_n^{m(r)}$ in (\ref{Lconslaw}) are generically not compatible with the coefficients $\beta_n^{m(r)}$ in (\ref{Jconslaw}) and therefore decomposing $L'^{gh (2)}_{-m}$ as in (\ref{L'def}), and applying the conservation laws creates terms which cannot be recombined back into $L'^{gh (r)}_n$. Attempting to do so by, leaves behind ghost current terms with nontrivial coefficients $ m \beta_n^{m(r)} + n \alpha_n^{m(r)}$. This means that for the computation of the singlet sector ghost vertex an auxiliary sector is required. Now the crucial observation is that the auxiliary sector formed by states of the form
\be\label{jLorder}
j_{-n}^{gh} L'^{gh}_{-I}  c_1 \ket{0}
\ee
is fully sufficient, in particular, that one does not need to deal with more ghost current generators. From the conservation law in the singlet sector one gets terms of the form
$j_{n}^{gh} L'^{gh}_{-I}  c_1 \ket{0}$. Commuting $j_{n}^{gh}$ via
\begin{equation}
\label{LJ commutator}
\left[ j_n^{gh},L'^{gh}_m \right]=n j^{gh}_{m+n}+\frac{1}{2}(n^2-3n)\delta_{m+n}
\end{equation}
all the way to the right generates terms with at most a single instance of $j_{\pm p}^{gh}$ in between a string of $L'^{gh}_{-m_i}$. Positive modes of the ghost current can be systematically eliminated as above, while the negative ones can be at the end reordered
into a combination of terms in the canonical ordering (\ref{jLorder}).

The outlined recursive procedure requires computation of the vertices of the form singlet--singlet--auxiliary only. In the next step of the recursion, to compute such vertices, one can apply the conservation law for the ghost current mode, so that the structure singlet--singlet--auxiliary is preserved. In every instance of the conservation law application the total level of the three states is reduced, so this method terminates in a finite number of steps. The huge number of vertices to be computed calls upon efficient parallelization of this algorithm. We discuss our approach and related interesting issues in appendix \ref{app: parallelization}.

As we have already mentioned, there does not seem to be a straightforward generalization of this algorithm to the non-singlet case. We can use the same conservation laws as before, but the ghost current produces new states like $L'^{gh}_{-I} j^{gh}_{k} \ket{j,0}$ which for nonsinglet representations might be nonzero. Such states are not in the Siegel gauge and would have to be decomposed into auxiliary sector states as above over various spin representations with spins up to $j$. This seems quite complicated so for the non-singlet sector we simply used the plain $bc$ basis.

\subsection{Observables and consistency checks}

Level truncation turns equation of motion of open string field theory into a system of a large number of coupled quadratic equations with exponentially large number of numerical solutions. To get any sense out of this and eliminate possible spurious solutions it is useful to have as many as possible gauge invariant observables and/or consistency requirements that one can verify.

\subsubsection{Energy and Ellwood invariant}

The most well known observable is given by the energy of the classical solution which for time independent solutions is simply given by minus the value of the action.
In this paper, for convenience, we measure the energy in units of the original D-brane energy for which we formulate the OSFT. This is equivalent to setting $2\pi g_o^2 =1$. Compared to previous works we also redefine the energy additively, so that the energy of the perturbative vacuum equals one, and the energy of tachyon vacuum equals zero.

When the equations of motion are satisfied, whether we gauge fix or not, the energy can be conveniently computed using the kinetic term only, simply by substituting the equations of motion to the cubic term. In our normalization it is given by
\begin{equation}\label{E_action}
E=\frac{\pi^2}{3}\aver{\Psi|Q\Psi} +1.
\end{equation}

In general, the energy is just one member out of a family of observables in OSFT describing coupling to closed strings. All these couplings can be nicely encoded in the boundary state.
In \cite{KMS} we gave a general construction based on Ellwood invariants applicable to solutions known numerically. For universal solutions however there is a single nontrivial independent Ellwood invariant which can be computed. That is because the matter Virasoros feel only conformal weight of the matter part of Ellwood state $\bra{E[V]} \equiv \bra{I} c\bar c V(i,-i)$, and for applicability of Ellwood conjecture we must require that the dimension of $V$ equals 1. Different choices for this operator give identical results, so in particular we may choose
$V=\del X^0\bar{\del}X^0$ with which the Ellwood invariant acquires the interpretation of energy
\begin{equation}
E_0=-4\pi i\la E[c\bar c\del X^0\bar{\del}X^0]|\Psi\ra+1.
\end{equation}
The normalization is chosen in such a way that the invariant matches energy. It is by no means obvious that it should coincide with the expression (\ref{E_action}). For analytic solutions this can be proved under some assumptions \cite{BabaIshibashi} but for level truncated solution the equivalence is nontrivial and in fact the two quantities seem to coincide only for some  solutions and only in the large level limit.

Numerically, we compute the invariant using conservation laws. For the matter Virasoros, ghost Virasoros and $b$ ghost we can use conservation laws from \cite{KMS}
\begin{eqnarray}
\la E[V]|K_n^m &=& -\frac{3}{4}n\left(i^n+(-i)^n\right)\la E[V]|,\\
\la E[V]|K_n^{gh} &=& \phantom{-}\frac{3}{4}n\left(i^n+(-i)^n\right)\la E[V]|, \label{Lgh conservation} \\
\la E[V]|B_n^{gh} &=& \phantom{-}0,
\end{eqnarray}
where $K_n=L_n-(-1)^n L_{-n}$ and $B_n=b_n-(-1)^n b_{-n}$.

To compute the $L'^{gh}$ conservation law we simply use conservation laws for ghost Virasoros and ghost current
\begin{equation}
\label{Jcons-law}
\la E[V]|J_n^{gh}=\left(-\frac{1}{2}\left(i^n+(-i)^n\right)+3\delta_{n,0}\right)\la E[V]|,
\end{equation}
where $J_n^{gh}=j_n^{gh}+(-1)^nj_{-n}^{gh}$, which we derive in appendix \ref{app: j ghost}.
By combining this conservation law with (\ref{Lgh conservation}) we get
\begin{equation}
\la E[V]|K'^{gh}_n=\frac{1}{4}n\left(i^n+(-i)^n\right)\la E[V]|.
\end{equation}
Using these conservation laws we obtain a very quick recursive algorithm to compute the Ellwood invariant.

\subsubsection{Out-of-Siegel equations}

Imposing the Siegel gauge ansatz for a string field before computing the variation of the action and setting it to zero we
find only projected equations of motion
\begin{equation}
\label{eom-projected}
c_0 b_0 (Q\Psi+\Psi\ast\Psi)=c_0 L_0 \Psi+c_0 b_0 (\Psi\ast \Psi)=0,
\end{equation}
which has the great advantage that the number of components matches the number of independent variables.
At any finite level the exact gauge symmetry of string field theory is broken, so the remaining equations of motion are not obeyed, but
for proper solutions they should be satisfied at least asymptotically \cite{HataShinohara}. Our code, which we optimized for Siegel gauge calculations, is not well suited for systematic study of these equations so we decided to check only the first nontrivial equation. In the singlet case we take advantage of fields in the auxiliary ghost sector and compute the contraction of $Q\Psi + \Psi * \Psi$ with the state $\bra{ 0} c_{-1} j^{gh}_2$. We define
\begin{equation}
\label{Delta}
-\Delta_S= \bra{0} c_{-1}j^{gh}_2\left|Q\Psi+\Psi*\Psi\right\ra = \bra{0} c_{-1} c_0 b_2\left|Q\Psi+\Psi*\Psi\right\ra.
\end{equation}
We have introduced a minus sign here, so that this quantity is positive for the tachyon vacuum. In the last equality we omitted a term proportional to $\bra{0}c_1=\bra{0}c_{-1}L'^{gh}_2$, because the corresponding equation is solved exactly when (\ref{eom-projected}) holds.

For our computations without imposing the singlet condition we have added a state $b_{-2}c_0 c_1 \ket{0}$ to our ghost basis for the vertex computation,  and have defined $\Delta_S$ using the last expression in (\ref{Delta}).

\subsubsection{Quadratic identities}

All solutions of OSFT equations should obey quadratic identities \cite{constraints}
\bea\label{quadids}
\aver{\Psi|[Q,L_n^{m}] |\Psi} &=& -\frac{65}{54} n (-1)^{n/2} \delta_{n \; \mathrm{even}} \aver{\Psi|Q\Psi},  \nonumber\\
\aver{\Psi|[Q,j_n^{gh}] |\Psi} &=& - (-1)^{n/2} \delta_{n \; \mathrm{even}} \aver{\Psi|Q\Psi},
\eea
which has been derived from conservation laws for anomalous derivations $K_{n}^m=L_{n}^m-(-1)^{n}L_{-n}^m$ and $H_{n}=j_{n}^{gh}+(-1)^{n}j_{-n}^{gh}$ in matter and ghost sectors respectively.
For classical solutions found via level truncation, however, these identities cannot be satisfied exactly. The reason is that the operators $K_n^m$ and $H_n^m$  do not preserve the maximum level of the string field, and as a consequence,  the quadratic identities are essentially testing how well a solution to level $L$ obeys equations of motion at level $L+n$ with the higher level fields set to zero.

In Siegel gauge these identities simplify considerably. Following \cite{GaiottoRastelli} we define for Siegel gauge solutions
\bea
\label{quadidsR}
R_n &=& (-1)^{n}\frac{54}{65}\frac{\la \Psi|c_0 L_{2n}^m|\Psi\ra}{\la \Psi|c_0 L_0|\Psi\ra},
\nonumber\\
\tilde R_n &=& (-1)^{n} \frac{\la \Psi|c_0 ({L'}_{2n}^{gh} + L_{2n}^m)|\Psi\ra}{\la \Psi|c_0 L_0|\Psi\ra},
\eea
which by the arguments of \cite{constraints} should equal to 1.  Similarly for odd generators we find
\bea\label{odd-ids}
\la \Psi|c_0 L_{2n+1}^m|\Psi\ra &=& 0,
\nonumber\\
\la \Psi|c_0 ({L'}_{2n+1}^{gh} + L_{2n+1}^m)|\Psi\ra &=& 0.
\eea
These identities are trivial for all twist even solutions. They are also obeyed exactly by all our convergent non-even solutions, as these solutions are invariant under $(-1)^j \Omega$ symmetry, which guarantees (\ref{odd-ids}) as well.

\section{Universal solutions in Siegel gauge}
\label{sec:Siegel}

The main achievement of this paper is a systematic exploration of universal solutions of OSFT in Siegel gauge in level truncation. Our strategy is first to find  with the homotopy continuation method {\em all} solutions, complex or real, at as high level as possible.  The second step is to take the viable ones and improve them by the standard Newton's method to as high level as possible and see whether they converge or not. We do not discard a priori the complex solutions for a number of reasons: some of the solutions become real at some high level, some seem to approach real solutions asymptotically. Moreover, the complex solutions might be interesting on their own from other points of view, just like instantons can be viewed as particular complex solutions of Yang-Mills equations of motion in Minkowski space.

\begin{figure}
\centering
\includegraphics[width=14cm]{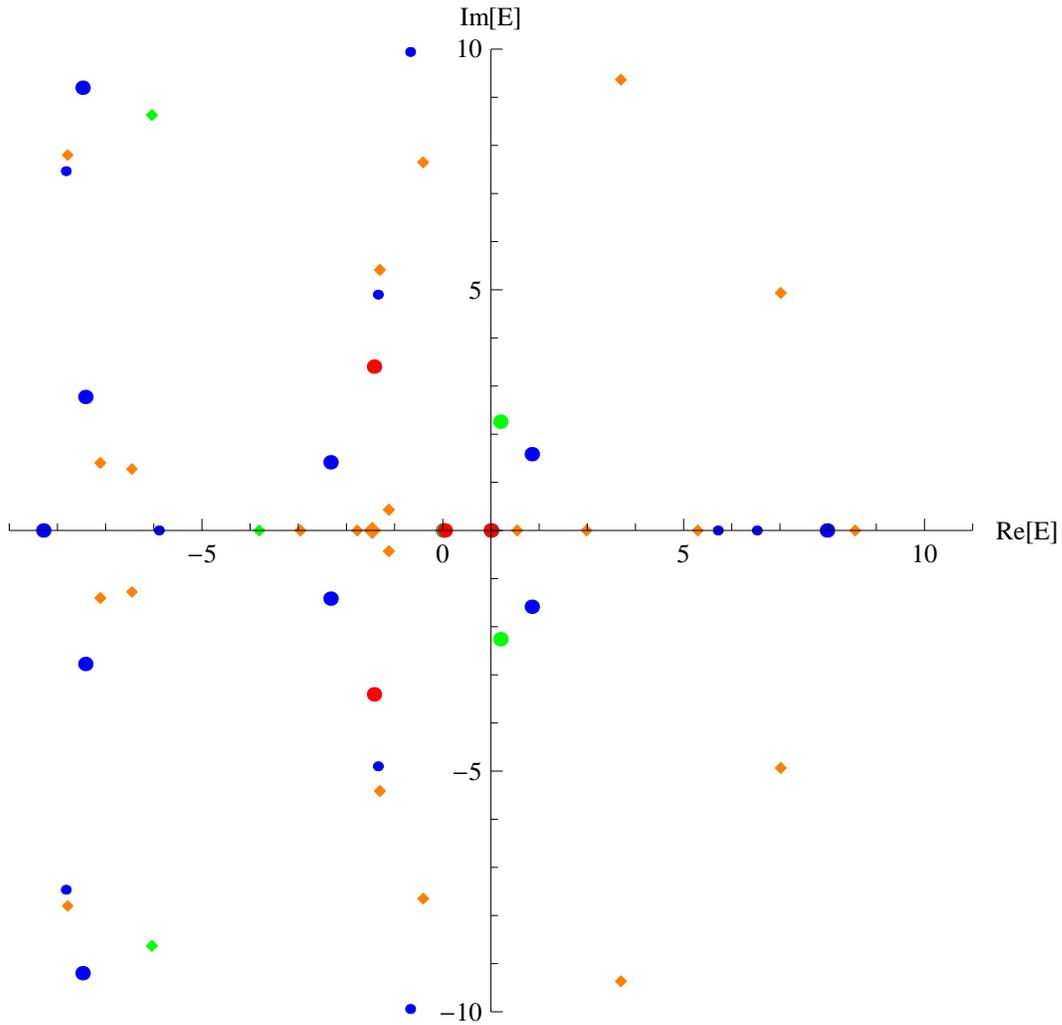}
\caption{Complete set of universal solutions in Siegel gauge at level 2 (red), level 4 (green), level 5 (orange) and level 6 (blue) within the indicated range of energies. Twist even solutions are denoted by circular dots, non-even solutions by diamonds. SU(1,1) singlet solutions use bigger symbols compared to non-singlet solution.}
\label{fig:universal sol lev6}
\end{figure}

The results of the first step can be summarized as follows. Imposing the twist even and SU(1,1) singlet condition which offers the greatest reduction in the number of states we had to solve
equations in 1, 3, 8 and 21 variables for levels 0, 2, 4, and 6 respectively (see table \ref{tab:states 2}). We found 2, 7, 250 and approximately 2096000 solutions respectively. As we already noticed in \cite{KRS} the number of solutions is smaller than the generic amount $2^N$, although not by much.\footnote{At level 6 we were not able to determine the number of solutions with absolute certainty. The exact degeneracy of the string field theory equations is broken by finite precision of our calculations and some solutions which the homotopy method should push to infinity are replaced by finite solutions with very high coefficients. Whether some of these fake solutions are accepted or not depends on various inner setting of the algorithm and the huge amount of solutions makes any detailed analysis impossible. For instance at level 6 in C++ double number format we find 2095858 solutions (only 1294 missing). If we switch to long double number format there is 2096079 solutions (1073 missing).} To visualize the results we plot these solutions as dots in the complex energy plane, see figure \ref{fig:universal sol lev6}.

Notice that among the millions of solutions there are only few with energy within a factor of $O(10)$ compared to the perturbative vacuum. Out of these only few survive the second step, i.e. turn out to provide stable starting points for the Newton's method.\footnote{By stable we mean that a solution converges within 20 iterations of Newton's method at every level. This is a relatively weak criterion, since good solutions usually converge within 4--6 iterations. When the Newton's method takes more iterations it often indicates a jump between two distant solutions.} Incidentally, none of the solutions except for the perturbative and tachyon vacuum seem particularly physical at level 6 in terms of other criteria, such as consistency between energy and $E_0$, and also out-of-Siegel-gauge equations seem to be grossly violated. These solutions should thus be viewed merely as starting points for Newton's method.
We have systematically investigated in detail all solutions with $|E|<50$. From several hundred possible starting points there are 17 twist even singlet solutions (up to complex conjugation) stable within the level truncation scheme (see table \ref{tab:sol table}). The tachyon vacuum is the only "nice" solution with unambiguous identification. There are however two more solutions which deserve extra attention and investigation which we nicknamed "double brane" and "ghost brane". For these solutions the various consistency conditions get satisfied reasonably---though not 100\% convincingly---well and we provide more insight into them below. Aside of these two solutions it is perhaps worth also drawing an attention to solution called No. 14, which in the infinite level limit seems to describe a Gribov copy of the perturbative vacuum.

To study the relevance of the twist non-even fields we were able to repeat this computation only up to level 5, where we had to solve completely 16  equations to find the total of 65106 solutions. However only one of these new solutions has energy in the correct range to appear in figure \ref{fig:universal sol lev6}. We have checked this as well as all other solutions with $|E|<50$ and found no new stable solution\footnote{Some solutions quickly converged to twist even solutions found above. Also there were some solutions convergent only at even levels --- computations at odd levels would terminate after 20 iterations without producing a valid solution --- but even those became eventually problematic at higher levels. }.

Finally we moved to the non-singlet case. In the twist even sector we start at level 4, where the non-singlet states first appear. We find 450 solutions for a system of 9 variables. With an increased effort we can reach again up to level 6, where now we get approximately $6.6\times 10^7$ solutions for the system of equations in 26 variables. None of these solutions however leads to stable solutions with reasonable action.

Without the twist condition we were able to go up to level 5 with approximately $2.1 \times 10^6$ solutions of the 21 equations.  The number of solutions has increased significantly compared to the singlet case, however once again most of them are ill-behaved, with the exception of two new interesting solutions which we nicknamed "half ghost brane" and "half brane".
Remarkably, all the stable non-even non-singlet solutions are even with respect to $(-1)^j\Omega$ symmetry, at least past some level. It seems to be in accord with the general pattern that symmetric solutions are more attractive not only aesthetically, but also from the perspective of Newton's method.

%Twist even: At level 4 we have 450 solutions out of 512, at level 6 26 equatios with approximately $6.6\times 10^7$ solutions. All stable solutions with reasonable action are the same as singlet basis.
%Without twist: At level 5 21 equations and approximately $2\times 10^6$ solutions. We found two nontrivial stable solutions, that are in tables \ref{tab: sol bc 21} and \ref{tab: sol bc 27}. The first one has real coefficients and action from level 22, although it is complex with respect to the string field complex conjugation.

\begin{table}[ht]\nonumber
\centering\footnotesize
\begin{tabular}{|c|c|llll|}
\multicolumn{6}{c}{\normalsize Twist even SU(1,1) singlets}                                                    \\\hline
Solution                            & level    & \ps Energy               & $\ps E_0$                 & $\ps |\Delta_S|    $ &  \ps Im/Re     \\\hline
perturbative vacuum                 &          & $\ps 1                 $ & $\ps 1                  $ & $\ps 0           $ & $\ps 0       $ \\\hline
\multirow{2}{*}{tachyon vacuum}     & 30       & $   -0.000627118       $ & $\ps 0.0120671          $ & $\ps 0.00090829  $ & $\ps 0       $ \\
                                    & $\infty$ & $   -8\times 10^{-6}   $ & $\ps 0.0004             $ & $-7\times 10^{-6}$ & $\ps 0       $ \\\hline
\multirow{2}{*}{"double brane"}     & 28       & $\ps 1.8832 -0.161337 i$ & $\ps 1.32953 +0.178426 i$ & $\ps 0.535827    $ & $\ps 0.51589 $ \\
                                    & $\infty$ & $\ps 1.40   +0.11     i$ & $\ps 1.23    +0.04     i$ & $\ps 0.20        $ & $   -0.05    $ \\\hline
\multirow{2}{*}{"ghost brane"}      & 28       & $   -2.11732-0.371832 i$ & $   -1.19063 +0.165908 i$ & $\ps 0.267977    $ & $\ps 0.398931$ \\
                                    & $\infty$ & $   -1.13   +0.024    i$ & $   -1.01    +0.11     i$ & $\ps 0.08        $ & $\ps 0.33    $ \\\hline
\multirow{2}{*}{No. 9}              & 24       & $   -0.35331           $ & $\ps 3.35886            $ & $\ps 1.97365     $ & $\ps 0       $ \\
                                    & $\infty$ & $\ps 0.4               $ & $\ps 2.9                $ & $\ps 0.5         $ & $\ps 0       $ \\\hline
\multirow{2}{*}{No. 10}             & 24       & $   -4.89552           $ & $\ps 2.46007            $ & $\ps 7.37201     $ & $\ps 0       $ \\
                                    & $\infty$ & $   -6.                $ & $\ps 2.3                $ & $\ps 10.         $ & $\ps 0       $ \\\hline
\multirow{2}{*}{No. 14}             & 28       & $   -0.61986-2.07194  i$ & $   -0.304394-0.125328 i$ & $\ps 0.51849     $ & $\ps 0.567639$ \\
                                    & $\infty$ & $   -0.059  -0.28     i$ & $   -0.18    -0.08     i$ & $\ps 0.16        $ & $\ps 0.4     $ \\\hline
No. 16                              & 24       & $\ps 19.1573+6.2523   i$ & $\ps 1.39521 +0.659265 i$ & $\ps 2.64483     $ & $\ps 1.40716 $ \\\hline
No. 49                              & 24       & $   -6.50268-8.39148  i$ & $\ps 2.15961 +0.077867 i$ & $\ps 1.79055     $ & $\ps 0.750635$ \\\hline
No. 51                              & 24       & $\ps 1.67067-4.96206  i$ & $   -0.091597-0.341405 i$ & $\ps 2.03866     $ & $\ps 1.31641 $ \\\hline
No. 55                              & 24       & $   -13.18  -1.514    i$ & $\ps 0.538205+0.192177 i$ & $\ps 1.35519     $ & $\ps 0.222453$ \\\hline
No. 65                              & 24       & $   -16.6534-5.7377   i$ & $\ps 0.541071-0.782319 i$ & $\ps 1.40638     $ & $\ps 0.756182$ \\\hline
\multirow{2}{*}{No. 77}             & 24       & $   -5.74905-4.10849  i$ & $\ps 0.508235+0.516736 i$ & $\ps 0.438803    $ & $\ps 0.844295$ \\
                                    & $\infty$ & $   -5.     -1.8      i$ & $\ps 0.4     +0.1      i$ & $\ps 0.2         $ & $\ps 0.8     $ \\\hline
\multirow{2}{*}{No. 81}             & 24       & $   -7.96846-3.62476  i$ & $\ps 0.95657 -0.64505  i$ & $\ps 1.63462     $ & $\ps 0.783216$ \\
                                    & $\infty$ & $   -5.     +1.       i$ & $\ps 0.8     -0.6      i$ & $\ps 1.9         $ & $\ps 0.8     $ \\\hline
\multirow{2}{*}{No. 91}             & 24       & $   -3.86278+0.78003  i$ & $   -0.75477 -0.028228 i$ & $\ps 1.49166     $ & $\ps 0.314904$ \\
                                    & $\infty$ & $   -1.     -1.       i$ & $   -0.5     +0.0      i$ & $\ps 0.9         $ & $\ps 0.4     $ \\\hline
No. 93                              & 24       & $   -6.60883-8.49812  i$ & $\ps 1.83907 -0.047786 i$ & $\ps 1.58305     $ & $\ps 0.532952$ \\\hline
\multirow{2}{*}{No. 95}             & 24       & $   -9.83474-8.05476  i$ & $\ps 0.24612 +0.462712 i$ & $\ps 1.02673     $ & $\ps 0.812354$ \\
                                    & $\infty$ & $\ps 0.     -8.       i$ & $\ps 0.4     +0.5      i$ & $   -0.7         $ & $\ps 0.6     $ \\\hline
\multicolumn{6}{l}{}                                                                                                                        \\
\multicolumn{6}{c}{\normalsize Twist even non-singlets}                                                                                     \\\hline
No. 231                             & 22       & $\ps 6.27071-30.8278  i$ & $   -1.35262 +1.22029  i$ & $\ps 6.83759     $ & $\ps 0.496046$ \\\hline
\multicolumn{6}{l}{}                                                                                                                        \\
\multicolumn{6}{c}{\normalsize Twist non-even non-singlets}                                                                                       \\\hline
\multirow{2}{*}{"half ghost brane"} & 26       & $   -0.88489           $ & $   -0.427091           $ & $\ps 0.105198    $ & $\ps 0.600394$ \\
                                    & $\infty$ & $   -0.51              $ & $   -0.66               $ & $\ps 0.17        $ & $\ps 0.31    $ \\\hline
\multirow{2}{*}{"half brane"}       & 24       & $\ps 0.47454-1.11238  i$ & $\ps 0.488976+0.107413 i$ & $\ps 0.565206    $ & $\ps 1.20649 $ \\
                                    & $\infty$ & $\ps 0.68   -0.010    i$ & $\ps 0.54    +0.10     i$ & $\ps 0.23        $ & $\ps 1.3     $ \\\hline
\multirow{2}{*}{No. 264}            & 22       & $   -10.5493           $ & $\ps 1.10974            $ & $\ps 7.58984     $ & $\ps 0.229845$ \\
                                    & $\infty$ & $   -11                $ & $\ps 0.9                $ & $\ps 11          $ & $   -0.3     $ \\\hline
\end{tabular}
\caption{List of convergent solutions from starting points with $|E|<50$ at the highest level computed. The second row  (when present) shows extrapolations to infinite level, see appendix \ref{app: fits} for more explanations.  We have to emphasize that many of the extrapolations of the more exotic solutions are quite unstable. When the estimated errors are of order 1 or bigger, the fits are not shown. Note that there are no stable twist non-even solutions in the singlet sector. }
\label{tab:sol table}
\end{table}

%In appendix \ref{app: coeff} we show first three coefficients of all the interesting solution and in appendix \ref{app: quadratic identities} first four quadratic identities $R_n$.

\FloatBarrier
\subsection{Tachyon vacuum}
\label{sec:Siegel TV}

Tachyon vacuum is the most famous solution of classical OSFT with a long history of numerical \cite{SZ,MT,GaiottoRastelli,Kishimoto} and analytical approaches \cite{Sch-Analytic,Okawa,FK,ES}. There is not much more which needs to be added to the story, except of course that it would be nice to find an analytic expression for the Siegel gauge which is still lacking. Numerical computations at higher levels might still in principle provide additional clues. From the perspective of this work, our main interest in the tachyon vacuum is that it provides a testbed for our complex algorithms.
\begin{table}[h]
\nonumber
\centering
\begin{tabular}{|l|ll|l|}\hline
Level    &  \ps Energy            & $E_0$       & $\Delta_S$             \\\hline
2        & $\ps 0.0406234       $ & $0.110138 $ & $\ps 0.0333299       $ \\
4        & $\ps 0.0121782       $ & $0.0680476$ & $\ps 0.0145013       $ \\
6        & $\ps 0.00482288      $ & $0.0489211$ & $\ps 0.00841347      $ \\
8        & $\ps 0.00206982      $ & $0.0388252$ & $\ps 0.00564143      $ \\
10       & $\ps 0.000817542     $ & $0.0318852$ & $\ps 0.00412431      $ \\
12       & $\ps 0.000177737     $ & $0.0274405$ & $\ps 0.00319231      $ \\
14       & $   -0.00017373      $ & $0.0238285$ & $\ps 0.00257255      $ \\
16       & $   -0.000375452     $ & $0.0213232$ & $\ps 0.00213597      $ \\
18       & $   -0.000493711     $ & $0.0190955$ & $\ps 0.00181467      $ \\
20       & $   -0.000562955     $ & $0.0174832$ & $\ps 0.00156995      $ \\
22       & $   -0.000602262     $ & $0.0159666$ & $\ps 0.00137834      $ \\
24       & $   -0.000622749     $ & $0.0148397$ & $\ps 0.00122487      $ \\
26       & $   -0.000631156     $ & $0.0137381$ & $\ps 0.0010996       $ \\
28       & $   -0.000631707     $ & $0.0129049$ & $\ps 0.00099569      $ \\
30       & $   -0.000627118     $ & $0.0120671$ & $\ps 0.00090829      $ \\\hline
$\infty$ & $   -8\times 10^{-6} $ & $0.0004   $ & $   -7\times 10^{-6} $ \\
$\sigma$ & $\ps 2\times 10^{-6} $ & $0.0013   $ & $\ps 2\times 10^{-6} $ \\\hline
%$\infty$ & $   -9\times 10^{-6} $ & $0.00015  $ & $   -7\times 10^{-6} $ \\                   % max level fit
%$\sigma$ & $\ps ?               $ & $0.00002  $ & $\ps ?               $ \\\hline
\end{tabular}
\caption{Energy, Ellwood invariant and out-of-Siegel equation for tachyon vacuum up to level 30.}
\label{tab:TV}
\end{table}

\begin{figure}
\centering
\includegraphics[width=10cm]{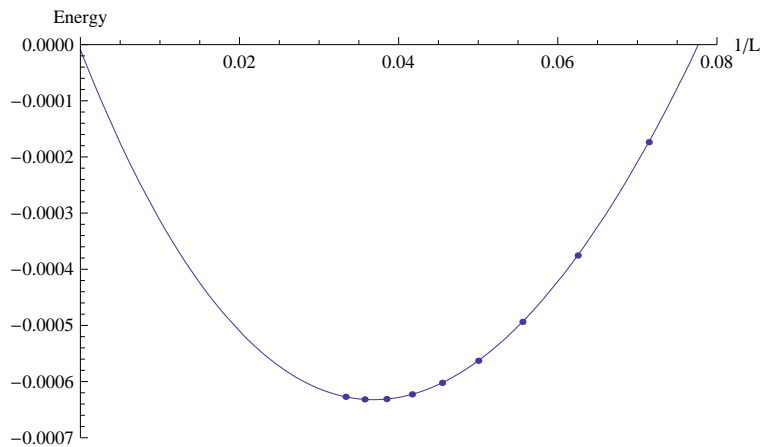}
\caption{Detail of the tachyon vacuum energy dependence from level 14 to 30 and extrapolation to infinite level of order 12. We clearly see the turning point at $L=27$.}
\label{fig:TV energy}
\end{figure}

\begin{figure}
\centering
\includegraphics[width=10cm]{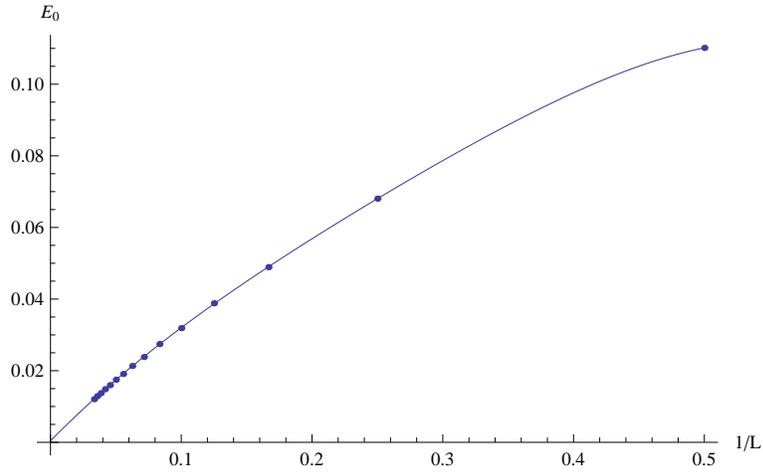}
\caption{$E_0$ invariant of the tachyon vacuum solution and order 4 extrapolation.}
\label{fig:TV E0}
\end{figure}

\begin{figure}
\centering
\includegraphics[width=10cm]{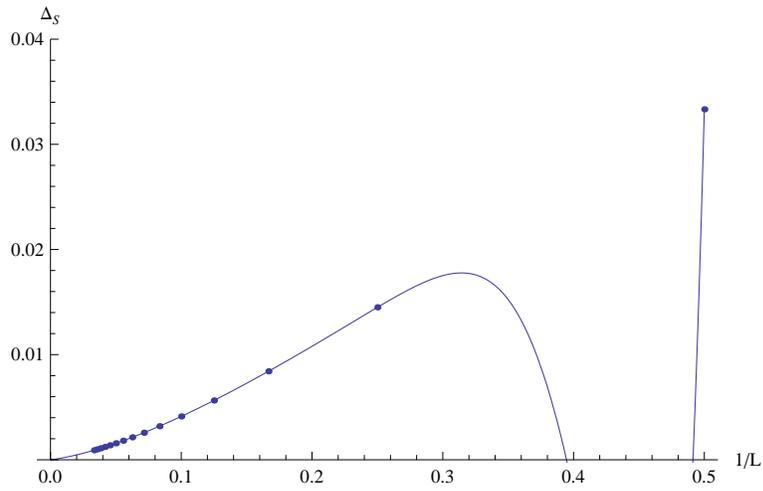}
\caption{Out-of-Siegel equation $\Delta_S$ for the tachyon vacuum and order 10 extrapolation.}
\label{fig:TV Delta}
\end{figure}

With our numerical codes, adopting a SU(1,1) singlet twist-even ansatz, we were able to reach level 30. In table \ref{tab:TV} we show our results for the energy, the Ellwood invariant $E_0$ and parameter $\Delta_S$ measuring the extent to which the first of the out-of-Siegel equations is satisfied. In figures \ref{fig:TV energy}, \ref{fig:TV E0} and \ref{fig:TV Delta} we show how these quantities behave with level and show the fitting polynomial which is used for the extrapolation to $L=\infty$. In appendix \ref{app: coeff} we give explicitly the first three coefficients of this solution and in appendix \ref{app: quadratic identities} we show first four ratios $R_n$ and $\tilde R_n$ associated the with quadratic identities. The ratios come indeed very close to the expected value of 1 with accuracy $10^{-5}$--$10^{-4}$ depending on the value of~$n$.

Having obtained the level 30 results we can finally prove explicitly the conjecture from \cite{Taylor-Pade,GaiottoRastelli} that the energy as a function of level is not monotonic, but has a minimum at level 28.  The extrapolated curve has minimum around level 27, however level 30 is needed to prove the existence of the minimum, because the energy at level 28 is still lower than at level 26. This is shown in detail in figure \ref{fig:TV energy}.

%{\bf I have added tables describing various possibilities of extrapolating the energy and $E_0$ (tables \ref{tab:TV fit}, \ref{tab:TV fit2}, \ref{tab:TV fitE}). The energy extrapolations work very well when we use high order (even maximal order) polynomials. The results we get are of order $10^{-6}$ to $10^{-5}$. The modified fit (\ref{fit 2}) shows gives similar results as the polynomial fit (\ref{fit 1}) because $\alpha$ is very small. On the other hand the Ellwood invariant works best with low order fits for both types of extrapolations.}

\FloatBarrier
\subsection{"Double brane"}
\label{sec:Siegel DB}

One of the main motivations for our systematic exploration of classical solutions in OSFT was to look for the conjectured double brane solution. The original reasoning was roughly the following:\footnote{The second author thanks Wati Taylor and Barton Zwiebach for inspiring discussions on this subject around year 2000.}  Just as we can find the tachyon vacuum from the theory on a single D-brane, we can find the D-brane from the theory around the tachyon vacuum. But why would the tachyon vacuum allow us to construct only a single D-brane and not more? And if we can construct multiple D-branes from tachyon vacuum, why not from the perturbative vacuum around a single D-brane?  Also the structure of the proposed form of solution in OSFT \cite{Sch-B} suggested that there should exist multiple D-brane solutions. More recent analytic proposals and detailed studies appeared in \cite{MS1,MS2,HK1,HK2,HK3,H-BV,MNT,EM}, but up to date there has been no numerical attempts.

Surprisingly, one of the seven solutions found at level 2, together with its complex complex conjugate, actually do give rise to a solution reminiscent of a double brane.
At level 2, its energy as measured by the Ellwood invariant is $E_0^{(2)} \approx 2.01 \pm 0.05 i$, tantalizingly close to the desired value $E^{exact} =2$ although when computed using the action it gives very different value with an opposite sign $E^{(2)} \approx -1.43 \pm 3.40 i$. At level 2, however, one should not realistically expect the match between these two values.

Taking this solution as a seed for the Newton's method and improving it repeatedly to higher levels, this time the energy computed from the action comes closer to the expected value. For instance at level 20 we find $E^{(2)} \approx 1.98 \pm 0.30i $ while unfortunately $E_0^{(2)} \approx 1.37 \pm 0.21i$ is departing from the wished for value. For some time we had hoped that these departures are an example of overshooting which we saw---albeit on a much smaller scale---for the tachyon vacuum above. This motivated us to go as high level as possible to obtain the best possible extrapolation to $L=\infty$. The numerical computations are a bit slower since we had to adapt our code to complex numbers, so we decided not to go past level 28.

\begin{table}[h]\nonumber
\centering
\begin{tabular}{|l|ll|ll|}\hline
Level    & \ps Energy               & $E_0$                & $|\Delta_S|  $ &  \ps Im/Re     \\\hline
2        &     $-1.42791-3.40442 i$ & $2.00934-0.054534 i$ & $2.9861  $ & $\ps 2.47302 $ \\
4        & \ps $1.19625-2.25966  i$ & $1.73651+0.117637 i$ & $1.65103 $ & $\ps 5.23177 $ \\
6        & \ps $1.84813-1.58507  i$ & $1.60634+0.195442 i$ & $1.2563  $ & $\ps 2.20828 $ \\
8        & \ps $2.04207-1.14971  i$ & $1.53973+0.217911 i$ & $1.05199 $ & $\ps 1.51157 $ \\
10       & \ps $2.08908-0.866428 i$ & $1.48598+0.228010 i$ & $0.921766$ & $\ps 1.20095 $ \\
12       & \ps $2.08515-0.674602 i$ & $1.45210+0.227059 i$ & $0.83043 $ & $\ps 1.01867 $ \\
14       & \ps $2.06302-0.53887  i$ & $1.42232+0.224184 i$ & $0.762358$ & $\ps 0.895048$ \\
16       & \ps $2.03499-0.439057 i$ & $1.40194+0.218266 i$ & $0.709389$ & $\ps 0.804018$ \\
18       & \ps $2.00593-0.363272 i$ & $1.38304+0.212332 i$ & $0.666812$ & $\ps 0.733269$ \\
20       & \ps $1.9778 -0.304197 i$ & $1.36942+0.205378 i$ & $0.631713$ & $\ps 0.675682$ \\
22       & \ps $1.95135-0.257139 i$ & $1.35632+0.198765 i$ & $0.602187$ & $\ps 0.627204$ \\
24       & \ps $1.92679-0.218971 i$ & $1.34654+0.191784 i$ & $0.576934$ & $\ps 0.585387$ \\
26       & \ps $1.90411-0.187545 i$ & $1.33691+0.185169 i$ & $0.555036$ & $\ps 0.548697$ \\
28       & \ps $1.8832 -0.161337 i$ & $1.32953+0.178426 i$ & $0.535827$ & $\ps 0.51589 $ \\\hline
$\infty$ & \ps $1.40   +0.11     i$ & $1.23   +0.04     i$ & $0.20    $ & $   -0.05    $ \\
$\sigma$ & \ps $0.02   +0.02     i$ & $0.02   +0.05     i$ & $0.01    $ & $\ps 0.16    $ \\\hline
%$\infty$ & \ps $1.40482+0.018686 i$ & $1.2098 -0.04     i$ & $0.25493 $ & $   -1.87627 $ \\
%$\sigma$ & \ps $?                i$ & $0.0008 +0.02     i$ & $?       $ & $\ps ?       $ \\\hline
\end{tabular}
\caption{Energy, Ellwood invariant, out-of-Siegel equation and reality of the "double brane" solution up to level 28 with extrapolations.}
\label{tab:sol DB}
\end{table}

Our main results for the "double brane" are summarized in table \ref{tab:sol DB}. What we do observe, is that the imaginary parts of both energies indeed tend to zero quite well, and,
more importantly the extrapolation of the (positive) norm\footnote{We use the standard Euclidean norm applied to the list of string field coefficients within a given basis. } of the imaginary part of the solution gives a negative value. This indicates, that at some finite level the solution becomes purely real, and this also explains why the fits of its imaginary part cannot be very stable, see the relatively large error $\sigma$ in the $\im/\re$ column and our discussion in appendix \ref{app: fits}. Looking closely at the various fits, we expect that the solution might become purely real plausibly already at level 50. This is nice and non-trivial.

%The seed for "double brane" solution appears at level 2 and we improved it up to level 28. The energy and other quantities are in table \ref{tab:sol DB} and the dependance of energy and $E_0$ on level is plotted in figure \ref{fig:sol DB}.

%At available level the solution is complex, but imaginary parts of all quantities in infinite level extrapolations (including $R_n$ in table \ref{tab:quadratic DB} and coefficients of the solution in table \ref{tab:coeff DB}) change sign or go to zero within estimated error. Therefore it is likely that the solution becomes real at some high level.

%The interpretation of the solution is far from clear. Energy and $E_0$ do not seem to converge to any integer value, but we can expect a non-analytic behavior when the solution becomes real like in the case of "half ghost brane" in section \ref{sec:Siegel HGB} or $\sigma$-brane solution in Ising model \cite{KRS}. At lower levels the real part of energy is close to 2, so we thought the solution could describe two copies of the initial D-brane. However the high levels make such interpretation unlikely. We can also consider a possibility that the energy decreases to 1 and the solution is large gauge transformation of perturbative vacuum. Finally the solution can be just an artefact of the Siegel gauge, since $\Delta_S$ does not seem to converge to 0.

\begin{figure}
   \centering
   \begin{subfigure}[t]{0.47\textwidth}
      \includegraphics[width=\textwidth]{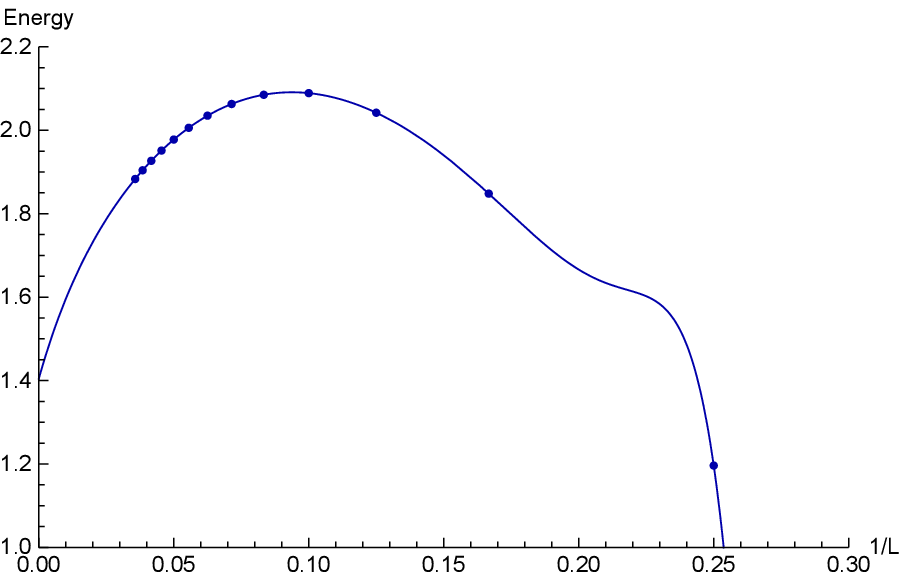}
   \end{subfigure}\qquad
   \begin{subfigure}[t]{0.47\textwidth}
      \includegraphics[width=\textwidth]{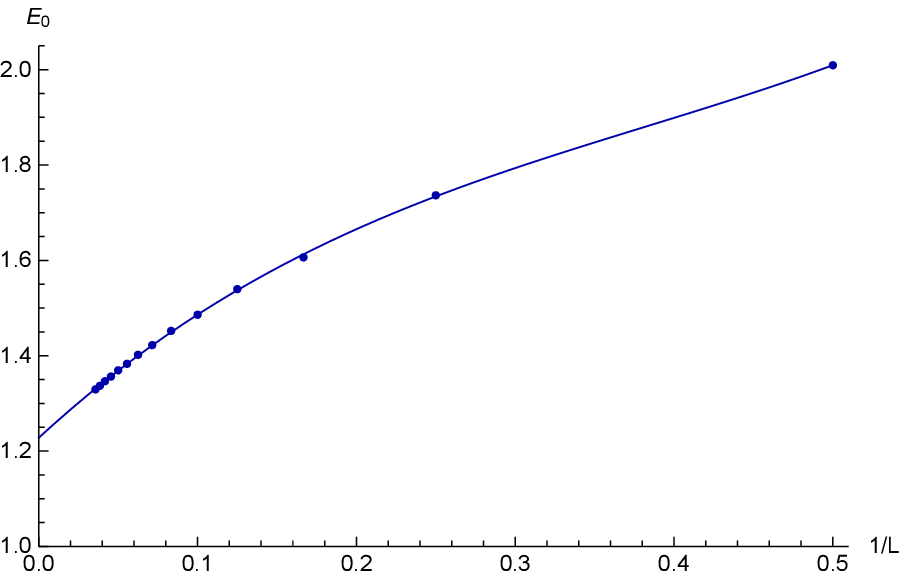}
   \end{subfigure}
\caption{Real part of energy (left) and Ellwood invariant $E_0$ (right) of "double brane" solution with extrapolations (order 12 and 3).}
\label{fig:sol DB}
\end{figure}

The extrapolated value of the energy is on the other hand quite puzzling, see figure \ref{fig:sol DB} where we show a typical fit for the energy. The two energies seem to asymptotically approach two different non-integer values 1.2 and 1.4 with relatively small errors. These values are indeed closer to 1 or 1.5  than 2, so the solution might as well be interpreted as a Gribov copy of the perturbative vacuum, or possibly "one-and-a-half-brane". A pessimist's point of view on this solution is that it presents a mere artefact of Siegel gauge, since the $\Delta_S$ does not seem to converge to zero.  On the other hand, an optimist can argue, that the fits might improve dramatically as soon as the solution becomes real. Such changes have been observed in the case of the "half ghost brane" in section \ref{sec:Siegel HGB} and $\sigma$-brane solution in Ising model \cite{KRS}. More realistic, but still optimistic attitude is to say, that the double brane solution cannot be constructed in Siegel gauge, which has limited validity \cite{ET}, but that it lives somewhere nearby in the field space.

The coefficients of $c_1 \ket{0}$, $L_{-2}^m c_1 \ket{0}$ and ${L'}_{-2}^{gh} c_1 \ket{0}$ are given for illustration in table \ref{tab:coeff DB} in appendix \ref{app: coeff}. Their extrapolations to infinite level have tiny imaginary part consistent with zero.
%{\bf [Idea: What would happen if took e.g. the level 20 solution and by force set the imaginary part to zero?]}
The table \ref{tab:quadratic DB} shows that the first few coefficients $R_n$ approach at large levels 0.9 instead of 1. Rather than to the inconsistency of the solution or of the quadratic identities this points to the fact that the coefficients of the solution do not decay as fast as in the case of the nicer solutions such as the tachyon vacuum.

\FloatBarrier
\subsection{"Ghost brane"} \label{sec:Siegel GB}

\begin{figure}
   \centering
   \begin{subfigure}[t]{0.47\textwidth}
      \includegraphics[width=\textwidth]{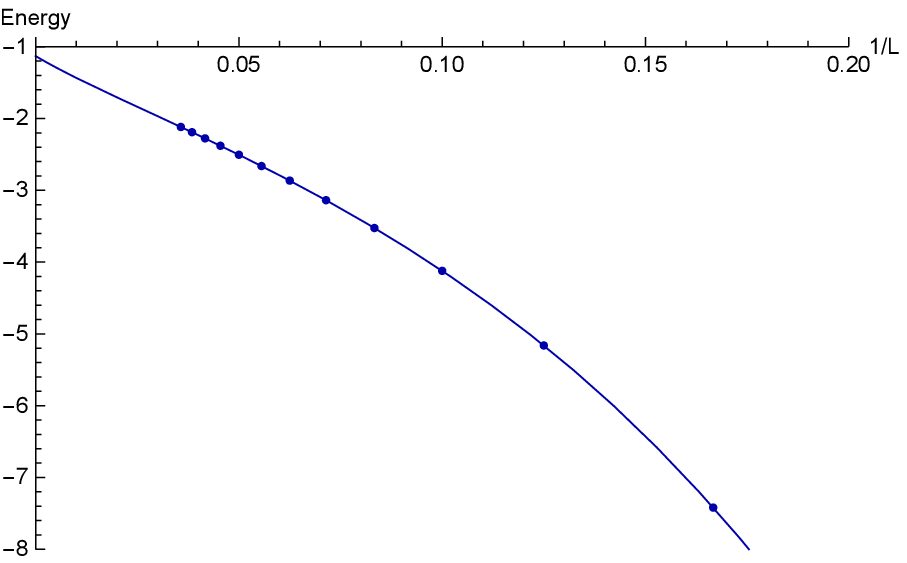}
   \end{subfigure}\qquad
   \begin{subfigure}[t]{0.47\textwidth}
      \includegraphics[width=\textwidth]{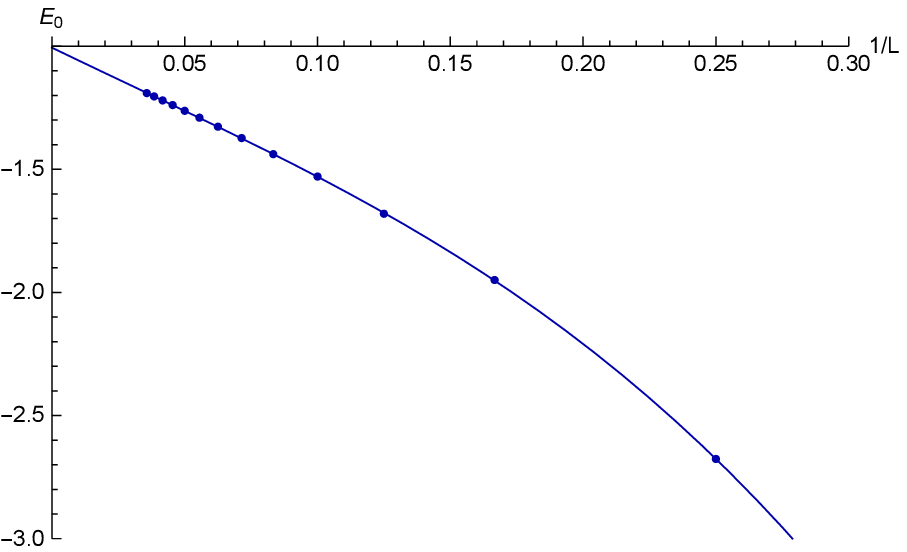}
   \end{subfigure}
\caption{Real part of energy (left) and Ellwood invariant $E_0$ (right) of the "ghost brane" solution with extrapolations of order 11 and 3 respectively.}
\label{fig:sol GB}
\end{figure}

A quite unexpected result of our systematic approach to numerical solutions is the existence of a "ghost brane" solution. One of the complex twist even singlet solutions at level 4 with
rather unphysical looking energy invariants turns out to give rise to a stable sequence of solutions. Asymptotically, after computing the solution to level 28, see table \ref{tab:sol GB}, the energy evaluated from the action and from the Ellwood invariant tend to $-1.13 + 0.02i$ and $-1.01   +0.11 i$ respectively. This is tantalizingly close to minus one, roughly within the precision of the extrapolations!\footnote{As we discussed in the section \ref{sec:Siegel TV} and appendix \ref{app: fits} the error estimates of the extrapolation for the energy computed from the action should allow for a factor of 4, just like for the tachyon vacuum. For the real part of $E_0$ one does not need any correction in neither case. For the imaginary part the error seems to be a factor of 8 smaller than what is required to claim consistency.} Possible extrapolations are shown in figure \ref{fig:sol GB}.

\begin{table}[h]\nonumber
\centering
\begin{tabular}{|l|ll|ll|}\hline
Level    &  \ps Energy              & $\ps E_0$                & $|\Delta_S|           $ &  Im/Re     \\\hline
4        & $   -15.534 -6.15021  i$ & $   -2.67655+0.349878 i$ & $2.06533 $ & $0.410863$ \\
6        & $   -7.41971-2.77358  i$ & $   -1.94969+0.275957 i$ & $1.02147 $ & $0.418622$ \\
8        & $   -5.16142-1.76196  i$ & $   -1.68072+0.250813 i$ & $0.71436 $ & $0.432211$ \\
10       & $   -4.12161-1.28659  i$ & $   -1.52959+0.229158 i$ & $0.56818 $ & $0.434482$ \\
12       & $   -3.52505-1.0123   i$ & $   -1.43859+0.216265 i$ & $0.482212$ & $0.432647$ \\
14       & $   -3.13749-0.834158 i$ & $   -1.37324+0.204977 i$ & $0.425217$ & $0.429121$ \\
16       & $   -2.86477-0.709192 i$ & $   -1.32713+0.19685  i$ & $0.384407$ & $0.424871$ \\
18       & $   -2.66193-0.616662 i$ & $   -1.29055+0.189495 i$ & $0.353578$ & $0.42024 $ \\
20       & $   -2.50477-0.545352 i$ & $   -1.26259+0.183675 i$ & $0.329357$ & $0.415469$ \\
22       & $   -2.37916-0.488679 i$ & $   -1.23917+0.178292 i$ & $0.309749$ & $0.410702$ \\
24       & $   -2.27628-0.442528 i$ & $   -1.22039+0.173791 i$ & $0.293495$ & $0.406007$ \\
26       & $   -2.19032-0.404195 i$ & $   -1.20411+0.169568 i$ & $0.279762$ & $0.402137$ \\
28       & $   -2.11732-0.371832 i$ & $   -1.19063+0.165908 i$ & $0.267977$ & $0.398931$ \\\hline
$\infty$ & $   -1.13   +0.024    i$ & $   -1.01   +0.11     i$ & $0.08    $ & $0.33    $ \\
$\sigma$ & $\ps 0.03   +0.003    i$ & $\ps 0.04   +0.02     i$ & $0.01    $ & $0.03    $ \\\hline
%$\infty$ & $   -1.13119+0.026235 i$ & $   -1.019  +0.087    i$ & $0.0868  $ & $        $ \\
%$\sigma$ & $\ps ?                i$ & $\ps 0.001  +0.005    i$ & $?       $ & $        $ \\\hline
\end{tabular} \caption{Energy, Ellwood invariant and out-of-Siegel equation and reality of the "ghost brane" solution up to level 28 with extrapolations.}
\label{tab:sol GB}
\end{table}

The proximity of both energy invariants to minus one is quite likely related to the fact, that this solution does not live far beyond the region of validity of the Siegel gauge. The violation of the out-of-Siegel equations $\Delta_S$ is the smallest of all the exotic solutions we have found in this paper. The quadratic identities, see table \ref{tab:coeff GB} in appendix
\ref{app: quadratic identities} are also obeyed with the best accuracy among all our exotic solutions, giving $R_n \approx 0.98$ and $\tilde R_n \approx 0.99$ which are again within $4\sigma$ of the predicted value. It thus seems that this is a best behaved exotic solution with fastest decay of higher level coefficients, beyond some level. The first three of them are given for illustration in table \ref{tab:coeff GB} in appendix \ref{app: coeff}.

An important aspect of the solution is that it is a truly complex solution, with the norm of the imaginary part being asymptotically at infinite level about a third of the real part. These solutions thus violate the reality condition of string field theory. Had this not been the case, the string field theory would have been inconsistent, as negative energy D-branes are not physical vacua.

Since our solution lives in the universal sector of the theory, the corresponding boundary state, see \cite{KMS}, is minus the one of the original D-brane. This is exactly the definition of the ghost D-brane by Okuda and Takaynagi \cite{OT}, see also \cite{Dijkgraaf} for more recent discussion. The field theory around the ghost D-brane differs from the original D-brane by an overall minus sign in front of the action. Stacks of D-branes and ghost D-branes possess supergroup gauge symmetries on their worldvolume and thus are described by a non-unitary field theory. Such objects better not correspond to real solutions of OSFT. They might however appear as non-trivial saddle points for the path integral of string field theory \cite{Aniceto}.

\FloatBarrier
\subsection{"Half ghost brane"}
\label{sec:Siegel HGB}

The biggest surprises of this work came when we decided to study the effect of twist non-even string fields. As we did not find then any stable solutions in the singlet sector, we relaxed also this condition. We have found two new stable solutions, which we nicknamed "half ghost brane" and "half brane", since their boundary states roughly correspond to $\pm \half$ the standard D-brane boundary state. Another surprise came when we looked at the coefficients. In the even SU(1,1) spin sectors the solutions were twist even, while in the odd sectors they were twist odd, so as a whole the solutions are symmetric under the $\Omega (-1)^J$ symmetry.

%When we analyze the solution with respect to the SU(1,1) we find that it is not generic. We find that the even part of the solution consists of even spin representations, while the odd part of the solutions consists of odd spin representations. The solutions are therefore symmetric under combination of twist $\Omega$ and $(-1)^j$ symmetry. This implies that the coefficients of purely matter states at odd levels are equal to zero.

The "half ghost brane" solution is quite unusual because it becomes "pseudoreal" at level 22, which caused the energy and the Ellwood invariant to become strictly real,
see table \ref{tab:sol HGB}. The coefficients also become real at this level, but under the usual definition of complex conjugation $\ast=bpz\circ hc$ the coefficients at odd levels should be purely imaginary. However it is formally possible to redefine complex conjugation by any $Z_2$ symmetry, in our case the twist symmetry $\Omega$ or SU(1,1) spin $(-1)^J$, and then the solution becomes real with respect to this new complex conjugation.
%Since it is not clear which definition of complex conjugation should be use we call the solution only pseudoreal.

\begin{table}[h]\nonumber
\centering
\begin{tabular}{|l|ll|ll|}\hline
Level    &  \ps Energy                & $\ps E_0$                  & $|\Delta_S|$  &   Im/Re     \\\hline
4        & $   -12.316 -3.03642    i$ & $   -1.67202+0.546917   i$ & $1.2313   $ & $ 0.865144$ \\
5        & $   -6.45268-1.27696    i$ & $   -0.906424+0.418332  i$ & $0.620185 $ & $ 1.16737 $ \\
6        & $   -4.35705-0.78598    i$ & $   -0.865572+0.35625   i$ & $0.415511 $ & $ 0.954718$ \\
7        & $   -3.33673-0.484238   i$ & $   -0.656885+0.311362  i$ & $0.312715 $ & $ 1.06523 $ \\
8        & $   -2.72264-0.346946   i$ & $   -0.649858+0.280771  i$ & $0.254214 $ & $ 0.941507$ \\
9        & $   -2.32595-0.239544   i$ & $   -0.548369+0.252446  i$ & $0.214425 $ & $ 1.00849 $ \\
10       & $   -2.03976-0.180921   i$ & $   -0.544639+0.229285  i$ & $0.187496 $ & $ 0.919346$ \\
11       & $   -1.83136-0.130289   i$ & $   -0.483438+0.20816   i$ & $0.166609 $ & $ 0.966213$ \\
12       & $   -1.6664 -0.0999362  i$ & $   -0.483896+0.190449  i$ & $0.151204 $ & $ 0.894168$ \\
13       & $   -1.53812-0.0724385  i$ & $   -0.442433+0.172804  i$ & $0.138347 $ & $ 0.930032$ \\
14       & $   -1.43071-0.0550496  i$ & $   -0.442547+0.157289  i$ & $0.128364 $ & $ 0.86812 $ \\
15       & $   -1.34368-0.0389625  i$ & $   -0.412389+0.141337  i$ & $0.119637 $ & $ 0.896007$ \\
16       & $   -1.26801-0.028504   i$ & $   -0.413658+0.127156  i$ & $0.112624 $ & $ 0.838677$ \\
17       & $   -1.20493-0.0188315  i$ & $   -0.390587+0.111523  i$ & $0.106296 $ & $ 0.860183$ \\
18       & $   -1.14861-0.0125338  i$ & $   -0.391441+0.0971558 i$ & $0.101085 $ & $ 0.804464$ \\
19       & $   -1.10069-0.00689461 i$ & $   -0.373155+0.0800877 i$ & $0.0962737$ & $ 0.819172$ \\
20       & $   -1.05703-0.00341677 i$ & $   -0.374456+0.0632733 i$ & $0.0922374$ & $ 0.760415$ \\
21       & $   -1.01932-0.00075642 i$ & $   -0.359549+0.0384798 i$ & $0.0884464$ & $ 0.760498$ \\
22       & $   -0.984567            $ & $   -0.383157            $ & $0.0927055$ & $ 0.65173 $ \\
23       & $   -0.955695            $ & $   -0.399688            $ & $0.0997222$ & $ 0.662377$ \\
24       & $   -0.929072            $ & $   -0.415166            $ & $0.102253 $ & $ 0.618092$ \\
25       & $   -0.906413            $ & $   -0.417504            $ & $0.104245 $ & $ 0.637956$ \\
26       & $   -0.884894            $ & $   -0.427091            $ & $0.105198 $ & $ 0.600394$ \\\hline
$\infty$ & $   -0.51                $ & $   -0.66                $ & $0.17     $ & $ 0.31    $ \\\hline
\end{tabular}
\caption{"Half ghost brane" solution. Since the character of the solution changes at level 22 we cannot do the usual extrapolations because we have too few points. We show quadratic extrapolation over even levels for energy and linear for other quantities.}
\label{tab:sol HGB}
\end{table}

\begin{figure}
   \centering
   \begin{subfigure}[t]{0.47\textwidth}
      \includegraphics[width=\textwidth]{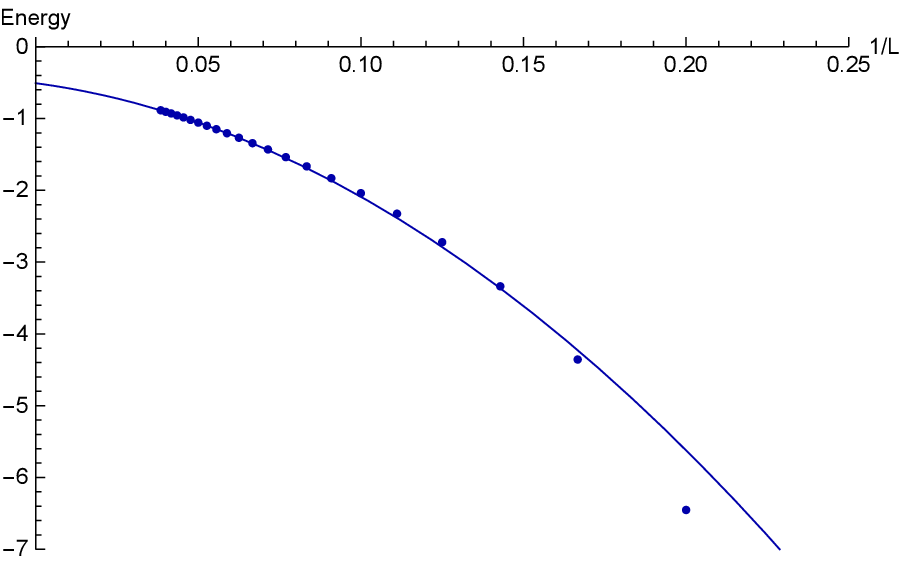}
   \end{subfigure}\qquad
   \begin{subfigure}[t]{0.47\textwidth}
      \includegraphics[width=\textwidth]{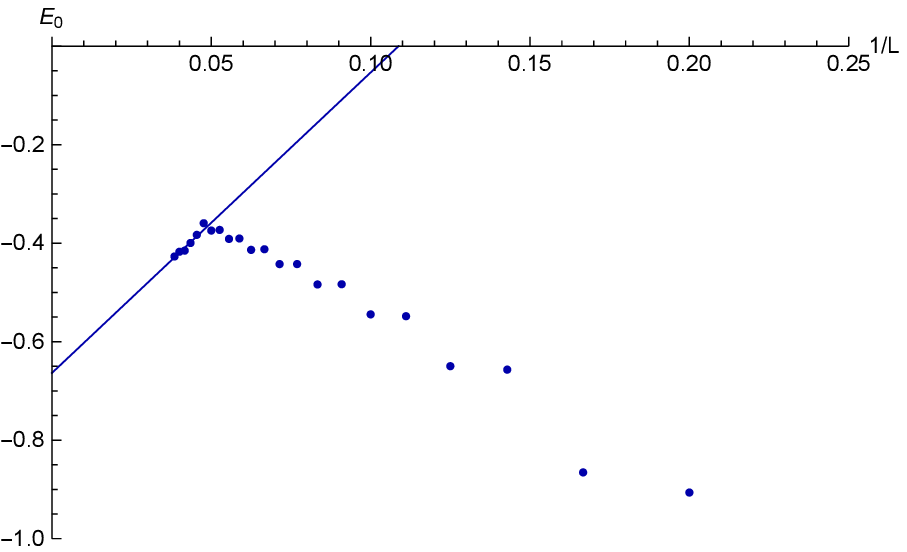}
   \end{subfigure}
\caption{Real part of energy (left) and Ellwood invariant $E_0$ (right) of "half ghost brane" solution with quadratic and linear extrapolations.}
\label{fig:sol HGB}
\end{figure}

The effect of the abrupt pseudoreality at level 22 introduces quite a non-analytic behaviour for most of the fits, see figure \ref{fig:sol HGB}. In order to obtain a reasonable extrapolation we had to compute this solution all the way to the maximum level of 26 allowed by our computational resources. Despite this, the extrapolations to infinite level did not come out very reliable because of the fact that even and odd levels follow different curves.

Within the limited accuracy it seems that both the energy as well as the Ellwood invariant go to value $-1/2$. The out-of-Siegel five data points can be fitted by a crude linear fit with the result 0.17
and quite nicely with a quadratic fit giving $-0.29$. Higher order fits are even more unstable. Given the absolute magnitude at level 26  and the fact that two best fits have opposite signs, we conclude that zero is not ruled out, and that the solution looks quite reasonable.

The quadratic identities, see table \ref{tab:coeff HGB} in appendix
\ref{app: quadratic identities} are obeyed with the second best accuracy among all our exotic solutions,
with $R_n \sim \tilde R_n \approx 0.9$ at level 26 and asymptotically approaching a value slightly above one.  First three coefficients are given again in appendix \ref{app: coeff} in table \ref{tab:coeff HGB} .

What could be the possible interpretation of half-branes? We might speculate that they are analogous to merons in gauge theory. Gauge theory merons carry one half of the topological charge, and can be thought of as half instantons. They interpolate between an ordinary perturbative vacuum and its Gribov copy. The exact half-brane solution of \cite{Erler-exotic} in some sense also interpolates the pure gauge perturbative vacuum with $f_+(0)<1$ and its nontrivial "Gribov copy" with $f_+(0)>1$. Pushing the analogy between half branes and merons farther is not easy since the latter are localized in spacetime, while the former are universal solutions.

\FloatBarrier
\subsection{"Half brane"} \label{sec:Siegel HB}

Finally we provide more details about the "half brane" solution which just like its ghosty cousin appears in the twist non-even and non-singlet sector starting at level 4.  And just as the
"half ghost brane" the "half brane" possess the accidental $\Omega (-1)^J$ symmetry.

\begin{figure}
   \centering
   \begin{subfigure}[t]{0.47\textwidth}
      \includegraphics[width=\textwidth]{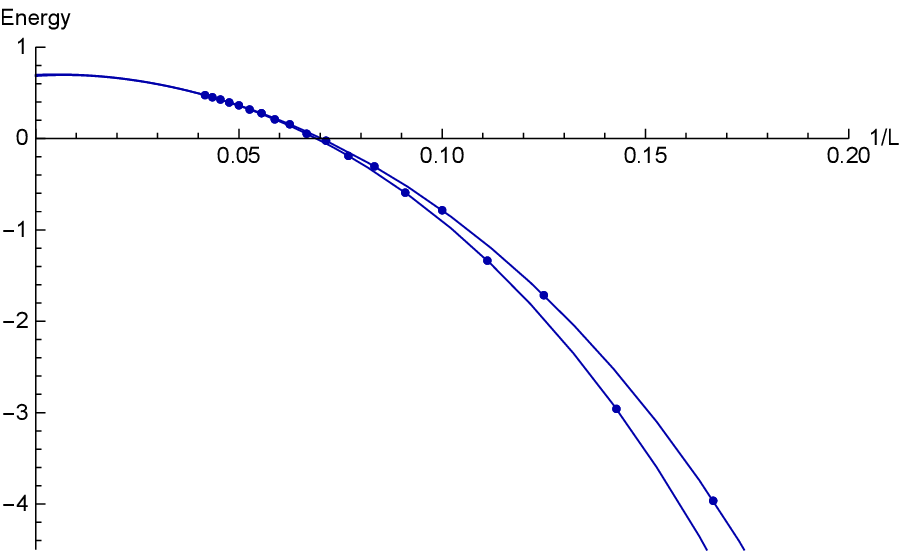}
   \end{subfigure}\qquad
   \begin{subfigure}[t]{0.47\textwidth}
      \includegraphics[width=\textwidth]{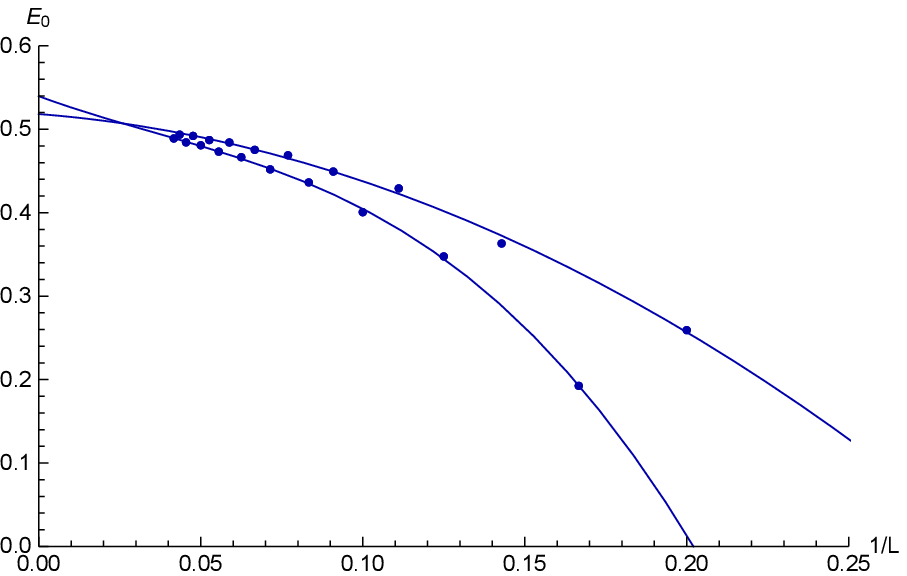}
   \end{subfigure}
\caption{Real part of energy (left) and Ellwood invariant $E_0$ (right) of "half brane" solution with extrapolations through even and odd levels (order 8 and 6 for energy, 4 and 2 for $E_0$).}
\label{fig:sol HB}
\end{figure}

The real parts of the energy computed from the action and the Ellwood invariant $E_0$ seem to converge to about 0.7 and 0.5 respectively, see table \ref{tab:sol HB}. The imaginary parts seem to go to zero or 0.1 respectively. Though not obvious from plots in figure \ref{fig:sol HB}, the extrapolations do not seem to be particularly stable.

The difference between the energy and $E_0$ is a measure of how well the solution behaves. It is correlated with the out-of-Siegel violation $\Delta_S$ which is the worst among all our considered solutions. Quadratic identities, see table \ref{tab:coeff HB}, are obeyed within 50--60\% at level 24 and asymptotically give about 70--140\% of the expected answer. This indicates quite a strong influence of higher level coefficients.

The solution remains complex as we go to the infinite level, as it should for consistency of string field theory. It is not the best one we have found, but still intriguing and fits the general pattern.

\begin{table}[h]\nonumber
\centering
\begin{tabular}{|l|ll|ll|}\hline
Level          &  \ps Energy              & $\ps E_0$                  & $|\Delta_S|$ &  Im/Re    \\\hline
4              & $   -13.6502 -11.1428 i$ & $   -0.371639-0.0630598 i$ & $2.33879 $ & $0.94069$ \\
5              & $   -7.78572 -7.79998 i$ & $\ps 0.259216-0.0735157 i$ & $2.02226 $ & $1.23991$ \\
6              & $   -3.96478 -5.50457 i$ & $\ps 0.192498-0.0106103 i$ & $1.47997 $ & $1.23173$ \\
7              & $   -2.95912 -4.81235 i$ & $\ps 0.363138+0.0030746 i$ & $1.40856 $ & $1.3664 $ \\
8              & $   -1.71495 -3.78855 i$ & $\ps 0.347577+0.0399685 i$ & $1.1451  $ & $1.25856$ \\
9              & $   -1.33624 -3.50822 i$ & $\ps 0.429148+0.0474    i$ & $1.11899 $ & $1.32883$ \\
10             & $   -0.784977-2.90236 i$ & $\ps 0.400591+0.0665808 i$ & $0.963064$ & $1.2333 $ \\
11             & $   -0.592433-2.75421 i$ & $\ps 0.449341+0.0714814 i$ & $0.950856$ & $1.27984$ \\
12             & $   -0.305058-2.35461 i$ & $\ps 0.436324+0.0818504 i$ & $0.847135$ & $1.21651$ \\
13             & $   -0.18996 -2.2641  i$ & $\ps 0.468871+0.0851479 i$ & $0.840552$ & $1.25027$ \\
14             & $   -0.023760-1.98195 i$ & $\ps 0.451942+0.0917189 i$ & $0.766144$ & $1.20675$ \\
15             & $\ps 0.052380-1.92132 i$ & $\ps 0.475369+0.0941379 i$ & $0.762248$ & $1.23276$ \\
16             & $\ps 0.155655-1.71203 i$ & $\ps 0.46651 +0.0976461 i$ & $0.705981$ & $1.20148$ \\
17             & $\ps 0.209629-1.66878 i$ & $\ps 0.484197+0.0994473 i$ & $0.703522$ & $1.2228 $ \\
18             & $\ps 0.277217-1.50758 i$ & $\ps 0.473199+0.101914  i$ & $0.659292$ & $1.19891$ \\
19             & $\ps 0.317445-1.47525 i$ & $\ps 0.487073+0.103319  i$ & $0.657667$ & $1.21717$ \\
20             & $\ps 0.363408-1.34736 i$ & $\ps 0.480852+0.104445  i$ & $0.621853$ & $1.19941$ \\
21             & $\ps 0.394545-1.32232 i$ & $\ps 0.492029+0.10555   i$ & $0.620743$ & $1.21622$ \\
22             & $\ps 0.426713-1.21841 i$ & $\ps 0.484336+0.10638   i$ & $0.591057$ & $1.20258$ \\
23             & $\ps 0.451536-1.19848 i$ & $\ps 0.493552+0.107274  i$ & $0.59028 $ & $1.21784$ \\
24             & $\ps 0.474535-1.11238 i$ & $\ps 0.488976+0.107413  i$ & $0.565206$ & $1.20649$ \\\hline
$\infty^{(e)}$ & $\ps 0.68    -0.010   i$ & $\ps 0.54    +0.10      i$ & $0.23    $ & $1.3    $ \\
$\sigma^{(e)}$ & $\ps 0.04    +0.007   i$ & $\ps 0.11    +0.03      i$ & $0.01    $ & $0.1    $ \\\hline
$\infty^{(o)}$ & $\ps 0.70    -0.02    i$ & $\ps 0.52    +0.09      i$ & $0.26    $ & $1.3    $ \\
$\sigma^{(o)}$ & $\ps 0.02    +0.08    i$ & $\ps 0.10    +0.03      i$ & $0.01    $ & $0.3    $ \\\hline
%$\infty$       & $\ps 0.672   -0.002   i$ & $\ps 0.515   +0.094     i$ & $0.225   $ & $       $ \\
%$\sigma$       & $\ps 0.002   +0.008   i$ & $\ps 0.001   +0.004     i$ & $0.004   $ & $       $ \\\hline
\end{tabular}
\caption{"Half brane" solution with infinite level fits using even and odd levels. The extrapolations with the $^{(e)}$ or $^{(o)}$ superscripts are obtained from results at even or odd levels only respectively. }
%while the last two rows are computed using the simple maximum polynomial fit for even and odd levels separately (for the Ellwood invariant we used mod 4) with average taken at the end. In this last case the extrapolation for the imaginary part is missing since it behaves unstably. }
\label{tab:sol HB}
\end{table}

\FloatBarrier
\section{Solutions without gauge fixing}
\label{sec:noGF}

In the previous section we saw a number of remarkable stable solutions with a clear correlation between the parameter $\Delta_S$ measuring the violation of the out-of-Siegel equations and the discrepancy in the energy computed from the action or from Ellwood invariant. With the exception of the tachyon vacuum these solutions are most likely located beyond the region of validity of the Siegel gauge. A great achievement would be to relax the Siegel gauge condition and find appropriate solution for the integer or half-integer D-branes of the previous section.
In this section we will mostly present approaches that do {\em not} work. We hope that this part of our work might inspire others to find a better solution.

The first idea that comes immediately to mind is to forget about the gauge fixing altogether. This has already been tested at level 2 in \cite{RastelliZwiebach}, where a tachyon vacuum was found providing 88\% of the expected result which is not as good as 96\% found in Siegel gauge \cite{SZ}. It was concluded, based on the suggestion of Ashoke Sen, that {\em uncontrolled lifting of flat directions is likely to make numerical work based on gauge invariant actions less reliable}. Eighteen years later, with new codes and much more computer power we wish to see whether there is a way around it.

So what happens, when one attempts to solve the OSFT equations of motion without imposing any gauge condition on the string field? The gauge symmetry $\delta \Psi=Q\Lambda+[\Psi,\Lambda]$ should in principle generate continuous spectra of solutions, but since at the nonlinear level it is completely broken by level truncation we get only a discrete set of solutions.
The major issue however is the fact, that solutions found at level $L$ cannot be reasonably used as starting points for the Newton's method for $L+2$ (or $L+1$ in twist non-even case). In such a case the Newton's method does not converge within reasonable number of iterations for majority of solutions, and when it does then the resulting solution is nowhere near the original one.
We have tried to solve this problem by modifying the homotopy continuation method (see appendix \ref{app: homotopy}) --- we have tried to continuously deform the equations at level $L$ to those at level $L+2$ and follow the individual solutions. Not only is this approach computationally much less efficient, but the resulting solutions are so far away in the coefficient space, that they cannot be thought of as corresponding to the same physical solution.

The linear homotopy calculations exhibit also a numerical instability caused by the kinetic term.
The BRST charge $Q$ annihilates a part of the string field and its matrix representation has zero eigenvalues. If the coefficients of the string field also have small values (which happens for solutions close to perturbative vacuum) then the Jacobian $Q_{ij}+2V_{ijk}t_k$ is a badly conditioned matrix and its inversion generates large numerical errors in the Newton's method. This prevents us from determining the exact degeneracy of the perturbative vacuum and of nearby solutions even when we use the long double number format and adapt various settings of the homotopy continuation method.

Our first approach to the problem of constructing solutions beyond the validity of Siegel gauge was therefore to give up the stability of Newton's method, and to construct all solutions using the homotopy continuation method at highest level possible and see whether there are some with the right properties. Our main criterion of viability was the proximity of the energy computed from the action and from the Ellwood invariant. We have also investigated the quadratic identities (\ref{quadids}). We have found only a qualitative correlation between these two consistency checks.

\subsection{Twist even solutions}

Let us now present our results for the twist even case. We have been able to find all solutions at levels 2 and 4. They correspond to the dots in figure \ref{fig:NG lev4 col} which shows the complex energy plane. The color of the dot corresponds to the absolute value of the difference between the energy and the Ellwood invariant, which should agree. The warmer the color is the better agreement there is. Surprisingly we find rather dense distribution of solutions throughout the energy plane quite far from its origin. We can see two clusters of solutions around the perturbative and tachyon vacuum that represent a discrete remnant of the gauge transformations. The perturbative vacuum turns out to have a nontrivial multiplicity, which does not happen in Siegel gauge. At level 2 we have confirmed analytically that it has multiplicity 3, and at level 4 we have estimated its multiplicity to be over 200.

\begin{figure}
\centering
\includegraphics[width=14cm]{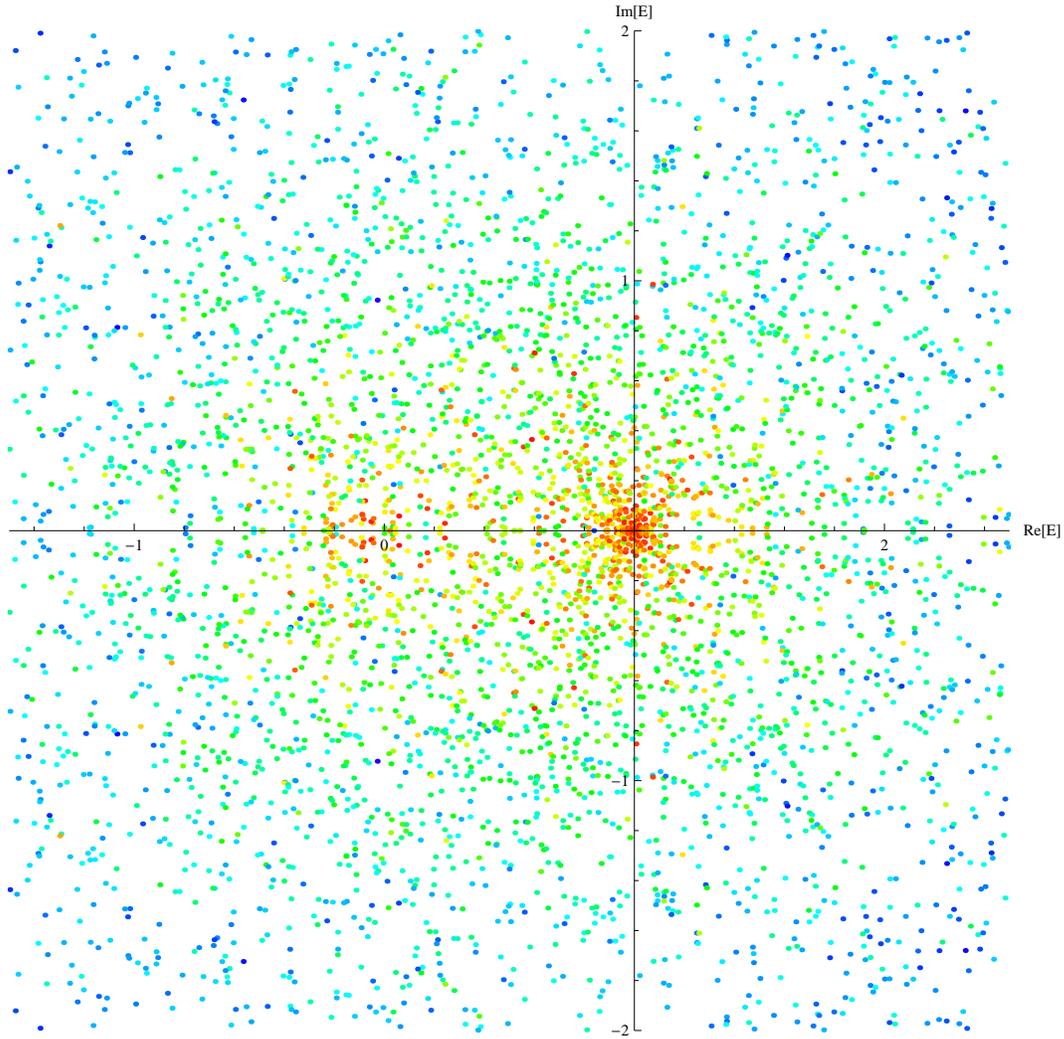}
\caption{A sample of $2\times 10^7$ fully randomly chosen twist even universal solutions at level 6 without imposing any gauge. Only a tiny fraction of these solutions fits the area shown in the figure.}\label{fig:NG lev6v1}
\end{figure}

\begin{figure}
\centering
\includegraphics[width=14cm]{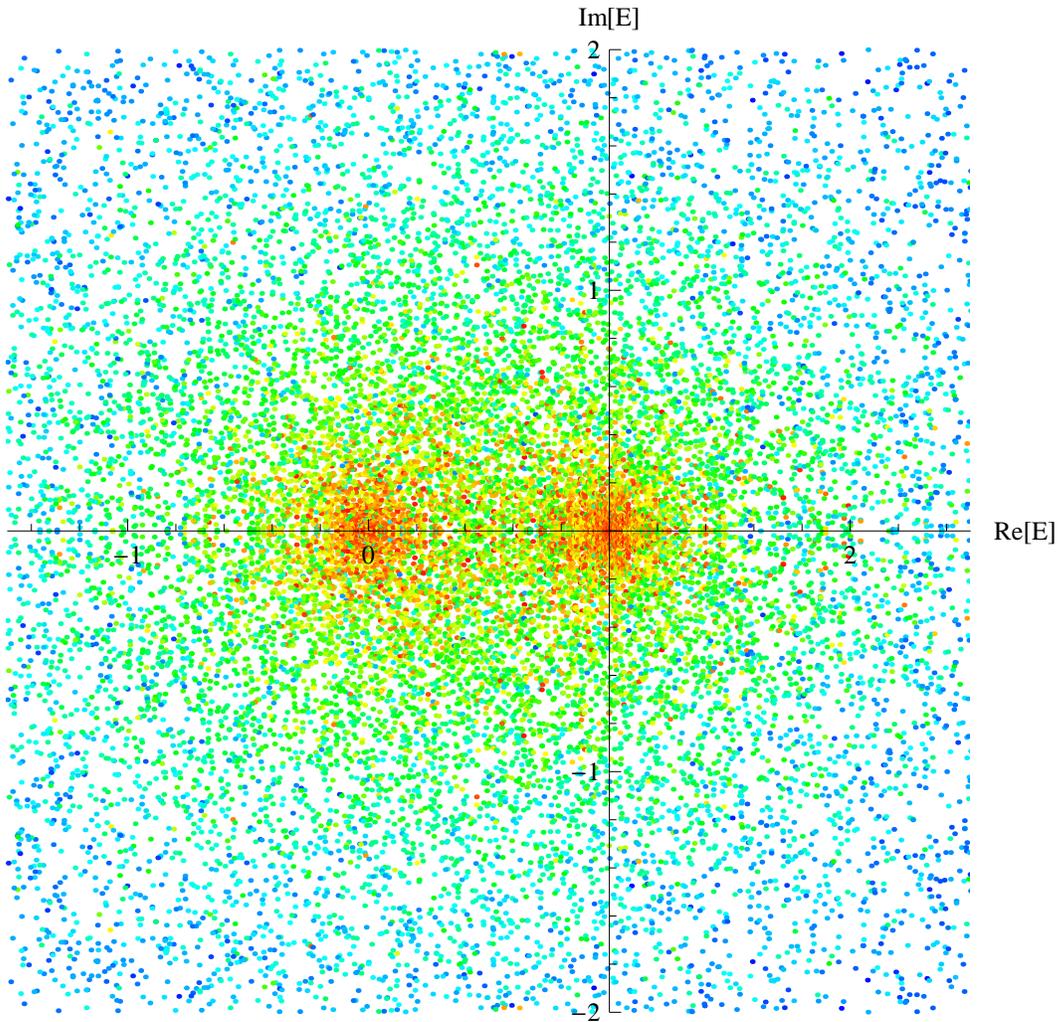}
\caption{A sample of first $10^7$ ordered twist even universal solutions without gauge fixing at level 6. The solutions are ordered  by systematically switching on and off the coefficients at the lowest levels in the starting exactly solvable system  (at $\alpha=0$, see appendix \ref{app: homotopy}) in the homotopy method.
Therefore the solutions appearing in this plot have most of their coefficients in the initial system equal to zero, and this appears to translate to a greater chance of ending up near a perturbative or tachyon vacuum.}\label{fig:NG lev6v2}
\end{figure}

Getting all solutions at level 6 is practically impossible. There is approximately $2^{43}\approx 8.8\times 10^{12}$ solutions. We estimated the required total CPU time to be around 500 000 years with the computers we used.   We have decided anyway to compute few millions of solutions by various methods. In figure \ref{fig:NG lev6v1} we give completely random solutions obtained by the homotopy method with random solutions to the initial system. The next approach we tried was to focus on solutions of the homotopy method wherein for the initial system of equations of the form $t_i (\rho_i-t_i)=0$ we take solutions which preferentially turn on the lower order coefficients. In practise, we have assembled a list of integer numbers from 1 to $10^7$ in binary digits. Reading the number from right to left, we set the initial $t_i$ to the digit on the $i$'th position. Note that the $t_i$ correspond to coefficients of the string field in some canonical order with non-decreasing levels. The results are given in figure \ref{fig:NG lev6v2}.

\begin{figure}
\centering
\includegraphics[width=14cm]{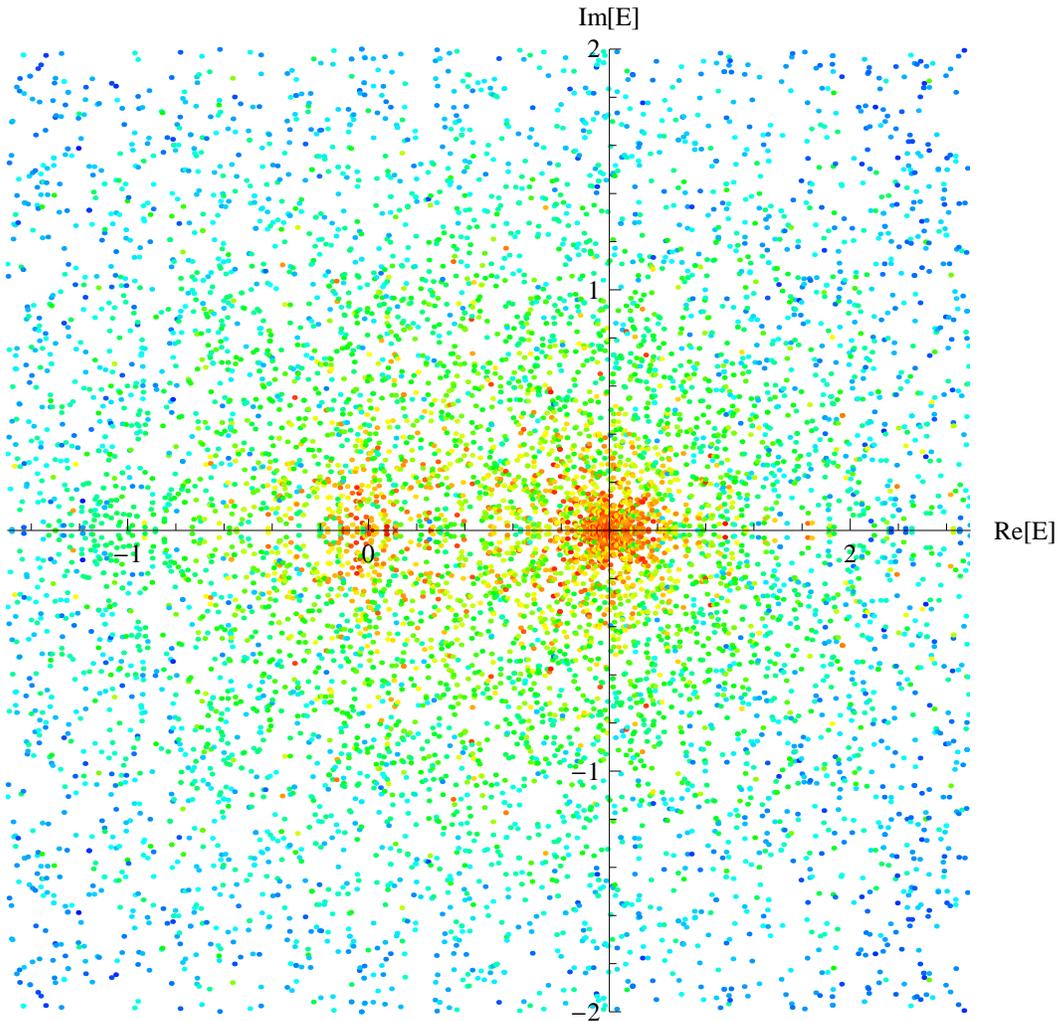}
\caption{A sample of $10^7$ twist even universal solutions at level 6 without gauge fixing which we obtained from 10 potentially interesting starting points at level 4. There are less solutions around the tachyon vacuum than around the perturbative vacuum compared to level 4. There are around 9000 solutions in this figure.}\label{fig:NG lev6v3}
\end{figure}

Finally we decided to see whether the notion of improving a solution can be given some sense using the modified homotopy continuation method, see appendix \ref{app: homotopy}. We have continuously deformed the system of equations from level 4 augmented by a trivial system $t_i (\rho_i-t_i)=0$ for the level 6 coefficients to the full set of level 6 equations and followed 10 promising solutions. For each of those we have constructed $10^6$  new solutions---out of the total of $2^{29} \approx 5.4 \times 10^{8}$ corresponding to the level 6 coefficients---by allowing different solutions for the first 20 coefficients at level 6 in the initial exactly solvable starting point of the homotopy method. The 10 original solutions that we chose at level 4 were our best bets for the tachyon vacuum, double brane and the ghost brane, and we chose also some solutions close to the perturbative vacuum. The results of this attempt are given in figure~\ref{fig:NG lev6v3}.

The three figures are quite different. Figure \ref{fig:NG lev6v2} shows most solutions, \ref{fig:NG lev6v1} the least. This indicates that the homotopy method to some extent preserves the character of the solution. Note that in figures  \ref{fig:NG lev6v1} and  \ref{fig:NG lev6v3} the cluster around the tachyon vacuum seems to be relatively smaller. This might be a generic feature of solutions at level 6, but it can be caused also by our selection of solutions.

Unfortunately we cannot make any decisive claim about the existence of multiple or ghost brane solutions outside the Siegel gauge. We do not see any "warm" clusters that would prove existence of these solutions. However it is quite possible, that at low levels the relevant solutions are not anywhere close to where they should be.  Analogously to Siegel gauge, see e.g. the tables \ref{tab:sol DB} and \ref{tab:sol GB}, we would not be able to recognize them among the landscape of unphysical solutions which are mere artifacts of the level truncation.

%\begin{figure}
%\centering
%\includegraphics[width=14cm]{NG_lev4.eps}
%\caption{Twist even universal solutions without Siegel gauge at level 2 (red) and level 4 (blue).}\label{fig:NG lev4}
%\end{figure}

\begin{figure}
\centering
\includegraphics[width=10cm]{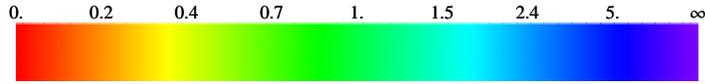}
\caption{Colors corresponding to difference between energy and Ellwood invariant.}\label{fig:colors}
\end{figure}

\begin{figure}
\centering
\includegraphics[width=12cm]{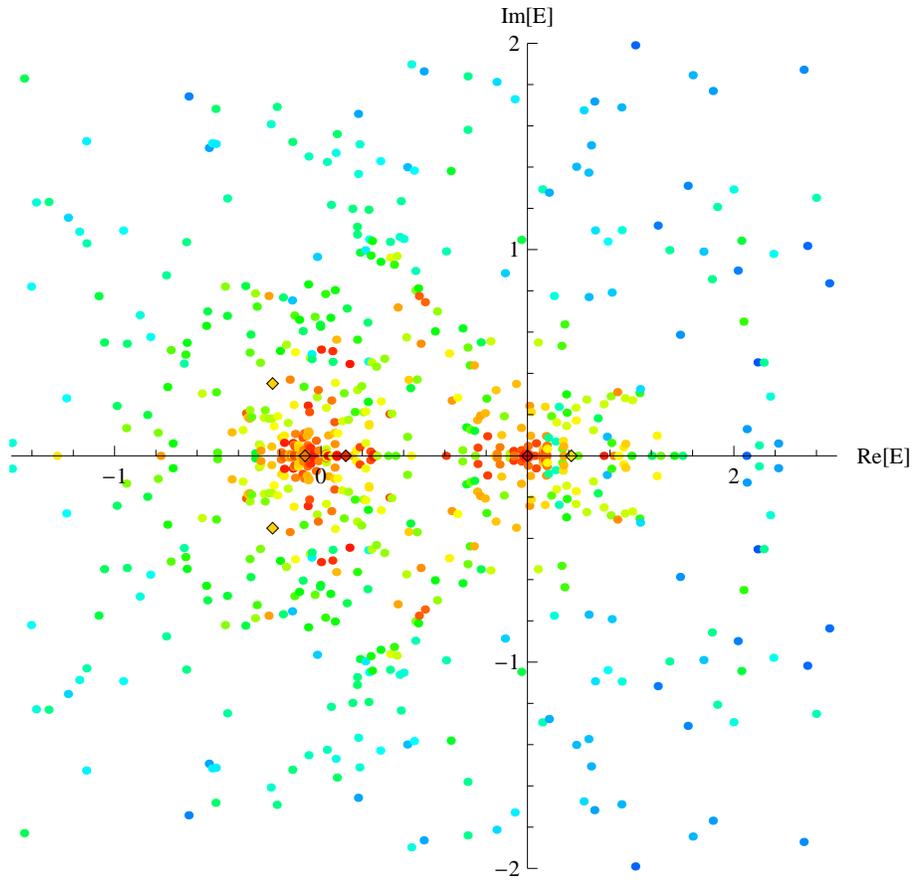}
\caption{Twist even universal solutions without gauge fixing at level 2 marked by framed diamonds and level 4 presented as dots. The color is given by the difference between energy and Ellwood invariant, see figure \ref{fig:colors}. The best solutions are red, the worst are blue.}\label{fig:NG lev4 col}
\end{figure}

\FloatBarrier
\subsection{Solutions without the twist symmetry}

\begin{figure}
\centering
\includegraphics[width=14cm]{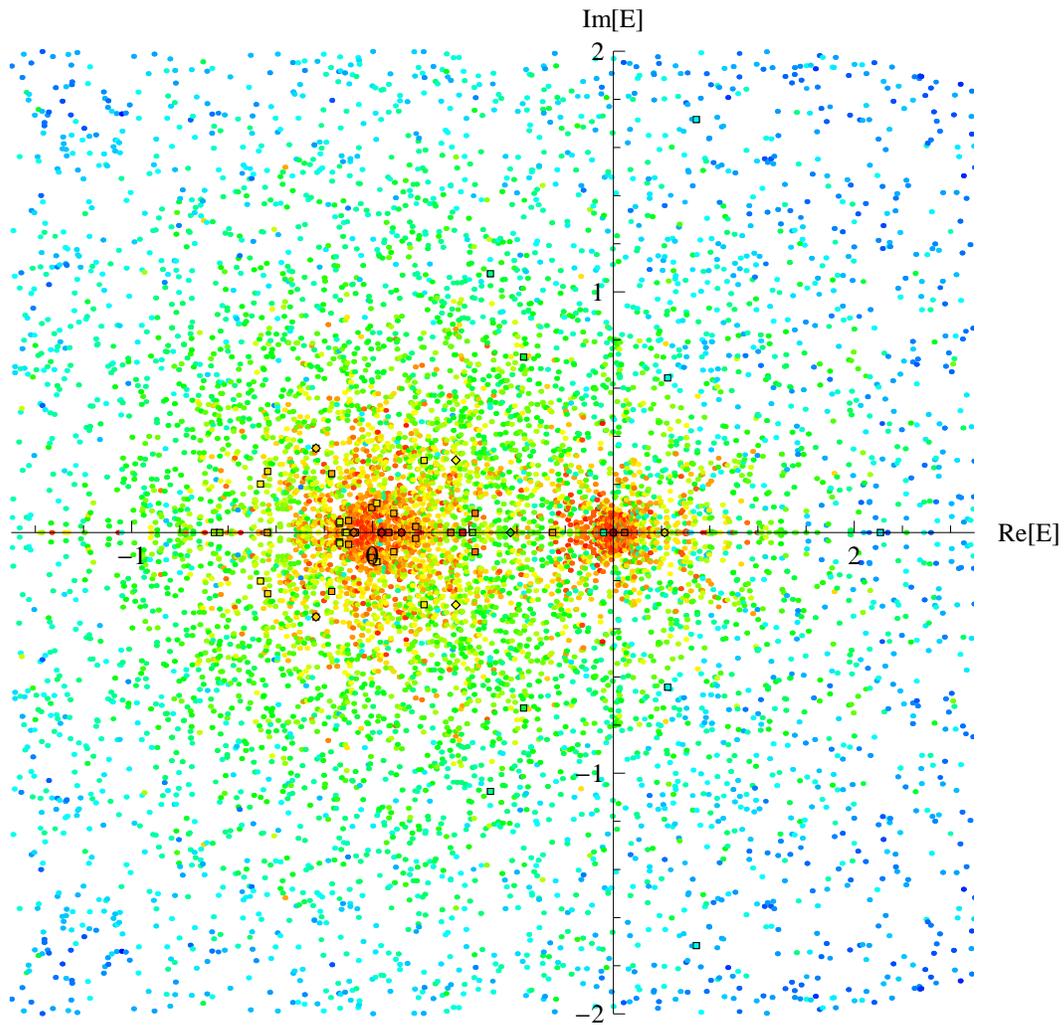}
\caption{Universal solutions without imposing the Siegel gauge and without imposing twist-even condition.  Solutions are shown at levels 2 (framed diamonds), 3 (framed squares) and 4 (circles). There are over 17000 solutions in this figure. }\label{fig:NGT lev4 col}
\end{figure}

In the previous section we saw that relaxing the twist symmetry (and the singlet condition) led in the gauge fixed case to new interesting solutions, namely the meronic branes. What does happen when we relax the twist symmetry for the gauge unfixed case? The number of string field components grows rapidly, but we were still able to construct all solutions up to level 4, same as for the twist even case. At level 4 we found around one million solutions, including of course the twist symmetric ones which take up about $1/64$ of the total number.  We have plotted them in figure \ref{fig:NGT lev4 col}. Clearly, there are many more solutions compared to the twist even case, but nothing substantially new emerges from the analysis. There are few reddish dots along the real axis, near integer or half-integer values $-3/2, -1,\ldots, 3/2$, but statistically this appears insignificant. One would need other criteria to ascertain the physical significance of these solutions and filter out the level-truncation artifact solutions.

\section{Conclusions, Discussion}

This work presents first comprehensive study of the landscape of universal classical solutions in open string field theory. Aside of the tachyon vacuum where we reached level 30 and demonstrated the non-monotonicity of the tachyon vacuum energy as a function of the level, we have found also a number of intriguing solutions which might correspond to multiple, ghost or meronic branes. With the exception of the double brane, which seems to become real at higher levels, the rest of the new exotic solutions violates the reality condition of string field theory, as they should.

The main question is whether these solutions are indeed true solutions of full string field theory. For our Siegel gauge solutions we have computed one parameter $\Delta_S$ which measures to what extent is the lowest level out-of-Siegel-gauge equation satisfied. This parameter got closest to zero for the ghost brane, where we also found the best match between energy computed from the action and the closed string overlap. Also for the ghost brane the agreement with the conjectured value was the best. What about the other solutions? Is it possible to deform these solutions slightly away from the Siegel gauge in a way which would lower the $\Delta_S$ parameter? One might consider deforming the set of the equations, to include some of the out-of-Siegel equations, and/or deform the gauge fixing condition for the solution. We expect that for the physical solutions the gauge invariants would change slightly towards the correct values, and in its vicinity they would not change significantly any more.

One possible approach, wherein all components of the string field and all equations are included without any gauge fixing, does not work as we have discussed in section \ref{sec:noGF}. There appear too big jumps in Newton's method when fields at the next level are included.  It is not clear to us why this happens, since it happens even when the Jacobian is not badly conditioned.  We have tried also the least modification of the Siegel gauge equations by including the lowest level out-of-Siegel component of the string field. Solutions to our equations then automatically satisfy $\Delta_S=0$. However, we found, that allowing even the minimal one-dimensional gauge orbit affects the Newton's method badly. Even the tachyon vacuum solution is not stable with this scheme. We clearly have to invent something better.

It is interesting to point out, that even in the Siegel gauge there are very few stable solutions found from a large number of solutions found at levels 4 to 6. Most solutions are unstable, which means that typically in the few steps of the Newton's method the solution wonders far away. All the solutions with $|E|<30$ that we found to be stable became at some level either singlet and twist even, or became symmetric under the $\Omega (-1)^J$ symmetry. It seems that stability likes symmetry, but we do not know why in our case this is so.

Most interesting future direction would be to study the cohomology around these universal solutions. Even for the tachyon vacuum some puzzles and controversies still remain \cite{ET2,GI,EFHM,EllS}. Can we see the $U(2)$ symmetry arising for the double brane? What about the ghost brane, and the meronic branes? Answering these questions in string field theory might elevate our understanding of these objects to a new level.

\section*{Acknowledgments}
\noindent

This research has been supported in part by the Czech Science Foundation (GA\v{C}R) grant 17-22899S and by the European Regional
Development Fund and the Czech Ministry of Education, Youth and Sports (M\v{S}MT), project No. CZ.02.1.01/0.0/0.0/15\_003/0000437.
Computational resources were provided by the CESNET LM2015042 and the CERIT Scientific Cloud LM2015085, provided under the programme "Projects of Large Research, Development, and Innovations Infrastructures".

\begin{appendix}

\section{Ghost current conservation law} \label{app: j ghost}

In this appendix we derive conservation laws (\ref{Jcons-law}) for the ghost current $j^{gh}$ which are helpful for computing the Ellwood invariant. We use the same approach and notation as in \cite{KMS} extending the earlier results of \cite{KKT, Kishimoto:2008}.
Let us define
\begin{equation}
J_n^{gh}=j_n^{gh}+(-1)^{n}j_{-n}^{gh}=\oint_0 \frac{dw}{2\pi i} g_n(w) j^{gh}(w),
\end{equation}
where we have introduced the function
\begin{equation}
g_n(w)=w^{n}+(-1)^{n}w^{-n}.
\end{equation}
We now wish to evaluate the action of $J_n^{gh}$ on the Ellwood state $\bra{E[V]} \equiv \bra{I} c\bar{c}V(i,-i)=\bra{0} U_f c\bar{c}V(i,-i)$, where $V$ is an arbitrary matter primary field with weights $(1,1)$ and $U_f$ represents conformal transformation given by $f(w)=\frac{2w}{1-w^2}$.

To follow the next steps, we remind the reader that the ghost current transforms anomalously
\begin{equation}\label{ghost current transformation}
\tilde j^{gh} (w)=\frac{dz}{dw}j^{gh}(z)+q\frac{d^2z}{dw^2}\left(\frac{dz}{dw}\right)^{-1},
\end{equation}
where $q=-\frac{3}{2}$ and $\tilde j^{gh}\equiv j^{gh}_w$ is the holomorphic component of the ghost current in the new $w$-coordinate.

To evaluate $\bra{E[V]} J_n^{gh}$ we begin by moving $U_f$ to the right
\begin{eqnarray}\label{ghost in}
\bra 0 U_f c\bar{c}V(i,-i)  J_n^{gh}&=& \oint_0 \frac{dw}{2\pi i} g_n(w)\bra 0 c\bar{c}V(i,-i)U_f j^{gh}(w) \\
&=& \oint_0 \frac{dw}{2\pi i} g_n(w)\bra 0 c\bar{c}V(i,-i)\left(f'(w)j^{gh}(f(w))+q\frac{f''(w)}{f'(w)}\right) U_f\nonumber.
\end{eqnarray}
%We emphasize that from now on the components of all fields are given in the $f(w)$ coordinate although we suppress the vector indices. {\bf[We never discussed vector indices. Are we suppressing tilda as well?]}
The anomalous term can be computed directly, since it does not contain the ghost current
\begin{eqnarray}
&&-\frac{3}{2}\oint_0 \frac{dw}{2\pi i} g_n(w)\frac{f''(w)}{f'(w)}=-3\oint_0 \frac{dw}{2\pi i}\left(w^{n}+(-1)^{n}w^{-n}\right)\frac{w(3+w^2)}{1-w^4} \nonumber\\
&=&-3\oint_0 \frac{dw}{2\pi i}\left(w^{n}+(-1)^{n}w^{-n}\right)w(3+w^2)\sum_{m=0}^\infty w^{4m} \nonumber\\
&=&
\begin{cases}
0  & \quad \text{if } n=0\ \text{or } n=2k+1,\\
-9 & \quad \text{if } n=4k+2,\qquad\qquad\qquad k\in \mathds{Z}\\
-3 & \quad \text{if } n=4k.\\
\end{cases} \label{ghost an1}
\end{eqnarray}

 Let us return to the first term in (\ref{ghost in}) and deform the integration contour, analogously to \cite{KMS}, so that it shrinks around the point at infinity. There are six points in the complex plane with possible singularities: 0, $\pm 1$, $\pm i$ and $\infty$. Observe that $\oint_{0}=\oint_{\infty}$ follows easily from the transformations of the functions appearing in (\ref{ghost in}) under $w \to -\frac{1}{w}$
\begin{eqnarray}
 f\left(-\frac{1}{w}\right)&=&f(w), \nonumber \\
 f'\left(-\frac{1}{w}\right)&=&w^2f'(w), \nonumber \\
 g_n\left(-\frac{1}{w}\right)&=&g_n(w).
\end{eqnarray}
We express the contour around 0 as
\begin{equation}
\oint_{0}=-\frac{1}{2}\oint_{(i,-i)}-\frac{1}{2}\oint_{(1,-1)}.
\end{equation}

The residues at $\pm i$ are easy to compute using the OPE between the ghost current and the $c$ ghost
\begin{eqnarray}
&&-\frac{1}{2}\oint_{(i,-i)} \frac{dw}{2\pi i} g_n(w)f'(w)\bra 0 c\bar{c}V(i,-i)j^{gh}(f(w)) U_f \nonumber \\
&=&-\frac{1}{2}\oint_{i} \frac{dw}{2\pi i} g_n(w)f'(w)\bra 0 \frac{c\bar{c}V(i,-i)}{f(w)-i} U_f
-\frac{1}{2}\oint_{-i} \frac{dw}{2\pi i} g_n(w)f'(w)\bra 0 \frac{c\bar{c}V(i,-i)}{f(w)+i} U_f  \nonumber \\
&=& -2\left(i^n+(-i)^n\right) \bra 0 c\bar{c}V(i,-i) U_f. \label{ghost reg}
\end{eqnarray}

Finally we compute the residues at $\pm 1$. These terms were not present in \cite{KMS} because they receive contributions only from the anomalous behavior of $j^{gh}$.
To compute the contour integrals we %first
introduce coordinate $z=f(w)$
% and then, to move the residues from $\infty$ to 0, we perform a second coordinate transformation $u=-\frac{1}{z}$
\begin{eqnarray}\label{ghost anom}
&&-\frac{1}{2}\oint_{(1,-1)} \frac{dw}{2\pi i} g_n(w)f'(w)\bra 0 c\bar{c}V(i,-i)j^{gh}(f(w)) U_f \nonumber \\
&=&-\oint_\infty \frac{dz}{2\pi i} g_n(f^{-1}(z))\bra 0 c\bar{c}V(i,-i)j^{gh}(z) U_f
%\nonumber \\
%&=&-\oint_0 \frac{du}{2\pi i} g_n\left(f^{-1}\left(-\frac{1}{u}\right)\right)\bra 0 c\bar{c}V(i,-i)\frac{j^{gh}(-\frac{1}{u})}{u^2} U_f.
\end{eqnarray}
%The inverse function $f^{-1}(z)$ contains a square root, but fortunately the functions $g_n(f^{-1}(z))$ turn out to be rational functions of $z$ without any branch cuts. We can contract the contour since the insertion from string field was moved to $u=\infty$ by the $-\frac{1}{u}$ transformation. As we pointed out the ghost current is now in the $z$ coordinate. In order to evaluate the integral we need to transform it to the $u$ coordinate using (\ref{ghost current transformation})
%\begin{equation}
%-\oint_0 \frac{du}{2\pi i} g_n\left(f^{-1}\left(-\frac{1}{u}\right)\right)\bra 0 c\bar{c}V(i,-i)\left(j^{gh}(u)+\frac{2q}{u}\right) U_f.
%\end{equation}
%Now the first term with $j^{gh}(u)$ disappears, since $g_n(w)$ is a regular function around $\pm 1$ and we are left only with the anomaly term. To simplify the expression we change the coordinates back to $w$ and we get
%\begin{eqnarray}
%&&-2q\oint_0 \frac{du}{2\pi i} g_n\left(f^{-1}\left(-\frac{1}{u}\right)\right)\frac{1}{u}
%=2q\oint_1 \frac{dw}{2\pi i} g_n(w)\frac{f'(w)}{f(w)}  \nonumber \\
%&=&2q\oint_1 \frac{dw}{2\pi i}\left(w^{n}+(-1)^{n}w^{-n}\right)\frac{1+w^2}{w(1-w^2)}   \nonumber \\
%&=& \begin{cases}
%6  & \quad \text{if } n \text{ is even}\\
%0  & \quad \text{if } n \text{ is odd}
%\end{cases}. \label{ghost an2}
%\end{eqnarray}
The functions $g_n(f^{-1}(z))$ turn out to be rational. For large $|z|$ they behave as $2+O(z^{-2})$ for $n$ even and $O(z^{-1})$ for $n$ odd. The vanishing terms do not contribute due to the conformal weight of the ghost current, while the constant term for $n$ even contributes thanks to the ghost number anomaly, note that the ghost charge obeys $\bra{0} Q^{gh} = 3 \bra{0}$ so we get
\be
\bra 0 c\bar{c}V(i,-i) U_f \times
\begin{cases}
6  & \quad \text{if } n \text{ is even}\\
0  & \quad \text{if } n \text{ is odd}
\end{cases}.
\label{ghost an2}
\ee
By combining the three contributions (\ref{ghost an1}), (\ref{ghost reg}) and (\ref{ghost an2}) we get the desired simple result
\begin{equation}
\bra{E[V]} J_n^{gh}=\biggl(-\frac{1}{2}\left(i^n+(-i)^n\right)  +3\delta_{n,0} \biggr) \bra{E[V]}.
\end{equation}

\section{Numerical algorithms} \label{app: numerics}

In this appendix we present various details about the numerical algorithms we have used and their implementation on multi-core computers. Most of this technology described below has been used also in our other works \cite{KRS,marginalKM}.

\FloatBarrier
\subsection{Solving the equations of motion}

In the string field theory we face two different problems that require two different approaches. One is to find a single solution of the high level equations from a convenient starting point, while the second problem is to find all solutions of low level equations.

\subsubsection{Newton's method}

The first case can be solved very efficiently by Newton's method. We expand the string field as $\ket{\Psi}=\sum_i t_i \ket{i}$. Then we define
\begin{eqnarray}
Q_{ij} &=& \la i|Q_B|j\ra, \\
V_{ijk} &=& \la V_3|i\ra|j\ra|k\ra,
\end{eqnarray}
so that the level-truncated action and equations of motion\footnote{In Siegel gauge these equations correspond only to a projection $b_0(Q|\Psi\ra+|\Psi\ra\ast |\Psi\ra) =0$ of the complete equations of motion.} are written as
\begin{eqnarray}
S&=&-\frac{1}{g_o^2}\left(\frac{1}{2} \sum_{ij}  Q_{ij}t_it_j+\frac{1}{3} \sum_{ijk}V_{ijk}t_it_jt_k\right),\\
f_i(t)& \equiv &\sum_{j} Q_{ij}t_j+\sum_{jk} V_{ijk} t_j t_k = 0. \label{eom truncated}
\end{eqnarray}
%where the repeated indices are summed over.
This system of quadratic equations can be solved efficiently by Newton's recursion
\begin{eqnarray}\label{Newton iterations}
t^{(n+1)}_i=t^{(n)}_i- \sum_{j} M^{-1}_{ij}(t^{(n)})f_j(t^{(n)}),
\end{eqnarray}
provided that we are given reasonable starting point. The Jacobian matrix $M_{ij}$ is given by
\begin{equation}\label{Jacobian matrix}
M_{ij}(t)=\frac{\del f_i(t)}{\del t_j}=-g_o^2\frac{\del^2 S(t)}{\del t_i\del t_j}=Q_{ij}+2 \sum_{k} V_{ijk}t_k.
\end{equation}

In practise we continue the Newton's method iterations until we reach the target precision for $\frac{\| t^{(n+1)}-t^{(n)}\|}{\|t^{(n)}\|}$ defined using the Euclidean norm.\footnote{For comparison we have tried several other norms for the vector $t_i$ and they all give essentially the same results.} We require precision of $10^{-12}$, which is usually reached within 4 or 5 iterations.

We solve the linear equations (\ref{Newton iterations}) by LU decomposition, which requires $O(N^3)$ operations, where $N$ is the total number of component fields we consider.  The calculation of the matrix $M_{ij}$ has the same complexity, but it takes more time because the vertices $V_{ijk}$ are factorized into the matter and ghost parts, so that there are more operations to be done. Naively it seems that evaluation of (\ref{eom truncated}) should also have the complexity  $O(N^3)$, however notice that it can be expressed in terms of $M_{ij}$ as
\begin{equation}\label{Newton eom from M}
f_i(t)=\frac{1}{2}\sum_{j} \left( M_{ij}+Q_{ij} \right) t_j.
\end{equation}
This formula requires only $O(N^2)$ operations and it needs only a negligible amount of time.

\subsubsection{Homotopy continuation method}
\label{app: homotopy}

The homotopy continuation method is a very efficient tool for finding {\em all} solutions of a system of polynomial equations. In string field theory we use it to find starting points for the Newton's method by fully solving the equations of motion at some low level.
%We will present our application of this method to the OSFT equations, the detailed description can be find e.g. in {\bf [References]}.

The homotopy continuation method deals with a system of $N$ polynomial equations
\begin{equation}
f_i(t_j)=0,\quad i,j=1\dots N
\end{equation}
in $N$ variables denoted here as $t_i$. In Witten's OSFT this system is given by quadratic equations (\ref{eom truncated}).
The basic idea of this algorithm is to take a simpler system of equations, which has explicitly known solutions, and continuously deform both the equations and their solutions until we reach the target system of equations. We will call the new system a start system and denote it as $g_i(t_j)=0$. The start system must have at least as many solutions as the target system, otherwise we could miss some of the solutions. The simplest choice for the start system is such that each equation is given by a polynomial in a single variable of the same order as the corresponding target equation. For the quadratic equations we encounter in OSFT a convenient start system is
\begin{equation}
g_i(t_j)=t_i(a_i t_i+b_i),
\end{equation}
where $a_i$, $b_i$ are some nonzero numbers. For numerical stability it is convenient to choose them of the same order as the coefficients in the target equations (\ref{eom truncated}), for example $a_i=\max\limits_{j,k \leq N}|V_{ijk}|$, $b_i=\max\limits_{j\leq N}|Q_{ij}|$.

In order to track the solutions from the start system to the target system we define a homotopy map
\begin{equation}
H_i(t_j,\alpha)=(1-\alpha) g_i(t_j) \gamma + \alpha f_i(t_j),
\end{equation}
where $0\leq\alpha\leq 1$ and $\gamma$ is some non-real complex number. Notice that at $\alpha=0$ the homotopy is equal to the start system and at $\alpha=1$ to the target system which we are interested in. The constant $\gamma$ is introduced so that we can reach complex solutions of the target system even when the initial solutions are real for convenience.

The path tracking of the solutions from $\alpha=0$ to 1 has to be done in finite steps. We use the predictor-corrector method with adaptive step-size in $\alpha$.
The algorithm proceeds via the following steps
\begin{enumerate}
   \item After $n$-th step of the algorithm we start with $\alpha_n$ and solution $t_i^{(n)}$.
   \item Increase the homotopy parameter by a given step $\Delta\alpha$ so that $\alpha_{n+1}=\min(\alpha_n+\Delta\alpha,1)$.
   \item Estimate the solution of the deformed equations $H_i(t_j,\alpha)=0$ at $\alpha_{n+1}$. The simplest possibility is $t_i^{(n+1)}=t_i^{(n)}$, however extrapolation of the previous path gives much better prediction. We use second order extrapolation in our code.
       % First order: $t_i^{(n+1)}=t_i^{(n)}+(\alpha_{n+1}-\alpha_n)\frac{t_i^{(n)}-t_i^{(n-1)}}{\alpha_n-\alpha_{n-1}}$
   \item Correct the predicted solution by Newton's method so that it satisfies $H_i(t_j^{(n+1)},\alpha_{n+1})=0$. However the Newton's method is allowed only given number of iterations $M$.
   \item If Newton's method converged within $M$ iterations and the solution satisfies $\|t_i^{(n+1)}-t_i^{(n)} \|<\epsilon \| t_i^{(n)} \|$ we accept the solution and move to the next step, possibly with increased $\Delta\alpha$ if the solution did not change significantly.
       Otherwise the solution is not accepted and we reduce the step-size $\Delta\alpha$ by a factor of 2.
   \item Return to point 1 and repeat these steps until we reach $\alpha=1$ meaning that we have found a regular solution, or $\Delta\alpha<(\Delta\alpha)_{min}$ showing that the solution of the deformed system fails to exist at some value of $\alpha$.
\end{enumerate}

The restrictions on the number of iterations of Newton's method $M$ and the relative change of the solution $\epsilon$ prevent the solution from jumping to a different path. We typically choose $M=5$ and $\epsilon\approx 0.1$. The step-size $\alpha$ is chosen initially at $0.01$ but it quickly adapts to a more convenient value controlled by the parameter $\epsilon$. We never let it increase beyond $(\Delta\alpha)_{max} =0.1$ and drop below $(\Delta\alpha)_{min} = 10^{-15}$ which would indicate that the target system has fewer solutions compared to the starting one. This does indeed happen generically for the OSFT equations of motion.
Solutions with multiplicity greater than 1 can be run again with smaller value of $\epsilon$ to verify their status. In practice, however, solutions with higher multiplicity in OSFT do not arise, except for the perturbative vacuum when no gauge is fixed.

The complexity of the above algorithm for quadratic equations is $N^3 2^N$, where $2^N$ comes from the number of solutions and $N^3$ from the Newton's method. Since the solutions are treated independently this algorithm can be parallelized in a very straightforward way. With our C++ code we were able to solve fully at most 26 equations with the available computer resources.
Considering that the number of states in OSFT grows exponentially with the level, we can find all solutions only at very low levels, that is level 6 for twist even solutions in Siegel gauge and less in all other cases. While this work has been in progress, a similar method was in the meantime implemented in Mathematica 10 for the NSolve function. Nonetheless, we still use our own C++ code since it has a direct communication with the rest of our codes and the optimalization for OSFT equations makes it faster.

\subsubsection{Modifications of the homotopy method}

We have also tried to modify the homotopy method to get a connection between OSFT equations at two different levels. Assume we have level $L_1$ with $N_1$ states and equations $f_i^{(1)}(t_j)=0$ and level $L_2>L_1$ with $N_2$ states and equations $f_i^{(2)}(t_j)=0$. Then we can take the start system of the homotopy to be $f_i^{(1)}(t_j)=0$ for $i=1,\dots, N_1$ and some $g_i(t_j)=0$ for $i=N_1+1,\dots, N_2$. We have tried two possibilities for $g_i$.

First possibility is to take simple linear equations $g_i(t_j)=a_i t_i$. In this case we get only a single solution at the higher level. This method can be used in case the Newton's method for the target system at level $L_2$ does not converge in reasonable amount of iterations. In Siegel gauge it usually reproduces the known solutions (although it takes significantly more time). For the calculations without gauge fixing it allows us to find solutions at level higher that 4, however the solutions are usually very distant from the lower level solutions. We can also use this approach to make a smooth interpolation of the energy (or some coefficient) between two different levels. However this interpolation depends on the choice of $g_i$.

The second possibility is to use $g_i(t_j)=t_i(a_i t_i+b_i)$. In this case we get $2^{N_2-N_1}$ new solutions for any given solution at the lower level. We have used it to find some solutions without gauge fixing at level 6 starting from nice solutions like perturbative or tachyon vacuum in a hope that there exists at least some preservation of the character of the solution.

\subsection{Parallelization}
\label{app: parallelization}

Nowadays, not only supercomputers, but also desktop machines are built with multicore processors. To take advantage of this fact we have designed most of our algorithms as parallel codes. Since the key parts of our codes were written in C++, we used OpenMP library.  We had to our disposal a computer time at CESNET and CERIT Scientific Cloud where we found most convenient for our purposes two machines with similar parameters, one of them {\tt ungu.cerit-sc.cz} with  46x6-core Intel Xeon E5-4617 processor and 5TB of shared memory and another one {\tt urga.cerit-sc.cz} with somewhat better parameters. For the computations presented in this paper we have typically used the optimal number of about 50 cores simultaneously. With more cores there was not a substantial gain in performance, but to the contrary we would have to wait longer for the assignment of the requested computer time.

There are three areas where we found parallelization indispensable: computation of the matter and ghost vertices, Newton's method and linear homotopy method.  The last of these is {\em embarrassingly parallel}, meaning that the computation of the roots proceeds by independent evolutions of the roots of the deformed system, see Appendix \ref{app: homotopy}.

In Newton's method we found convenient to speed up the calculations of the Jacobian matrix (\ref{Jacobian matrix}) as well as calculations of its inverse.
At the highest levels we were able to reach (level 30 for tachyon vacuum, or level 26 for the "half-ghost brane") where there are about $O(10^5)$ nontrivial coefficients, we had to compute
for every iteration of Newton's method --- usually four or five of them --- $O(10^9)$ entries of the Jacobian matrix, each requiring about $O(10^{-3})$ seconds to compute. To reduce the single iteration time under one day we used 50 cores in parallel with dynamic scheduling. Once a given core finished computing the assigned entry of the Jacobian matrix, it was assigned a new one. We have also parallelized the computation of the inverse via LU decomposition by dividing the matrix row operations between the individual cores.

In the computation of the ghost and matter vertices we formed a three dimensional array of vertices, where the three directions correspond to the states in the given sector ordered by level (and some quite random ordering within a given level). We moved through the array with an obvious triple {\tt for}-cycle, collapsed for parallel evaluation. It was not guaranteed that when a given vertex was assigned for computation, all the required lower ones were already computed. In such a case the given core started computing recursively also the lower level vertex, which introduced slight inefficiency. We could have in principle broken every cycle over the basis fields into two: one over the level, and another over fields at that level. This would have ensured that each vertex is computed exactly ones. However this approach effectively prevents us from using parallelization due to the overhead time. It would also require computing all of the auxiliary vertices, while empirically we need only about half of them. A curious technicality we had to deal with were large errors arising when one core was writing down a value of a just computed vertex, while another core was already attempting to read its value. For that matter we have introduced a boolean array data structure for all the vertices keeping track whether that particular vertex has been already computed or not.

\subsection{Time and memory requirements}

The maximal level that can be reached in level truncation is determined by the time and memory requirements and the availability of suitable computer.
In the universal subsector of the string field theory the memory requirements are more constraining, so we will discuss them first.

Most of the needed memory is required to store the cubic vertices, even though we kept them only in the factorized form, in matter and ghost sectors separately. For the initial calculation we considered only vertices ordered by the cyclic and twist symmetry, so the required number of vertices in the matter sector with $N^{\mathrm{matter}}$ states was roughly $\frac{1}{6} \left(N^{\mathrm{matter}}\right)^3$, and analogously in the ghost sector when no gauge fixing is imposed. In Siegel gauge, however, we also need a set of auxiliary vertices both in singlet and non-singlet cases. The number of auxiliary vertices for the singlet case equals to $\frac{1}{2} {\tilde N}^{\mathrm{aux}} ({\tilde N}^{\mathrm{singlet}})^2$, while in the non-singlet case we need states at ghost numbers 0 and 2, so we need  ${\tilde N}^\mathrm{Siegel}_0 {\tilde N}^\mathrm{Siegel}_1 {\tilde N}^\mathrm{Siegel}_2$ vertices.  We use tildes to distinguish purely ghost sector number of states from the combined matter plus ghost theory.  The numbers of states in the individual subspaces are given in table \ref{tab:states 1}, note that in Siegel gauge ${\tilde N}^\mathrm{Siegel}_0  =  {\tilde N}^\mathrm{Siegel}_2$.  Each vertex is a rational number times a power of $\sqrt{3}$, but for practical purposes we represent it by a real number with 15 digit precision taking up 8 bytes of memory. The memory needed to store the vertices is given in table \ref{tab:memory}.  The actual memory requirements are higher by another $1/8$th, because for every vertex we need an auxiliary boolean number recording its status, see appendix \ref{app: parallelization} for explanation. In C++ the boolean variables take up full one byte of memory.

After we evaluate all the matter and ghost vertices and deallocate the auxiliary structures we restore the full set of vertices in both sectors using the cyclic and twist symmetry. This process does not require much time and, as we are going to show, it does not increase the overall memory requirements by much. Storing the full set of vertices is beneficial for the Newton's method, because if we had to reorder the indices of the vertices every time we need to access them, it would slow down the Newton's method method approximately by a factor of four.
In the Newton's method itself most memory is needed for the Jacobian matrix, which scales as $N^2$. Asymptotically in the singlet case $N \sim \left(N^{\mathrm{matter}}\right)^{\sqrt{2}}$, so the Jacobian matrix is using less space than the the cubic vertices.

At level 30 in the SU(1,1) singlet basis one needs approximately 2.6 TB of memory for the reduced ghost vertices and the auxiliary vertices. The memory for the full set of matter and ghost vertices is given by twelve times the first column in table \ref{tab:memory} and it equals also approximately 2.6~TB. In reality we need about 3~TB for some intermediate manipulations and storage of various smaller objects, but the overall memory increase due to considering full set of vertices is minimal.
To demonstrate the usefulness of the matter-ghost factorization we can compare this number with the memory that would be required for the non-factorized vertices,  which is approximately
4.5~PB, so the factorization reduces the memory requirements by three orders of magnitude.
In the generic Siegel gauge basis we were able to reach level 26. At this level we needed approximately 1.3~TB for the evaluation of the reduced ghost vertices and approximately 1.9~TB for the full vertices. This time the memory requirements grow a bit more, but not by a drastic amount.

When it comes to time, which is needed to execute the calculations, both the cubic vertices and the Newton's method play a significant role. The time needed for the cubic vertices scales primarily with the number of vertices, that is by $\left(N^{\mathrm{matter}}\right)^3$ and likewise in the ghost sector. The evaluation of the matter vertices takes significantly less time than the ghost vertices, for which one needs to evaluate the auxiliary sector. A closer look reveals that the time to compute a single vertex grows with the level, because the conservation laws like (\ref{Lconslaw}) have more nontrivial terms. So $L \left(N^{\mathrm{matter}}\right)^3$ is a better estimate. The time also depends quite significantly on the complexity of the operator algebra, for example a single vertex in the basis of $b$ and $c$ ghosts takes much less time than a vertex in the Virasoro basis. Thanks to that, the evaluation of SU(1,1) singlet vertices and all Siegel gauge vertices at the same level takes a similar amount of time, despite the fact that the second set of vertices is several times larger.

In the Newton's method both the evaluation time of the Jacobian matrix and solving the corresponding system of linear equations by LU decomposition scale as $N^3$. The Jacobian matrix requires 3--4 times more time, because we have to put together the factorized vertices, so introducing a more sophisticated method to solve the linear equations would not help us significantly.

Asymptotically at very large level we can expect that the time requirements for Newton's method are going to be dominant, because $N^3\gg N_\mathrm{ghost}^3$, but at the available levels the evaluation of cubic vertices takes similar amount of time. This is caused by several factors: the auxiliary vertices in ghost sector, large time needed to evaluate a single vertex due to complexity of the conservation laws and lesser efficiency of parallelization of the cubic vertex recurrent algorithm.
%The exact ratio depends on the chosen basis and on number and properties of the solutions.

Now we can discuss the advantages of using the SU(1,1) singlet string field compared to the generic $b$ and $c$ ansatz. As we mentioned before, the time needed for the cubic vertices is comparable, but from table \ref{tab:memory} we can see that required memory at level 30 is lower approximately by one order of magnitude. In the Newton's method the use of SU(1,1) singlet string field reduces the required time also approximately by a factor of 10. If we wanted to use SU(1,1) singlet string field together with non-singlet basis of vertices, we would have to multiply the vertices with a transformation matrix, which would take $O\left(\left(N^\mathrm{ghost}\right)^4\right)$ operations.

\begin{table}\nonumber
\centering
\begin{tabular}{|l|ll|ll|l|}\hline
$L$ & ${\tilde N}^\mathrm{singlet}$ & ${\tilde N}^\mathrm{aux}$ & ${\tilde N}^\mathrm{Siegel}_1$ & ${\tilde N}^\mathrm{Siegel}_{0,2}$ & ${\tilde N}^\mathrm{generic}$
\rule{0pt}{14pt}
\rule[-1.2ex]{0pt}{0pt}
\\\hline
2  & 2     & 1     & 2     & 1     & 4     \\
4  & 5     & 4     & 7     & 3     & 12    \\
6  & 11    & 12    & 17    & 8     & 30    \\
8  & 22    & 30    & 37    & 20    & 67    \\
10 & 42    & 67    & 76    & 44    & 139   \\
12 & 77    & 139   & 148   & 89    & 272   \\
14 & 135   & 272   & 275   & 171   & 508   \\
16 & 231   & 508   & 493   & 315   & 915   \\
18 & 385   & 915   & 857   & 561   & 1597  \\
20 & 627   & 1597  & 1451  & 970   & 2714  \\
22 & 1002  & 2714  & 2403  & 1635  & 4508  \\
24 & 1575  & 4508  & 3902  & 2696  & 7338  \\
26 & 2436  & 7338  & 6224  & 4360  & 11732 \\
28 & 3718  & 11732 & 9774  & 6930  & 18460 \\
30 & 5604  & 18460 & 15131 & 10847 & 28629 \\
32 & 8349  & 28629 & 23119 & 16742 & 43820 \\
34 & 12310 & 43820 & 34907 & 25511 & 66273 \\\hline
\end{tabular}
\caption{Number of states in the ghost BCFT {\em up to} a given level with various conditions imposed. Starting from the left we show the number of states in the SU(1,1) singlet subspace (which also equals the number of states in the universal matter BCFT, i.e. ${\tilde N}^\mathrm{singlet} = N^\mathrm{matter}$) and in the associated auxiliary space with a single $j_{-k}^{gh}$ operator. Then we count all states in Siegel gauge with ghost number one and the auxiliary states with ghost numbers 0 or 2 which happen to be equal. In the last column we impose no gauge condition, so the number of states is the same as in the generic Virasoro representation. In the auxiliary spaces we count only states up to one level less compared to the physical space, because states at the highest level are not needed for the cubic vertices thanks to the structure of the conservation laws.}
\label{tab:states 1}
\end{table}

\begin{table}\nonumber
\centering
\begin{tabular}{|l|ll|ll|ll|}\hline
$L$  & $N^\mathrm{singlet}_\mathrm{even}$ & $N^\mathrm{singlet}$ & $N^\mathrm{Siegel}_\mathrm{even}$ & $N^\mathrm{Siegel}$ & $N^\mathrm{generic}_\mathrm{even}$ & $N^\mathrm{generic}$ \rule{0pt}{14pt}
\rule[-1.2ex]{0pt}{0pt}
\\\hline
2  & 3      & 3      & 3      & 3      & 4       & 5       \\
4  & 8      & 10     & 9      & 12     & 14      & 20      \\
6  & 21     & 29     & 26     & 38     & 43      & 65      \\
8  & 51     & 75     & 69     & 106    & 118     & 185     \\
10 & 117    & 181    & 171    & 272    & 299     & 481     \\
12 & 259    & 413    & 402    & 653    & 712     & 1165    \\
14 & 549    & 895    & 898    & 1482   & 1607    & 2665    \\
16 & 1124   & 1866   & 1925   & 3218   & 3473    & 5822    \\
18 & 2236   & 3760   & 3985   & 6726   & 7233    & 12230   \\
20 & 4328   & 7352   & 7995   & 13602  & 14585   & 24842   \\
22 & 8176   & 14008  & 15606  & 26733  & 28593   & 49010   \\
24 & 15121  & 26085  & 29736  & 51232  & 54678   & 94235   \\
26 & 27419  & 47575  & 55433  & 95989  & 102253  & 177087  \\
28 & 48841  & 85175  & 101323 & 176246 & 187428  & 326015  \\
30 & 85604  & 149938 & 181927 & 317724 & 337366  & 589128  \\
32 & 147809 & 259891 & 321352 & 563264 & 597257  & 1046705 \\
34 & 251719 & 444135 & 559168 & 983373 & 1041392 & 1831065 \\\hline
\end{tabular}
\caption{Number of states in the full string field with various conditions imposed up to a given level. Left column counts the SU(1,1) singlet states, the middle column gives number of states in the Siegel gauge string field and finally the right column shows number of states present in a string field without any gauge fixing. In all three cases we show the effect of imposing the twist even condition.}
\label{tab:states 2}
\end{table}

\begin{table}\nonumber
\centering
\begin{tabular}{|l|rr|rr|r|}\hline
$L$  & $V^\mathrm{singlet}$ & $V^\mathrm{singlet}_\mathrm{aux}$ & $V^{bc}$ & $V^{bc}_\mathrm{aux}$ & $V^\mathrm{Vir}$
\rule{0pt}{14pt}
\rule[-1.2ex]{0pt}{0pt}
\\\hline
2  & 32    B  & 24    B  & 32    B  & 16    B  & 160   B  \\
4  & 280   B  & 480   B  & 672   B  & 504   B  & 2.844 kB \\
6  & 2.234 kB & 6.188 kB & 7.57  kB & 8.5   kB & 38.75 kB \\
8  & 15.81 kB & 59.3  kB & 71.4  kB & 115.6 kB & 409.3 kB \\
10 & 103.5 kB & 472.7 kB & 594.3 kB & 1.123 MB & 25.87 MB \\
12 & 617.8 kB & 3.185 MB & 4.206 MB & 8.944 MB & 29.11 MB \\
14 & 3.198 MB & 19.05 MB & 26.73 MB & 61.35 MB & 167.7 MB \\
16 & 15.88 MB & 103.9 MB & 153.3 MB & 373.2 MB & 977.3 MB \\
18 & 73.13 MB & 518.7 MB & 803.2 MB & 2.01  GB & 5.067 GB \\
20 & 314.9 MB & 2.343 GB & 3.801 GB & 10.17 GB & 24.85 GB \\
22 & 1.253 GB & 10.16 GB & 17.25 GB & 47.86 GB & 113.8 GB \\
24 & 4.861 GB & 41.69 GB & 73.83 GB & 211.3 GB & 490.9 GB \\
26 & 17.97 GB & 162.3 GB & 299.5 GB & 881.5 GB & 1.959 TB \\
28 & 63.87 GB & 604.3 GB & 1.133 TB & 3.415 TB & 7.63  TB \\
30 & 218.7 GB & 2.109 TB & 4.202 TB & 12.95 TB & 28.46 TB \\
32 & 722.9 GB & 7.261 TB & 14.99 TB & 47.15 TB & 102.  TB \\
34 & 2.263 TB & 24.16 TB & 51.58 TB & 165.3 TB & 353.  TB \\\hline
\end{tabular}
\caption{Minimum memory requirements for storing various sets of ghost vertices. From the left we give the memory needed for SU(1,1) singlet vertices reduced by cyclic and twist symmetry (which is the same as for the matter vertices), and the requirements for the corresponding auxiliary vertices, then the memory needed for the ordered vertices in Siegel gauge in $bc$ basis, and for the corresponding auxiliary vertices, and finally memory requirements for ordered vertices without any gauge condition. We assume 8~B of memory for one number in C++ double format. Note that for the evaluation of the ordered vertices we need additional 1~B for an auxiliary boolean array. The total memory requirements for the full unordered set of vertices is given approximately by six times the memory needed for the ordered vertices. }
\label{tab:memory}
\end{table}

\subsection{Extrapolations to infinite level}
\label{app: fits}

It has been observed by Taylor \cite{Taylor-Pade} that one can obtain remarkably accurate results for the energy difference in tachyon condensation at higher levels from extrapolating results at lower truncation levels. Taylor applied Pad\'{e} approximation method to the tachyon potential (since the tachyon vacuum is located beyond the radius of its convergence), where all the coefficients were obtained by extrapolation from level truncated string field theory. This extrapolation is nicely justified by results from his previous paper \cite{Taylor-Pert} where he provided an evidence that the coefficients of the tachyon potential at large levels can be expanded in a series in $1/L$.

Gaiotto and Rastelli \cite{GaiottoRastelli} have successfully verified Taylor's predictions up to level 18 by intensive numerical computations, but they also noticed that it is possible to obtain results for the energy and other observables at higher levels by plainly fitting the results at lower levels with a polynomial in $1/L$ of maximum level.\footnote{Obviously one has to exclude the result at $L=0$, or fit with a polynomial in $1/(L+a)$, where $a$ is an $O(1)$ constant. For tachyon condensation the results do not depend significantly on $a$.  For the computations in this work it is conceptually more logical to exclude the results at $L=0$ and the first few lower levels as well, since the more exotic solutions do not exist at those levels.}

In this work we have studied many classical solutions in the universal sector which we obtained from different starting points. For each of those solutions we have computed several observables, to learn about their relevance from physics point of view. The truth is that, there is no unique simple universal extrapolation procedure that would work in all the cases. On the other hand we wanted to avoid a case-by-case analysis, since we could run into a danger of selecting the fits that confirm our hypotheses.

After some experimentation we have finally adopted the following extrapolation algorithm: For a given solution and a given observable, we compute extrapolations to $L=\infty$ by considering fitting functions of different order, from linear, i.e. $a+b/L$, to the maximum order polynomial fit.  Then we have repeated the computation with the lowest level result excluded and checked how much it affected the result. As the "predicted" value at $L=\infty$ we took the one coming from the extrapolation order with the least dependence on the lowest level included. We had estimated an error of the extrapolation procedure by considering five best orders and computed the statistical deviation $\sigma$. Admittedly, considering different fitting orders does not correspond to fully random choices, so the error estimated cannot be entirely trusted. For the tachyon vacuum, the error computed by this prescription is clearly underestimated, see Table \ref{tab:TV}, the correct value lives within $4\sigma$ from the obtained fit.

For the Ellwood invariant extrapolation there is another option which works quite nicely. In the twist-even case one can extrapolate separately the results obtained at levels $L=4k$ and $L=4k+2$, and in twist non-even case one can split the results into four sets $L = 0,1,2,3 \mod 4$ and obtain 4 different extrapolations. We do not show the results of this procedure in this work.

%What is the convenient form of the function and number of its parameters depend on the solution and the quantity we wish to extrapolate. Usually the energy allows us to use fit with large number of parameters, while the Ellwood invariants, which tend to oscillate with level, require more robust low order fits.
%
%In this article we use polynomial in $1/L$ as an ansatz for the extrapolations.
%\begin{equation}\label{fit 1}
%E^M(L)=\sum_{k=0}^M\frac{a_k}{L^k}.
%\end{equation}
%We also experimented with a more general formula
%\begin{equation}\label{fit 2}
%E^M(L)=a_0+\sum_{k=1}^M\frac{a_k}{L^{k+\alpha}},
%\end{equation}
%that takes into account the leading behavior can be different from $1/L$.
%
%It is difficult to estimate the precision of the extrapolations. The errors given by Mathematica are clearly too small. We usually try to vary either the order of the fit or the number of data points and then take an average and standard deviation from these numbers. Unfortunately this does not take into account systematic errors like a wrong choice of the fitting function.
%
%{\bf Note about the extrapolations: The order of infinite level fit is chosen such that it has minimal difference between using all data and data without the lowest level. $\sigma$ is a standard deviation of 5 best orders using this criteria.}  {\bf [Describe this better!!!]}

Admittedly for some solutions the fits do not quite work, which means that the extrapolation to $L=\infty$ is very sensitive to the number of data point included or the order of the fitting polynomial chosen. We expect that this happens when the dependence of the family of level truncated solutions on $1/L$ is nonanalytic. First, such a non-analyticity might happen at some fixed level, when the solution changes abruptly its character. This is the case of the "half ghost brane" of this paper, or the positive energy solution for the Ising model in \cite{KRS}.
Another possibility is that the approximate level truncation solution is jumping between different exact but unstable solutions. This would be the case if the solution changed significantly every time a level is increased.
%{\bf [We should have paid attention to the norm of the difference in the first iteration of the Newton's method at every level. We did later. The nice solutions are OK.]}
Final possibility is that the dependence of the observable in question is non-analytic in $1/L$ around zero. This can be due to a benign periodic modulation in $L$ as we see happening in the case of Ellwood invariants. More serious issue would be if the series in $1/L$ were only asymptotic. This possibility was discussed already in \cite{GaiottoRastelli} where it was concluded that this is probably not the case for the tachyon vacuum in Siegel gauge. On the other hand, for the simple solution \cite{ES} studied in ${\cal L}_0$-level expansion, the energy is indeed given by a divergent asymptotic sum which has to be resummed.

\FloatBarrier
\section{First coefficients of the solutions}
\label{app: coeff}

In this section we show the first three coefficients $c_1|0\ra$, $L_{-2}^m c_1|0\ra$ and $L'^{gh}_{-2}c_1|0\ra \equiv c_{-1}|0\ra$ for the interesting Siegel gauge solutions. It would be nice to compare them with possible analytic solutions, should they be found.

\begin{table}[h]\nonumber
\centering
\begin{tabular}{|l|lll|}\hline
Level    & $c_1|0\ra$ & $L_{-2}^m c_1|0\ra$ & $L'^{gh}_{-2}c_1|0\ra$ \\\hline
2        & 0.544204   & 0.0559637           & 0.19019                \\
4        & 0.548399   & 0.0569235           & 0.205673               \\
6        & 0.547932   & 0.0571435           & 0.211815               \\
8        & 0.547052   & 0.0572141           & 0.215025               \\
10       & 0.546261   & 0.0572411           & 0.216983               \\
12       & 0.545608   & 0.057252            & 0.218296               \\
14       & 0.545075   & 0.0572562           & 0.219236               \\
16       & 0.544637   & 0.0572573           & 0.219942               \\
18       & 0.544272   & 0.0572569           & 0.220491               \\
20       & 0.543964   & 0.0572558           & 0.22093                \\
22       & 0.543702   & 0.0572543           & 0.221288               \\
24       & 0.543476   & 0.0572528           & 0.221587               \\
26       & 0.54328    & 0.0572512           & 0.22184                \\
28       & 0.543107   & 0.0572496           & 0.222056               \\
30       & 0.542955   & 0.0572481           & 0.222243               \\\hline
$\infty$ & 0.540493   & 0.0572118           & 0.224830               \\
$\sigma$ & 0.000002   & 0.0000004           & 0.000001               \\\hline
$\infty$ & 0.540493   & 0.0572127           & 0.224840               \\\hline
\end{tabular}
\caption{First three coefficients for the tachyon vacuum solution.}
\label{tab:coeff TV}
\end{table}

\begin{table}\nonumber
\centering
\begin{tabular}{|l|lll|}\hline
Level    & $\ps c_1|0\ra$            & $L_{-2}^m c_1|0\ra$    & $\ps L'^{gh}_{-2}c_1|0\ra$ \\\hline
2        & $   -0.248369-0.583223 i$ & $0.085433-0.300309  i$ & $   -0.137895-0.282985 i$ \\
4        & $   -0.009731-0.482707 i$ & $0.124141-0.15809   i$ & $\ps 0.008769-0.250395 i$ \\
6        & $\ps 0.109579-0.432809 i$ & $0.130541-0.10662   i$ & $\ps 0.071990-0.22202  i$ \\
8        & $\ps 0.176236-0.389536 i$ & $0.128308-0.0785356 i$ & $\ps 0.103823-0.196227 i$ \\
10       & $\ps 0.216513-0.353096 i$ & $0.124039-0.0611306 i$ & $\ps 0.121856-0.175035 i$ \\
12       & $\ps 0.242778-0.322863 i$ & $0.119651-0.0495016 i$ & $\ps 0.13296 -0.157939 i$ \\
14       & $\ps 0.260981-0.297571 i$ & $0.115649-0.04127   i$ & $\ps 0.140258-0.143975 i$ \\
16       & $\ps 0.274219-0.276094 i$ & $0.112125-0.035172  i$ & $\ps 0.145302-0.132351 i$ \\
18       & $\ps 0.284219-0.257563 i$ & $0.109046-0.0304875 i$ & $\ps 0.148929-0.12249  i$ \\
20       & $\ps 0.292009-0.241335 i$ & $0.106352-0.0267817 i$ & $\ps 0.151621-0.11398  i$ \\
22       & $\ps 0.298231-0.226934 i$ & $0.103984-0.0237783 i$ & $\ps 0.153669-0.106524 i$ \\
24       & $\ps 0.303307-0.214002 i$ & $0.101889-0.0212947 i$ & $\ps 0.155261-0.099906 i$ \\
26       & $\ps 0.307519-0.202268 i$ & $0.100023-0.0192054 i$ & $\ps 0.156518-0.093965 i$ \\
28       & $\ps 0.311068-0.191522 i$ & $0.098352-0.0174218 i$ & $\ps 0.157527-0.088575 i$ \\\hline
$\infty$ & $\ps 0.3522  -0.04     i$ & $0.066   +0.009     i$ & $\ps 0.164   -0.01     i$ \\
$\sigma$ & $\ps 0.0004  +0.04     i$ & $0.002   +0.002     i$ & $\ps 0.002   +0.02     i$ \\\hline
%$\infty$ & $\ps 0.352194+0.159836 i$ & $0.066048+0.0091391 i$ & $\ps 0.161077+0.071191 i$ \\\hline
\end{tabular}
\caption{First coefficients for the "double brane" solution. Note that the coefficients are likely to become asymptotically real.}
\label{tab:coeff DB}
\end{table}

\begin{table}\nonumber
\centering
\begin{tabular}{|l|lll|}\hline
Level    & $\ps c_1|0\ra$             & $L_{-2}^m c_1|0\ra$     & $L'^{gh}_{-2}c_1|0\ra$ \\\hline
4        & $   -0.13981  -0.782315 i$ & $0.0448139-0.081877  i$ & $1.41615 +0.0787593 i$ \\
6        & $\ps 0.0065121-0.563642 i$ & $0.0471727-0.0550217 i$ & $0.939357-0.0158866 i$ \\
8        & $\ps 0.0545914-0.469837 i$ & $0.0460908-0.0456129 i$ & $0.766149-0.048512  i$ \\
10       & $\ps 0.0786389-0.41484  i$ & $0.0448967-0.0405147 i$ & $0.673741-0.0641042 i$ \\
12       & $\ps 0.0930916-0.377582 i$ & $0.043875 -0.0371817 i$ & $0.615214-0.07279   i$ \\
14       & $\ps 0.102743 -0.35013  i$ & $0.043023 -0.0347678 i$ & $0.574291-0.0780674 i$ \\
16       & $\ps 0.109647 -0.328761 i$ & $0.042307 -0.0329044 i$ & $0.543773-0.0814451 i$ \\
18       & $\ps 0.114829 -0.311471 i$ & $0.0416966-0.031402  i$ & $0.519957-0.0836723 i$ \\
20       & $\ps 0.118861 -0.297076 i$ & $0.0411687-0.030152  i$ & $0.500738-0.0851597 i$ \\
22       & $\ps 0.122085 -0.284824 i$ & $0.0407063-0.0290871 i$ & $0.484822-0.086149  i$ \\
24       & $\ps 0.124721 -0.27421  i$ & $0.0402965-0.028163  i$ & $0.471369-0.0867912 i$ \\
26       & $\ps 0.126914 -0.264886 i$ & $0.0399299-0.027349  i$ & $0.459807-0.0871844 i$ \\
28       & $\ps 0.128766 -0.256597 i$ & $0.0395991-0.0266232 i$ & $0.449734-0.0873951 i$ \\\hline
$\infty$ & $\ps 0.1503   -0.07     i$ & $0.035    -0.009     i$ & $0.27    -0.09      i$ \\
$\sigma$ & $\ps 0.0006   +0.01     i$ & $0.002    +0.003     i$ & $0.02    +0.01      i$ \\\hline
$\infty$ & $\ps 0.1478   -0.071851 i$ & $0.0314867-0.0093631 i$ & $0.269869-0.0572439 i$ \\\hline
\end{tabular}
\caption{First coefficients for the "ghost brane" solution. Note the relative importance of the third coefficient.}
\label{tab:coeff GB}
\end{table}

\begin{table}\nonumber
\centering
\begin{tabular}{|l|lll|}\hline
Level    & $\ps c_1|0\ra$             & $L_{-2}^m c_1|0\ra$      & $\ps c_{-1}|0\ra$       \\\hline
4        & $   -0.044501-0.572857  i$ & $0.0326938-0.0644381  i$ & $-1.02516 +0.0482026 i$ \\
5        & $\ps 0.092967-0.421043  i$ & $0.0353087-0.0436127  i$ & $-0.693577+0.0431765 i$ \\
6        & $\ps 0.122358-0.346867  i$ & $0.036831 -0.0357198  i$ & $-0.576494+0.0611767 i$ \\
7        & $\ps 0.150092-0.293272  i$ & $0.0370089-0.0295124  i$ & $-0.501029+0.0561691 i$ \\
8        & $\ps 0.159049-0.258534  i$ & $0.0369315-0.0261961  i$ & $-0.45759 +0.0590126 i$ \\
9        & $\ps 0.17103 -0.227718  i$ & $0.0367733-0.0228909  i$ & $-0.423374+0.0543592 i$ \\
10       & $\ps 0.175249-0.205506  i$ & $0.0365268-0.0207998  i$ & $-0.400175+0.053743  i$ \\
11       & $\ps 0.181938-0.183777  i$ & $0.0363366-0.0185415  i$ & $-0.380316+0.0494988 i$ \\
12       & $\ps 0.184361-0.167179  i$ & $0.0360953-0.0169636  i$ & $-0.365661+0.0476819 i$ \\
13       & $\ps 0.188629-0.149913  i$ & $0.0359206-0.0151925  i$ & $-0.352527+0.0436856 i$ \\
14       & $\ps 0.190189-0.136172  i$ & $0.0357096-0.0138652  i$ & $-0.342316+0.0412969 i$ \\
15       & $\ps 0.193148-0.121183  i$ & $0.0355565-0.0123346  i$ & $-0.332897+0.0373681 i$ \\
16       & $\ps 0.194231-0.108798  i$ & $0.0353751-0.0111178  i$ & $-0.32531 +0.0345623 i$ \\
17       & $\ps 0.196402-0.0946463 i$ & $0.0352417-0.00967257 i$ & $-0.318172+0.0304738 i$ \\
18       & $\ps 0.197193-0.0823945 i$ & $0.0350853-0.00844877 i$ & $-0.312272+0.0271527 i$ \\
19       & $\ps 0.198853-0.0674423 i$ & $0.0349684-0.00691792 i$ & $-0.306642+0.0224758 i$ \\
20       & $\ps 0.199455-0.0532407 i$ & $0.0348324-0.00547702 i$ & $-0.301897+0.0180795 i$ \\
21       & $\ps 0.200764-0.0321788 i$ & $0.034729 -0.0033119  i$ & $-0.29732 +0.0110322 i$ \\
22       & $\ps 0.220148            $ & $0.0365609             $ & $-0.299988            $ \\
23       & $\ps 0.24517             $ & $0.0389438             $ & $-0.304649            $ \\
24       & $\ps 0.257413            $ & $0.0400753             $ & $-0.305768            $ \\
25       & $\ps 0.268656            $ & $0.0410695             $ & $-0.306397            $ \\
26       & $\ps 0.276103            $ & $0.0417286             $ & $-0.306422            $ \\\hline
$\infty$ & $\ps 0.58                $ & $0.07                  $ & $-0.34                $ \\\hline
\end{tabular}
\caption{First coefficients for the "half ghost brane" solution.}
\label{tab:coeff HGB}
\end{table}

\begin{table}\nonumber
\centering
\begin{tabular}{|l|lll|}\hline
Level    & $\ps c_1|0\ra$            & $L_{-2}^m c_1|0\ra$    & $\ps c_{-1}|0\ra$         \\\hline
4        & $   -0.337305-0.864196 i$ & $0.203606-0.218207  i$ & $   -1.1627  +0.155229 i$ \\
5        & $   -0.223613-0.761672 i$ & $0.184722-0.207311  i$ & $   -0.809587+0.190767 i$ \\
6        & $   -0.045184-0.679141 i$ & $0.179075-0.138666  i$ & $   -0.681504+0.194112 i$ \\
7        & $   -0.027395-0.647622 i$ & $0.173714-0.13678   i$ & $   -0.600435+0.20938  i$ \\
8        & $\ps 0.060692-0.596611 i$ & $0.166383-0.101405  i$ & $   -0.562729+0.197846 i$ \\
9        & $\ps 0.067399-0.58064  i$ & $0.163745-0.100943  i$ & $   -0.525435+0.20574  i$ \\
10       & $\ps 0.119007-0.541597 i$ & $0.156378-0.0793106 i$ & $   -0.506944+0.192101 i$ \\
11       & $\ps 0.122334-0.531723 i$ & $0.154835-0.0792157 i$ & $   -0.485236+0.197    i$ \\
12       & $\ps 0.155776-0.500668 i$ & $0.148223-0.0647733 i$ & $   -0.473431+0.184408 i$ \\
13       & $\ps 0.157675-0.493862 i$ & $0.147226-0.0647985 i$ & $   -0.459108+0.187791 i$ \\
14       & $\ps 0.180903-0.468566 i$ & $0.141496-0.0545498 i$ & $   -0.450464+0.176796 i$ \\
15       & $\ps 0.182085-0.463539 i$ & $0.140808-0.054617  i$ & $   -0.440241+0.179298 i$ \\
16       & $\ps 0.199067-0.442506 i$ & $0.135877-0.047002  i$ & $   -0.433418+0.169802 i$ \\
17       & $\ps 0.199846-0.438611 i$ & $0.135378-0.0470823 i$ & $   -0.425716+0.171743 i$ \\
18       & $\ps 0.21276 -0.420806 i$ & $0.13112 -0.0412175 i$ & $   -0.420085+0.163518 i$ \\
19       & $\ps 0.213295-0.417679 i$ & $0.130745-0.0412992 i$ & $   -0.414049+0.165076 i$ \\
20       & $\ps 0.223424-0.402371 i$ & $0.127043-0.0366513 i$ & $   -0.409263+0.157899 i$ \\
21       & $\ps 0.223802-0.399793 i$ & $0.126753-0.0367297 i$ & $   -0.40439 +0.159184 i$ \\
22       & $\ps 0.231949-0.386456 i$ & $0.123507-0.0329594 i$ & $   -0.400239+0.152868 i$ \\
23       & $\ps 0.232221-0.384285 i$ & $0.123278-0.0330327 i$ & $   -0.396211+0.15395  i$ \\
24       & $\ps 0.23891 -0.372531 i$ & $0.12041 -0.0299149 i$ & $   -0.392556+0.148346 i$ \\\hline
$\infty^{(e)}$ & $\ps 0.3057  -0.13     i$ & $0.076   +0.0014    i$ & $   -0.276   -0.06     i$ \\
$\sigma^{(e)}$ & $\ps 0.0007  +0.02     i$ & $0.002   +0.0003    i$ & $\ps 0.005   +0.01     i$ \\\hline
$\infty^{(o)}$ & $\ps 0.3060  -0.123    i$ & $0.0693  +0.0018    i$ & $   -0.275   +0.057    i$ \\
$\sigma^{(o)}$ & $\ps 0.0006  +0.004    i$ & $0.0007  +0.0005    i$ & $\ps 0.001   +0.002    i$ \\\hline
\end{tabular}
\caption{First coefficients for the "half brane" solution. The extrapolations with the $^{(e)}$ or $^{(o)}$ superscripts are obtained from results at even or odd levels only respectively. }
\label{tab:coeff HB}
\end{table}

\FloatBarrier
\section{Quadratic identities}
\label{app: quadratic identities}

In this appendix we show how well are the first four quadratic identities (\ref{quadids}) for both the Virasoro and ghost current generators obeyed for our most interesting Siegel gauge solutions. We show the ratios $R_n$ and $\tilde R_n$ introduced in (\ref{quadidsR}).

\begin{table}[h]\nonumber
\centering
\begin{tabular}{|l|llll|}\hline
Level    & $R_1$    & $R_2$    & $R_3$    & $R_4$    \\\hline
2        & 1.12793  & 0        & 0        & 0        \\
4        & 1.06964  & 1.07986  & 0        & 0        \\
6        & 1.04647  & 1.0519   & 1.05352  & 0        \\
8        & 1.03459  & 1.03755  & 1.04077  & 1.03698  \\
10       & 1.02744  & 1.0293   & 1.03137  & 1.03308  \\
12       & 1.02269  & 1.02398  & 1.02537  & 1.0268   \\
14       & 1.01931  & 1.02026  & 1.02126  & 1.02232  \\
16       & 1.01679  & 1.01752  & 1.01828  & 1.01908  \\
18       & 1.01485  & 1.01542  & 1.01602  & 1.01664  \\
20       & 1.0133   & 1.01377  & 1.01425  & 1.01474  \\
22       & 1.01204  & 1.01243  & 1.01282  & 1.01323  \\
24       & 1.01099  & 1.01132  & 1.01165  & 1.01199  \\
26       & 1.01011  & 1.01039  & 1.01067  & 1.01096  \\
28       & 1.00936  & 1.0096   & 1.00985  & 1.01009  \\
30       & 1.00871  & 1.00892  & 1.00913  & 1.00935  \\\hline
$\infty$ & 0.999972 & 0.999974 & 0.99998  & 1.000    \\
$\sigma$ & 0.000009 & 0.000006 & 0.00005  & 0.002    \\\hline
%$\infty$ & 0.999971 & 0.999977 & 0.999986 & 0.999945 \\\hline
\multicolumn{5}{c}{}\\\hline
Level    & $J_1$    & $J_2$    & $J_3$    & $J_4$    \\\hline
2        & 1.00276  & 0        & 0        & 0        \\
4        & 1.00581  & 0.973577 & 0        & 0        \\
6        & 1.00304  & 0.996478 & 0.961485 & 0        \\
8        & 1.00173  & 0.998889 & 0.991382 & 0.955352 \\
10       & 1.00105  & 0.999493 & 0.996215 & 0.988281 \\
12       & 1.00066  & 0.999696 & 0.997878 & 0.994403 \\
14       & 1.00042  & 0.999774 & 0.998633 & 0.996696 \\
16       & 1.00026  & 0.999807 & 0.999032 & 0.997807 \\
18       & 1.00016  & 0.999822 & 0.999266 & 0.998428 \\
20       & 1.00009  & 0.999829 & 0.999413 & 0.998808 \\
22       & 1.00003  & 0.999832 & 0.999512 & 0.999057 \\
24       & 0.999997 & 0.999835 & 0.999582 & 0.999228 \\
26       & 0.999969 & 0.999837 & 0.999633 & 0.999352 \\
28       & 0.999948 & 0.999839 & 0.999671 & 0.999443 \\
30       & 0.999933 & 0.999841 & 0.999701 & 0.999513 \\\hline
$\infty$ & 0.999995 & 0.999996 & 1.0000   & 0.999    \\
$\sigma$ & 0.000002 & 0.000003 & 0.0002   & 0.005    \\\hline
\end{tabular}
\caption{First quadratic identities of tachyon vacuum.
%{\bf Last line uses maximum order polynomial fit, whereas the previous one use the maximum-stability order.}
}
\label{tab:quadratic TV}
\end{table}

\begin{table}\nonumber
\centering
\footnotesize{
\begin{tabular}{|l|llll|}\hline
Level    & $\ps R_1                $ & $R_2                $ & $R_3                $     & $R_4                    $ \\\hline
2        & $   -0.797517+1.48063  i$ & $0                  $ & $0                  $     & $\ps 0                  $ \\
4        & $\ps 0.38078 +0.694967 i$ & $0.062054+0.235578 i$ & $0                  $     & $\ps 0                  $ \\
6        & $\ps 0.612704+0.519635 i$ & $0.451043+0.309178 i$ & $0.014913+0.290702 i$     & $\ps 0                  $ \\
8        & $\ps 0.697104+0.402875 i$ & $0.60919 +0.328165 i$ & $0.4011  +0.346463 i$     & $   -0.021659+0.321987 i$ \\
10       & $\ps 0.738174+0.326964 i$ & $0.691165+0.290894 i$ & $0.561506+0.350386 i$     & $\ps 0.348972+0.370768 i$ \\
12       & $\ps 0.763124+0.276214 i$ & $0.735675+0.252711 i$ & $0.652812+0.308207 i$     & $\ps 0.513092+0.367442 i$ \\
14       & $\ps 0.780214+0.24022  i$ & $0.763086+0.222535 i$ & $0.705686+0.266364 i$     & $\ps 0.613335+0.322764 i$ \\
16       & $\ps 0.792791+0.213325 i$ & $0.781631+0.19908  i$ & $0.739118+0.233158 i$     & $\ps 0.674195+0.278458 i$ \\
18       & $\ps 0.802501+0.192361 i$ & $0.795041+0.180466 i$ & $0.761949+0.207317 i$     & $\ps 0.713584+0.242893 i$ \\
20       & $\ps 0.81026 +0.175452 i$ & $0.805218+0.165305 i$ & $0.778484+0.186855 i$     & $\ps 0.74078 +0.21504  i$ \\
22       & $\ps 0.816624+0.161432 i$ & $0.813226+0.152649 i$ & $0.791006+0.170257 i$     & $\ps 0.760569+0.192937 i$ \\
24       & $\ps 0.821952+0.149539 i$ & $0.81971 +0.141856 i$ & $0.800826+0.156472 i$     & $\ps 0.775575+0.175017 i$ \\
26       & $\ps 0.826488+0.139254 i$ & $0.825077+0.132478 i$ & $0.80874 +0.144779 i$     & $\ps 0.787333+0.160166 i$ \\
28       & $\ps 0.830401+0.130215 i$ & $0.829604+0.124199 i$ & $0.815261+0.134673 i$     & $\ps 0.796793+0.147606 i$ \\\hline
$\infty$ & $\ps 0.8882  -0.01     i$ & $0.89    -0.07     i$ & $0.87    -0.10     i$     & $\ps 0.77    -0.2      i$ \\
$\sigma$ & $\ps 0.0007  +0.04     i$ & $0.01    +0.04     i$ & $0.02    +0.06     i$     & $\ps 0.07    +0.2      i$ \\\hline
\multicolumn{5}{c}{}\\\hline
Level    & $\ps J_1                $ & $J_2                $ & $\ps J_3                $ & $\ps J_4                $ \\\hline
2        & $   -0.996762+1.62962  i$ & $0                  $ & $\ps 0                  $ & $\ps 0                  $ \\
4        & $\ps 0.344201+0.751959 i$ & $0.039738+0.246206 i$ & $\ps 0                  $ & $\ps 0                  $ \\
6        & $\ps 0.604029+0.553983 i$ & $0.448032+0.324584 i$ & $   -0.010686+0.295371 i$ & $\ps 0                  $ \\
8        & $\ps 0.698777+0.42207  i$ & $0.615048+0.340496 i$ & $\ps 0.398205+0.358732 i$ & $   -0.046084+0.322118 i$ \\
10       & $\ps 0.745722+0.336541 i$ & $0.702378+0.297104 i$ & $\ps 0.567956+0.359751 i$ & $\ps 0.346131+0.380163 i$ \\
12       & $\ps 0.774568+0.279701 i$ & $0.750376+0.253957 i$ & $\ps 0.664918+0.312357 i$ & $\ps 0.519779+0.374602 i$ \\
14       & $\ps 0.794444+0.23971  i$ & $0.780209+0.220289 i$ & $\ps 0.721408+0.26624  i$ & $\ps 0.625903+0.325478 i$ \\
16       & $\ps 0.80911 +0.210102 i$ & $0.800529+0.194414 i$ & $\ps 0.757285+0.229975 i$ & $\ps 0.69052 +0.277423 i$ \\
18       & $\ps 0.82044 +0.187247 i$ & $0.815294+0.174113 i$ & $\ps 0.781865+0.201992 i$ & $\ps 0.732422+0.239121 i$ \\
20       & $\ps 0.829488+0.168999 i$ & $0.826536+0.157766 i$ & $\ps 0.799707+0.180028 i$ & $\ps 0.761391+0.209326 i$ \\
22       & $\ps 0.836899+0.15402  i$ & $0.835404+0.144275 i$ & $\ps 0.813243+0.162375 i$ & $\ps 0.782486+0.185855 i$ \\
24       & $\ps 0.843095+0.141441 i$ & $0.842595+0.132899 i$ & $\ps 0.823869+0.147854 i$ & $\ps 0.798491+0.166974 i$ \\
26       & $\ps 0.84836 +0.130673 i$ & $0.848555+0.123125 i$ & $\ps 0.832441+0.135653 i$ & $\ps 0.811036+0.151455 i$ \\
28       & $\ps 0.852896+0.121303 i$ & $0.853584+0.114591 i$ & $\ps 0.839509+0.125212 i$ & $\ps 0.821131+0.138441 i$ \\\hline
$\infty$ & $\ps 0.919  +0.01      i$ & $0.916   -0.06     i$ & $\ps 0.91    -0.1      i$ & $\ps 0.8     -0.2      i$ \\
$\sigma$ & $\ps 0.001  +0.02      i$ & $0.001   +0.03     i$ & $\ps 0.03    +0.1      i$ & $\ps 0.2     +0.3      i$ \\\hline
\end{tabular} }
\caption{First quadratic identities of "double brane".}
\label{tab:quadratic DB}
\end{table}

\begin{table}\nonumber
\centering
\footnotesize{
\begin{tabular}{|l|llll|}\hline
Level    & $R_1                 $     & $\ps R_2                 $ & $\ps R_3                 $ & $\ps R_4                 $ \\\hline
4        & $0.161263-0.0157066 i$     & $   -0.11755+0.0603157  i$ & $\ps 0                   $ & $\ps 0                   $ \\
6        & $0.494824-0.0682754 i$     & $\ps 0.114839-0.0514479 i$ & $   -0.12916+0.0266329  i$ & $\ps 0                   $ \\
8        & $0.630135-0.0701986 i$     & $\ps 0.442368-0.08277   i$ & $\ps 0.099833-0.0793153 i$ & $   -0.12582+0.0012891  i$ \\
10       & $0.702753-0.0664846 i$     & $\ps 0.595454-0.078676  i$ & $\ps 0.412188-0.097738  i$ & $\ps 0.094414-0.0977405 i$ \\
12       & $0.748378-0.0620671 i$     & $\ps 0.679395-0.0719109 i$ & $\ps 0.571049-0.0885736 i$ & $\ps 0.391861-0.109436  i$ \\
14       & $0.779907-0.0579203 i$     & $\ps 0.731796-0.065763  i$ & $\ps 0.661129-0.0788049 i$ & $\ps 0.552207-0.0971392 i$ \\
16       & $0.803121-0.0542292 i$     & $\ps 0.767567-0.0605603 i$ & $\ps 0.718003-0.0707628 i$ & $\ps 0.645876-0.0852009 i$ \\
18       & $0.821001-0.050986  i$     & $\ps 0.793578-0.0561871 i$ & $\ps 0.756899-0.0643102 i$ & $\ps 0.705879-0.0756421 i$ \\
20       & $0.835247-0.0481371 i$     & $\ps 0.813391-0.0524828 i$ & $\ps 0.78512 -0.0590795 i$ & $\ps 0.747184-0.0681139 i$ \\
22       & $0.846899-0.0456236 i$     & $\ps 0.829026-0.0493099 i$ & $\ps 0.806536-0.0547663 i$ & $\ps 0.777221-0.0621042 i$ \\
24       & $0.856631-0.043393  i$     & $\ps 0.841708-0.0465616 i$ & $\ps 0.82336 -0.0511485 i$ & $\ps 0.80001 -0.0572144 i$ \\
26       & $0.864899-0.0414011 i$     & $\ps 0.852227-0.0441563 i$ & $\ps 0.836947-0.0480669 i$ & $\ps 0.817891-0.0531608 i$ \\
28       & $0.872024-0.0396115 i$     & $\ps 0.861109-0.0420314 i$ & $\ps 0.848168-0.0454063 i$ & $\ps 0.832301-0.0497433 i$ \\\hline
$\infty$ & $0.980   -0.010     i$     & $\ps 0.976   -0.006     i$ & $\ps 0.98    -0.003     i$ & $\ps 1.0     -0.01      i$ \\
$\sigma$ & $0.005   +0.002     i$     & $\ps 0.003   +0.001     i$ & $\ps 0.01    +0.008     i$ & $\ps 0.1     +0.03      i$ \\\hline
\multicolumn{5}{c}{}\\\hline
Level    & $\ps J_1                 $ & $\ps J_2                 $ & $\ps J_3                 $ & $\ps J_4                 $ \\\hline
4        & $   -0.0200887+0.122136 i$ & $   -0.059915+0.153637  i$ & $\ps 0                   $ & $\ps 0                   $ \\
6        & $\ps 0.416668-0.0166129 i$ & $\ps 0.034192+0.0150489 i$ & $   -0.094000+0.0822858 i$ & $\ps 0                   $ \\
8        & $\ps 0.598124-0.0388519 i$ & $\ps 0.404203-0.0514278 i$ & $\ps 0.051587-0.0359164 i$ & $   -0.101888+0.0414975 i$ \\
10       & $\ps 0.691364-0.0444604 i$ & $\ps 0.582434-0.0575168 i$ & $\ps 0.389374-0.0747689 i$ & $\ps 0.06144 -0.0655705 i$ \\
12       & $\ps 0.747734-0.0451485 i$ & $\ps 0.679136-0.0557842 i$ & $\ps 0.565648-0.0720715 i$ & $\ps 0.377155-0.0911473 i$ \\
14       & $\ps 0.785473-0.0441157 i$ & $\ps 0.738547-0.0525818 i$ & $\ps 0.665299-0.065595  i$ & $\ps 0.551012-0.0834295 i$ \\
16       & $\ps 0.812533-0.0424754 i$ & $\ps 0.778449-0.0492812 i$ & $\ps 0.727649-0.0595459 i$ & $\ps 0.652501-0.0738387 i$ \\
18       & $\ps 0.832913-0.0406636 i$ & $\ps 0.807023-0.046224  i$ & $\ps 0.769835-0.0544249 i$ & $\ps 0.717113-0.0657248 i$ \\
20       & $\ps 0.848837-0.0388605 i$ & $\ps 0.828482-0.0434784 i$ & $\ps 0.800111-0.0501454 i$ & $\ps 0.761233-0.0591835 i$ \\
22       & $\ps 0.861641-0.0371402 i$ & $\ps 0.845195-0.041033  i$ & $\ps 0.822839-0.046546  i$ & $\ps 0.793038-0.0538958 i$ \\
24       & $\ps 0.872174-0.0355298 i$ & $\ps 0.858591-0.0388553 i$ & $\ps 0.840512-0.0434846 i$ & $\ps 0.816958-0.0495606 i$ \\
26       & $\ps 0.881   -0.0340355 i$ & $\ps 0.869578-0.0369096 i$ & $\ps 0.854646-0.0408499 i$ & $\ps 0.835561-0.0459489 i$ \\
28       & $\ps 0.888513-0.0326539 i$ & $\ps 0.878762-0.0351635 i$ & $\ps 0.86621 -0.0385574 i$ & $\ps 0.850429-0.0428939 i$ \\\hline
$\infty$ & $\ps 0.990   -0.004     i$ & $\ps 0.989   -0.003     i$ & $\ps 0.98    -0.01      i$ & $\ps 1.0     -0.1       i$ \\
$\sigma$ & $\ps 0.008   +0.001     i$ & $\ps 0.002   +0.005     i$ & $\ps 0.02    +0.04      i$ & $\ps 0.2     +0.2       i$ \\\hline
\end{tabular}}
\caption{First quadratic identities of "ghost brane".}
\label{tab:quadratic GB}
\end{table}

\begin{table}\nonumber
\centering
\footnotesize{
\begin{tabular}{|l|llll|}\hline
Level    & $R_1                 $ & $\ps R_2                  $ & $\ps R_3                 $ & $\ps R_4                 $ \\\hline
4        & $0.090457-0.0460301 i$ & $   -0.0899649+0.0173802 i$ & $\ps 0                   $ & $\ps 0                   $ \\
5        & $0.301218-0.0654693 i$ & $   -0.0866234-0.0365613 i$ & $\ps 0                   $ & $\ps 0                   $ \\
6        & $0.465538-0.080373  i$ & $\ps 0.0718491-0.0935994 i$ & $   -0.076282-0.0510005 i$ & $\ps 0                   $ \\
7        & $0.565639-0.0718453 i$ & $\ps 0.270247 -0.0894059 i$ & $   -0.058409-0.0699729 i$ & $\ps 0                   $ \\
8        & $0.63279 -0.067889  i$ & $\ps 0.427954 -0.0908638 i$ & $\ps 0.089607-0.110411  i$ & $   -0.045421-0.0751681 i$ \\
9        & $0.680313-0.0596202 i$ & $\ps 0.531944 -0.0764609 i$ & $\ps 0.264976-0.0979118 i$ & $   -0.029185-0.0813073 i$ \\
10       & $0.715689-0.0546    i$ & $\ps 0.60491  -0.0692425 i$ & $\ps 0.411804-0.0933335 i$ & $\ps 0.108545-0.111606  i$ \\
11       & $0.74306 -0.0480294 i$ & $\ps 0.657003 -0.058987  i$ & $\ps 0.512538-0.0768525 i$ & $\ps 0.265642-0.0967383 i$ \\
12       & $0.764829-0.0435743 i$ & $\ps 0.696456 -0.0527597 i$ & $\ps 0.586407-0.0679437 i$ & $\ps 0.402852-0.089734  i$ \\
13       & $0.782601-0.0383064 i$ & $\ps 0.726734 -0.0453693 i$ & $\ps 0.639922-0.0567514 i$ & $\ps 0.499345-0.0731041 i$ \\
14       & $0.79737 -0.0344736 i$ & $\ps 0.751037 -0.0403597 i$ & $\ps 0.681392-0.0497334 i$ & $\ps 0.57262 -0.0635558 i$ \\
15       & $0.809862-0.0300677 i$ & $\ps 0.77064  -0.0346569 i$ & $\ps 0.713297-0.0417992 i$ & $\ps 0.626371-0.0521306 i$ \\
16       & $0.820569-0.0266763 i$ & $\ps 0.787034 -0.0304723 i$ & $\ps 0.739248-0.0362901 i$ & $\ps 0.668887-0.044676  i$ \\
17       & $0.829855-0.0227611 i$ & $\ps 0.80073  -0.0257045 i$ & $\ps 0.760121-0.0301303 i$ & $\ps 0.701749-0.0363905 i$ \\
18       & $0.837999-0.0195636 i$ & $\ps 0.81253  -0.0219386 i$ & $\ps 0.777729-0.0254579 i$ & $\ps 0.728826-0.0303896 i$ \\
19       & $0.845193-0.0157231 i$ & $\ps 0.822644 -0.0174798 i$ & $\ps 0.792364-0.0200423 i$ & $\ps 0.750617-0.0235758 i$ \\
20       & $0.851614-0.0122536 i$ & $\ps 0.831554 -0.0135467 i$ & $\ps 0.80505 -0.0154086 i$ & $\ps 0.769159-0.0179504 i$ \\
21       & $0.857368-0.0072802 i$ & $\ps 0.839341 -0.0079943 i$ & $\ps 0.815859-0.0090091 i$ & $\ps 0.784543-0.010376  i$ \\
22       & $0.8668              $ & $\ps 0.85094              $ & $\ps 0.830599            $ & $\ps 0.803872            $ \\
23       & $0.87672             $ & $\ps 0.862762             $ & $\ps 0.845128            $ & $\ps 0.822261            $ \\
24       & $0.883501            $ & $\ps 0.871019             $ & $\ps 0.855453            $ & $\ps 0.835521            $ \\
25       & $0.889491            $ & $\ps 0.878216             $ & $\ps 0.86432             $ & $\ps 0.846717            $ \\
26       & $0.894485            $ & $\ps 0.88426              $ & $\ps 0.871793            $ & $\ps 0.856168            $ \\\hline
$\infty$ & $1.05                $ & $\ps 1.07                 $ & $\ps 1.10                $ & $\ps 1.14                $ \\\hline
\end{tabular}}
\caption{First quadratic identities of "half ghost brane". The extrapolation to infinite level is made from data starting at level 22, where the solution turns real, and due to small number of data points we refrain from estimating the error. }
\label{tab:quadratic HGB}
\end{table}

\begin{table}\nonumber
\centering
\footnotesize{
\begin{tabular}{|l|llll|}\hline
Level    & $\ps J_1                 $ & $\ps J_2                 $ & $\ps J_3                 $ & $\ps J_4                 $ \\\hline
4        & $   -0.029033+0.0588577 i$ & $   -0.062089+0.0840112 i$ & $\ps 0                   $ & $\ps 0                   $ \\
5        & $\ps 0.157641-0.0019578 i$ & $   -0.082296+0.0227561 i$ & $\ps 0                   $ & $\ps 0                   $ \\
6        & $\ps 0.36921 -0.0483997 i$ & $\ps 0.013411-0.0534062 i$ & $   -0.073446-0.0123467 i$ & $\ps 0                   $ \\
7        & $\ps 0.503818-0.0474286 i$ & $\ps 0.201283-0.0603903 i$ & $   -0.062089-0.0337981 i$ & $\ps 0                   $ \\
8        & $\ps 0.592633-0.0504506 i$ & $\ps 0.37934 -0.0741804 i$ & $\ps 0.051782-0.086731  i$ & $   -0.0492229-0.048218 i$ \\
9        & $\ps 0.65401 -0.0449088 i$ & $\ps 0.49951 -0.0620217 i$ & $\ps 0.221449-0.0795994 i$ & $   -0.0361502-0.056114 i$ \\
10       & $\ps 0.699072-0.0430237 i$ & $\ps 0.584579-0.0583628 i$ & $\ps 0.380953-0.0820671 i$ & $\ps 0.080647-0.0955275 i$ \\
11       & $\ps 0.733041-0.037879  i$ & $\ps 0.644752-0.0491875 i$ & $\ps 0.491874-0.0666385 i$ & $\ps 0.234439-0.0838117 i$ \\
12       & $\ps 0.759885-0.0351904 i$ & $\ps 0.690398-0.0447765 i$ & $\ps 0.574293-0.0600443 i$ & $\ps 0.381096-0.0814582 i$ \\
13       & $\ps 0.781224-0.0308573 i$ & $\ps 0.72489 -0.0381386 i$ & $\ps 0.633574-0.0494779 i$ & $\ps 0.485016-0.0654704 i$ \\
14       & $\ps 0.798958-0.0281743 i$ & $\ps 0.752653-0.0342751 i$ & $\ps 0.679789-0.0437306 i$ & $\ps 0.565044-0.0575804 i$ \\
15       & $\ps 0.813552-0.0244856 i$ & $\ps 0.774633-0.0291939 i$ & $\ps 0.714895-0.0363398 i$ & $\ps 0.623335-0.0466177 i$ \\
16       & $\ps 0.826127-0.021933  i$ & $\ps 0.793117-0.0258385 i$ & $\ps 0.743619-0.0316996 i$ & $\ps 0.669794-0.0401322 i$ \\
17       & $\ps 0.836727-0.0186425 i$ & $\ps 0.808239-0.0216445 i$ & $\ps 0.76636 -0.0260717 i$ & $\ps 0.705291-0.0323255 i$ \\
18       & $\ps 0.846112-0.0161313 i$ & $\ps 0.821378-0.018557  i$ & $\ps 0.785686-0.0220926 i$ & $\ps 0.734747-0.0270492 i$ \\
19       & $\ps 0.854162-0.0129158 i$ & $\ps 0.832385-0.0146965 i$ & $\ps 0.801461-0.0172548 i$ & $\ps 0.758116-0.0207854 i$ \\
20       & $\ps 0.861441-0.0101148 i$ & $\ps 0.842192-0.0114264 i$ & $\ps 0.815264-0.0132904 i$ & $\ps 0.778161-0.0158399 i$ \\
21       & $\ps 0.867767-0.0059878 i$ & $\ps 0.850555-0.0067073 i$ & $\ps 0.826791-0.0077177 i$ & $\ps 0.794524-0.0090816 i$ \\
22       & $\ps 0.877072            $ & $\ps 0.862037            $ & $\ps 0.841547            $ & $\ps 0.814105            $ \\
23       & $\ps 0.886448            $ & $\ps 0.873301            $ & $\ps 0.855595            $ & $\ps 0.832165            $ \\
24       & $\ps 0.893261            $ & $\ps 0.881584            $ & $\ps 0.866029            $ & $\ps 0.845686            $ \\
25       & $\ps 0.899125            $ & $\ps 0.88863             $ & $\ps 0.874786            $ & $\ps 0.856865            $ \\
26       & $\ps 0.904233            $ & $\ps 0.894773            $ & $\ps 0.882412            $ & $\ps 0.866568            $ \\\hline
$\infty$ & $\ps 1.05                $ & $\ps 1.07                $ & $\ps 1.11                $ & $\ps 1.15                $ \\\hline
\end{tabular}}
\caption{First ghost current quadratic identities for the "half ghost brane". The extrapolation to infinite level is made from data starting at level 22, where the solution turns real, and due to small number of data points we refrain from estimating the error. }
\label{tab:quadraticJ HGB}
\end{table}

\begin{table}\nonumber
\centering
\footnotesize{
\begin{tabular}{|l|llll|}\hline
Level    & $\ps R_1                 $ & $\ps R_2                 $ & $\ps R_3                 $ & $\ps R_4                 $ \\\hline
4        & $   -0.0254668+0.152203 i$ & $\ps 0.010316+0.14997   i$ & $\ps 0                   $ & $\ps 0                   $ \\
5        & $\ps 0.097196+0.169702  i$ & $   -0.003524+0.153827  i$ & $\ps 0                   $ & $\ps 0                   $ \\
6        & $\ps 0.228046+0.130156  i$ & $\ps 0.013731+0.17523   i$ & $   -0.081197+0.154249  i$ & $\ps 0                   $ \\
7        & $\ps 0.283705+0.135413  i$ & $\ps 0.14829 +0.113421  i$ & $   -0.087436+0.164814  i$ & $\ps 0                   $ \\
8        & $\ps 0.353008+0.102987  i$ & $\ps 0.214195+0.116476  i$ & $   -0.089385+0.207705  i$ & $   -0.145047+0.162232  i$ \\
9        & $\ps 0.377892+0.107193  i$ & $\ps 0.288029+0.0857203 i$ & $\ps 0.033057+0.122652  i$ & $   -0.150934+0.171074  i$ \\
10       & $\ps 0.421476+0.0774607 i$ & $\ps 0.343314+0.0835458 i$ & $\ps 0.103074+0.124746  i$ & $   -0.181707+0.216821  i$ \\
11       & $\ps 0.435588+0.0796393 i$ & $\ps 0.379335+0.0705448 i$ & $\ps 0.181364+0.0807859 i$ & $   -0.067732+0.11432   i$ \\
12       & $\ps 0.465534+0.0554158 i$ & $\ps 0.417725+0.0596881 i$ & $\ps 0.250181+0.081199  i$ & $\ps 0.001883+0.116451  i$ \\
13       & $\ps 0.474552+0.0564077 i$ & $\ps 0.438632+0.0523908 i$ & $\ps 0.291544+0.0606679 i$ & $\ps 0.082549+0.0617309 i$ \\
14       & $\ps 0.496571+0.0364125 i$ & $\ps 0.465623+0.0399908 i$ & $\ps 0.342339+0.0526671 i$ & $\ps 0.161229+0.0668901 i$ \\
15       & $\ps 0.50277 +0.0367129 i$ & $\ps 0.479109+0.0351281 i$ & $\ps 0.367206+0.0408396 i$ & $\ps 0.206479+0.0399367 i$ \\
16       & $\ps 0.519855+0.019696  i$ & $\ps 0.499022+0.0229798 i$ & $\ps 0.403927+0.0305051 i$ & $\ps 0.267914+0.0364017 i$ \\
17       & $\ps 0.524333+0.019568  i$ & $\ps 0.50836 +0.0193619 i$ & $\ps 0.420253+0.0226855 i$ & $\ps 0.295996+0.0205896 i$ \\
18       & $\ps 0.538193+0.0046751 i$ & $\ps 0.52379 +0.0078462 i$ & $\ps 0.447657+0.0120293 i$ & $\ps 0.341745+0.0135492 i$ \\
19       & $\ps 0.541548+0.0042647 i$ & $\ps 0.530592+0.0049416 i$ & $\ps 0.459073+0.0063521 i$ & $\ps 0.360535+0.0030780 i$ \\
20       & $\ps 0.553226-0.0090611 i$ & $\ps 0.543094-0.0059147 i$ & $\ps 0.480348-0.0040256 i$ & $\ps 0.395226-0.0051522 i$ \\
21       & $\ps 0.555811-0.0096677 i$ & $\ps 0.548242-0.0083778 i$ & $\ps 0.488717-0.0084371 i$ & $\ps 0.408528-0.0127    i$ \\
22       & $\ps 0.565973-0.0217918 i$ & $\ps 0.558777-0.0186288 i$ & $\ps 0.505875-0.0183743 i$ & $\ps 0.435669-0.0212189 i$ \\
23       & $\ps 0.568012-0.0225408 i$ & $\ps 0.562791-0.0208028 i$ & $\ps 0.512238-0.0219828 i$ & $\ps 0.445504-0.0270113 i$ \\
24       & $\ps 0.577102-0.0337083 i$ & $\ps 0.571978-0.030511  i$ & $\ps 0.52656 -0.0314544 i$ & $\ps 0.467432-0.0354602 i$ \\\hline
$\infty$ & $\ps 0.77    -0.3       i$ & $\ps 0.7     -0.2       i$ & $\ps 0.7     -0.5       i$ & $\ps 1.3     -0.3       i$ \\
$\sigma$ & $\ps 0.08    +0.1       i$ & $\ps 0.2     +0.1       i$ & $\ps 0.3     +0.2       i$ & $\ps 1.0     +0.8       i$ \\\hline
\end{tabular}}
\caption{First quadratic identities for the "half brane".}
\label{tab:quadratic HB}
\end{table}

\begin{table}\nonumber
\centering
\footnotesize{
\begin{tabular}{|l|llll|}\hline
Level    & $\ps J_1                 $ & $\ps J_2                 $ & $\ps J_3                 $ & $\ps J_4                 $ \\\hline
4        & $   -0.055701+0.245633  i$ & $\ps 0.052169+0.16128   i$ & $\ps 0                   $ & $\ps 0                   $ \\
5        & $\ps 0.041887+0.242881  i$ & $\ps 0.023412+0.156761  i$ & $\ps 0                   $ & $\ps 0                   $ \\
6        & $\ps 0.183144+0.187835  i$ & $   -0.001256+0.196035  i$ & $   -0.058820+0.147982  i$ & $\ps 0                   $ \\
7        & $\ps 0.258308+0.171568  i$ & $\ps 0.117977+0.129273  i$ & $   -0.068422+0.154839  i$ & $\ps 0                   $ \\
8        & $\ps 0.344791+0.135071  i$ & $\ps 0.191007+0.135739  i$ & $   -0.096603+0.218656  i$ & $   -0.124789+0.148995  i$ \\
9        & $\ps 0.375522+0.129992  i$ & $\ps 0.273188+0.0929409 i$ & $\ps 0.015732+0.131073  i$ & $   -0.132311+0.155428  i$ \\
10       & $\ps 0.429112+0.0976352 i$ & $\ps 0.341968+0.0934396 i$ & $\ps 0.090717+0.136677  i$ & $   -0.184916+0.222345  i$ \\
11       & $\ps 0.445719+0.0944911 i$ & $\ps 0.381381+0.0746219 i$ & $\ps 0.174395+0.0835684 i$ & $   -0.078113+0.117994  i$ \\
12       & $\ps 0.48227 +0.0691501 i$ & $\ps 0.428946+0.0651746 i$ & $\ps 0.254169+0.0869332 i$ & $   -0.004893+0.124189  i$ \\
13       & $\ps 0.492599+0.0666558 i$ & $\ps 0.451535+0.0544234 i$ & $\ps 0.298183+0.0616357 i$ & $\ps 0.079965+0.0617566 i$ \\
14       & $\ps 0.519375+0.0465129 i$ & $\ps 0.484931+0.0433061 i$ & $\ps 0.357108+0.0555869 i$ & $\ps 0.168148+0.0703378 i$ \\
15       & $\ps 0.526376+0.0443563 i$ & $\ps 0.499373+0.0361418 i$ & $\ps 0.383441+0.0408757 i$ & $\ps 0.215736+0.0391807 i$ \\
16       & $\ps 0.547079+0.0277806 i$ & $\ps 0.523992+0.0253508 i$ & $\ps 0.426059+0.0322467 i$ & $\ps 0.284677+0.0380715 i$ \\
17       & $\ps 0.552109+0.0258395 i$ & $\ps 0.533937+0.0200956 i$ & $\ps 0.443276+0.0225003 i$ & $\ps 0.314154+0.0196965 i$ \\
18       & $\ps 0.56882 +0.0117481 i$ & $\ps 0.55297 +0.010018  i$ & $\ps 0.475086+0.0134993 i$ & $\ps 0.365502+0.0147073 i$ \\
19       & $\ps 0.572591+0.0099581 i$ & $\ps 0.560196+0.0059010 i$ & $\ps 0.487091+0.0064564 i$ & $\ps 0.385183+0.0025303 i$ \\
20       & $\ps 0.586569-0.0023416 i$ & $\ps 0.575537-0.0034609 i$ & $\ps 0.511759-0.0022734 i$ & $\ps 0.424111-0.0038056 i$ \\
21       & $\ps 0.58949 -0.0040201 i$ & $\ps 0.581005-0.0068450 i$ & $\ps 0.520547-0.0076888 i$ & $\ps 0.438019-0.0125519 i$ \\
22       & $\ps 0.601535-0.0149767 i$ & $\ps 0.593824-0.015569  i$ & $\ps 0.540377-0.0159818 i$ & $\ps 0.468434-0.0192572 i$ \\
23       & $\ps 0.603859-0.0165699 i$ & $\ps 0.598097-0.0184537 i$ & $\ps 0.547059-0.0203449 i$ & $\ps 0.478707-0.0259203 i$ \\
24       & $\ps 0.614504-0.0264809 i$ & $\ps 0.609144-0.0266206 i$ & $\ps 0.563521-0.028181  i$ & $\ps 0.503212-0.032609  i$ \\\hline
$\infty$ & $\ps 0.77    -0.1       i$ & $\ps 1.0     -0.1       i$ & $\ps 1.1     -0.22      i$ & $\ps 1.4     +0.2       i$ \\
$\sigma$ & $\ps 0.10    +0.1       i$ & $\ps 0.3     +0.2       i$ & $\ps 0.8     +0.08      i$ & $\ps 0.9     +0.3       i$ \\\hline
\end{tabular}}
\caption{First ghost current quadratic identities for the "half brane".}
\label{tab:quadraticJ HB}
\end{table}

\end{appendix}

\end{document}